

\def\ub#1{{\underbar{#1}}}

\def\id{{1\kern-.235em \text{I}}}


\catcode`\@=11
\loadbold
\catcode`\@=\active

\def\ub#1{{\underbar{#1}}}

\def\id{{1\kern-.235em \text{I}}}


\catcode`\@=11
\loadbold
\catcode`\@=\active

\input vanilla.sty
\overfullrule=0pt
\magnification=\magstep1
\vsize=23.5truecm
\hsize=16.5truecm
\baselineskip=.6truecm

\def\ub#1{{\underbar{#1}}}

\font\ti=cmbx10 scaled\magstep1
\TagsOnRight
\loadmsam
\loadmsbm
\UseAMSsymbols

$\;$

\pageno=0

\rightline{\bf ETH-TH/94-11\rm\quad}
\vskip 1truecm
\centerline{\ti{THE \ CHERN-SIMONS \ ACTION IN}}

\centerline{\ti{NON-COMMUTATIVE GEOMETRY}}\rm
\vskip 3truecm
\bf
\centerline{A.H. Chamseddine \ and \ J. Fr\"ohlich}

\centerline{Theoretical Physics}

\centerline{ETH-H\"onggerberg}

\centerline{CH-8093 \ Z\"urich}

\vskip 5truecm

\noindent Abstract\rm . \ A general definition of Chern-Simons actions
in non-commutative geometry is proposed and illustrated in several
examples. These examples are based on ``space-times'' which are
products of even-dimensional, Riemannian spin manifolds by a discrete
(two-point) set. If the $^*$algebras of operators describing the
non-commutative spaces are generated by functions over such
``space-times'' with values in certain Clifford algebras the
Chern-Simons actions turn out to be the actions of topological gravity
on the even-dimensional spin manifolds. By constraining the space of
field configurations in these examples in an appropriate manner one is
able to extract dynamical actions from Chern-Simons actions.

\vskip 1truecm

\noindent PACS No. \ A0210 / A0240 / A0350K / A0420C / A0420F
\vfill\eject

\pageno=1

\bf
\noindent 1. \ub{Introduction}\rm

During the past several years, topological field theories have been
the subject of a lot of interesting work. For example, deep connections
between three-dimensional, topological Chern-Simons theories [1], or,
equivalently, two-dimensional, chiral conformal field theories [2], on
one hand, and a large family of invariants of links, including the
famous Jones polynomial, and of three-manifolds [1], on the other
hand, have been discovered. Other topological field theories have been
invented to analyze e.g. the moduli space of flat connections on
vector bundles over Riemann surfaces or to elucidate the Donaldson
invariants of four-manifolds.
These topological field theories are formulated as theories over some
classical (topological or differentiable) manifolds.

Connes has proposed notions of non-commutative spaces generalizing,
for example, the notion of a classical differentiable manifold [4].
His theory is known under the name of ``non-commutative geometry''.
Dubois-Violette [3] and Connes have proposed to study field theories
over non-commutative spaces. In joint work with J. Lott [4], Connes
has found a construction of the classical action of the standard
model, using tools of non-commutative geometry, which yields a
geometrical interpretation of the scalar Higgs field responsible for
the ``spontaneous breaking of the electroweak gauge symmetry''. In
fact, the Higgs field appears as a component of a generalized gauge
field (connection 1-form) associated with the gauge group,
SU(2)$_w$\;$\times$\;U(1)$_{em}$,  of electroweak interactions. This is
accomplished by formulating gauge theory on a generalized space
consisting of two copies of standard Euclidian space-time the
``distance'' between which is determined by the weak scale. Although
the space-time model underlying the Connes-Lott construction is a
commutative space, it is not a classical manifold, and analysis on
space-times of the Connes-Lott type requires some of the tools of
non-commutative geometry.

The results of Connes and Lott have been reformulated and refined in
[6,7] and extended to grand-unified theories in [8]. In [9], G. Felder
and the authors have proposed some form of non-commutative Riemannian
geometry and applied it to derive an analogue of the Einstein-Hilbert
action in non-commutative geometry.

Our aim in this article is to attempt to do some steps towards a
synthesis between the different developments just described. Some of
our results have been described in our review paper [10]. We start by
presenting a general definition of the Chern-Simons action in
non-commutative geometry, (Section 2). Our definition is motivated by
some results of Quillen [11] and is based on joint work with O.
Grandjean [12]. In Section 3, we discuss a first family of examples.
In these examples, the non-commutative space is described in terms of
a $^*$algebra of matrix-valued functions over a Connes-Lott type
``space-time'', i.e., over a commutative space consisting of two
copies of an even-dimensional, differentiable spin manifold. The
Chern-Simons actions on such non-commutative spaces turn out to be
actions of topological gauge- and gravity theories, as studied in
[13,14]. In Section 3, the dimension of the continuous, differentiable
spin manifold is two, i.e., we consider products of Riemann surfaces
by discrete sets, and our Chern-Simons action is based on the
Chern-Simons 3-form.

In Section 4, we consider two- and four-dimensional topological
theories derived from a Chern-Simons action based on the Chern-Simons
5-form.

In Section 5, we describe connections of the theories found in Section
4 with four-dimensional gravity and supergravity theories.

In Section 6, we suggest applications of our ideas to string field
theory [15], and we draw some conclusions.


\vskip 1.5truecm

\bf
\ub{Acknowledgements}\rm\;. \ We thank G. Felder, K. Gaw\c edzki and
D. Kastler for their stimulating interest and O. Grandjean for very
helpful discussions on the definition of the Chern-Simons action in
non-commutative geometry and for collaboration on related matters, [12].

\vskip 1.5truecm

\bf
2. \ub{Elements of non-commutative geometry}.\rm
\medskip
This section is based on Connes' theory of non-commutative geometry,
as described in [4], and on results in [9,10,11,12].

We start by recalling the definition of a special case of Connes'
general definition of non-commutative spaces. A \ub{real, compact
non-commutative space} is defined by the data $({\Cal A}, \pi, H,
D)$, where ${\Cal A}$ is a $^*$algebra of bounded operators containing
an identity element, $\pi$ is a $^*$representation of ${\Cal A}$ on
$H$, where $H$ is a separable Hilbert space, and $D$ is a
\ub{selfadjoint} operator on $H$, with the following properties:
\medskip
\item{(i)} $\bigl[D,\pi(a)\bigr]$ is a bounded operator on $H$, for
all $a\in{\Cal A}$. [This condition determines the analogue of a
differentiable structure on the non-commutative space described by
${\Cal A}$.]

\item{} In the following, we shall usually identify the
algebra ${\Cal A}$ with the $^*$subalgebra $\pi({\Cal A})$ of the
algebra $B(H)$ of all bounded operators on $H$; (we shall thus assume
that the kernel of the representation $\pi$ in ${\Cal A}$ is trivial).
We shall often write ``$a$'' for both, the element $a$ of ${\Cal A}$
and the operator $\pi(a)$ on $H$.

\item{(ii)} $(D^2+\id)^{-1}$ is a compact operator on $H$. More
precisely, $\exp (-\varepsilon D^2)$ is trace-class, for any
$\varepsilon > 0$.
\medskip
Given a real, compact non-commutative space $({\Cal A}, \pi, H, D)$,
one defines a differential algebra, $\Omega_D({\Cal A})$, of forms as
follows: 0-forms (``scalars'') form a $^*$algebra with identity,
$\Omega_D^0({\Cal A})$, given by $\pi({\Cal A})$; $n$-forms form a
linear space, $\Omega_D^n({\Cal A})$, spanned by equivalence classes
of operators on $H$,
$$
\Omega_D^n({\Cal A})\;:=\;\Omega^n({\Cal A}) / Aux^n\;,\tag 2.1
$$
where the linear space $\Omega^n({\Cal A})$ is spanned by the
operators
$$
\bigl\{ \sum_i a_0^i [D,a_1^i]\cdots[D,a_n^i]\;:\; a_j^i \in\;
{\Cal A}\;\equiv \pi({\Cal A}),\;\forall i,j\bigr\},
\tag 2.2
$$
and $Aux^n$, the space of ``auxiliary fields'' [5], is spanned by
operators of the form
$$
\align
Aux^n\;:=\;\bigl\{\;&\sum_i [D,a_0^i][D,a_1^i]\cdots[D,a_n^i]\;:\\
&\sum_i a_0^i [D,a_1^i]\cdots [D,a_n^i]\;=\;0,\; a_j^i\in {\Cal
A}\bigr\}\;. \tag 2.3
\endalign
$$
Using the Leibniz rule
$$
[D,ab]\;=\;[D,a]b\;+\;a[D,b],\; a,b\in {\Cal A},
\tag 2.4
$$
and
$$
[D,a]^*\;=\;-\;[D,a^*],\quad a\in {\Cal A},
\tag 2.5
$$
we see that the spaces $\Omega^n({\Cal A})$ are ${\Cal A}$-bimodules
closed under the involution $^*$ and that
$$
Aux\;:=\;\oplus \;Aux^n
\tag 2.6
$$
is a two-sided ideal in
$$
\Omega ({\Cal A})\;:=\;\oplus\;\Omega^n ({\Cal A}),
\tag 2.7
$$
closed under the operation $^*$. Thus, for each $n$, $\Omega_D^n
({\Cal A})$ is an ${\Cal A}$-bimodule closed under $^*$. It follows
that
$$
\Omega_D ({\Cal A})\;:=\;\oplus\;\Omega_D^n({\Cal A})
\tag 2.8
$$
is a $^*$algebra of equivalence classes of bounded operators on $H$,
with multiplication defined as the multiplication of operators on $H$.
Since ${\Cal A} = \Omega^0({\Cal A})=\Omega_D^0({\Cal A})$ is a
$^*$subalgebra of $\Omega_D({\Cal A})$ containing an identity element,
$\Omega_D({\Cal A})$ is a unital $^*$algebra of equivalence classes \
(mod $Aux)$ of bounded operators on $H$ which is an ${\Cal
A}$-bimodule.

The degree of a form $\alpha \in \Omega_D^n({\Cal A})$ is defined by
$$
deg (\alpha)\;=\;n,\quad n\;=\;0,1,2,\cdots .
\tag 2.9
$$
Clearly, $deg (\alpha^*) = deg (\alpha)$, by (2.4), (2.5). With this
definition of $deg$, $\Omega_D({\Cal A})$ is $\Bbb Z$-graded. If
$\alpha$ is given by
$$
\alpha\;=\;\sum_i a_0^i [D,a_1^i]\cdots [D,a_n^i] (\text{mod} Aux^n) \in
\Omega_D^n ({\Cal A})
$$
we set
$$
d\alpha\;:=\;\sum_i [D,a_0^i] [D,a_1^i]\cdots [D,a_n^i] \in
\Omega_D^{n+1} ({\Cal A})\;. \tag 2.10
$$
The map
$$
d\;:\;\Omega_D^n ({\Cal A}) \;\rightarrow\;\Omega_D^{n+1} ({\Cal
A}), \quad \alpha\;\mapsto\; d\alpha
\tag 2.11
$$
is a $\Bbb C$-linear map from $\Omega_D({\Cal A})$ to itself which
increases the degree of a form by one and satisfies
$$
d(\alpha\cdot\beta)\;=\;(d\alpha)\cdot\beta\;+\;(-1)^{deg\,\alpha}\;
\alpha\cdot(d\beta),
\tag 2.12
$$
for any homogeneous element $\alpha$ of $\Omega_D({\Cal A})$ (Leibniz
rule) and
$$
d^2\;=\;0\;.
\tag 2.13
$$
Hence $\Omega_D({\Cal A})$ is a differential algebra which is a $\Bbb
Z$-graded complex.

These notions are described in detail (and in a more general setting)
in [4].

In non-commutative geometry,\  \ub{vector bundles} \ over a
non-commutative space described by a $^*$algebra ${\Cal A}$ are
defined as \ub{finitely generated}, \ub{projective left} ${\Cal
A}$-\ub{modules}. Let $E$ denote (the ``space of sections'' of) a
vector bundle over ${\Cal A}$. A connection $\nabla$ on $E$ is a
$\Bbb C$-linear map
$$
\nabla\;:\;E\;\rightarrow\;\Omega_D^1 ({\Cal A}) \;\otimes_{{\Cal
A}}\;E
\tag 2.14
$$
with the property that (with $da = [D,a]$, for all $a\in{\Cal A}$)
$$
\nabla (as)\;=\;da\;\otimes_{{\Cal A}}\;s + a\;\nabla s,
\tag 2.15
$$
for arbitrary $a\in{\Cal A}$, $s\in E$. The definition of $\nabla$
can be extended to the space
$$
\Omega_D(E)\;=\;\Omega_D({\Cal A})\;\otimes_{{\Cal A}}\;E
\tag 2.16
$$
in a canonical way, and, for $s\in \Omega_D(E)$ and a homogeneous form
$\alpha\in\Omega_D({\Cal A})$,
$$
\nabla(\alpha s)\;=\;(d\alpha) s\;+\;(-1)^{deg\,\alpha}\;\alpha \;
\nabla s .\tag 2.17
$$
Thanks to (2.14) - (2.17), it makes sense to define the
\ub{curvature}, $R(\nabla)$, of the connection $\nabla$ as the $\Bbb
C$-linear map
$$
R(\nabla)\;:=\;- \nabla^2
\tag 2.18
$$
from $\Omega_D(E)$ to $\Omega_D(E)$. Actually, it is easy to check
that $R(\nabla)$ is ${\Cal A}$-\ub{linear}, i.e. $R(\nabla)$ is a
\ub{tensor}.

A \ub{trivial} vector bundle, $E^{(N)}$, corresponds to a finitely
generated, \ub{free} left ${\Cal A}$-module, i.e., one that has a
basis $\{ s_1,\cdots, s_N\}$, for some finite $N$ called its
dimension. Then
$$
E^{(N)}\; \simeq\; {\Cal A} \oplus \cdots \oplus {\Cal A}\;\equiv\;
{\Cal A}^n,
$$
(with $N$ summands). The affine space of connections on $E^{(N)}$ can
be characterized as follows: Given a basis $\{ s_1,\cdots, s_N\}$ of
$E^{(N)}$, there are $N^2$ 1-forms $\rho_\alpha^\beta \in \Omega_D^1
({\Cal A})$, the \ub{components} of the connection $\nabla$, such that
$$
\nabla s_\alpha\;=\;-\;\rho_\alpha^\beta \otimes_{{\Cal A}} s_\beta,
\tag 2.19
$$
(where, here and in the following, we are using the summation
convention). Then
$$
\nabla (a^\alpha s_\alpha)\;=\;da^\alpha \otimes_{{\Cal A}} s_\alpha -
a^\alpha \rho_\alpha^\beta \otimes_{{\Cal A}} s_\beta,
\tag 2.20
$$
by (2.15). Furthermore, by (2.18) and (2.20),
$$
\align
R(\nabla)(a^\alpha s_\alpha)\;=\;
&-\;\nabla \bigl( da^\alpha \otimes_{{\Cal A}} s_\alpha - a^\alpha
\rho_\alpha^\beta \otimes_{{\Cal A}} s_\beta\bigr) \\
=\;&-\;\bigl(d^2 a^\alpha \otimes_{{\Cal A}} s_\alpha + da^\alpha
\rho_\alpha^\beta \otimes_{{\Cal A}} s_\beta \\
&-\;da^\alpha \rho_\alpha^\beta \otimes_{{\Cal A}} s_\beta - a^\alpha
d \rho_\alpha^\beta \otimes_{{\Cal A}} s_\beta \\
&-\; a^\alpha \rho_\alpha^\gamma \rho_\gamma^\beta \otimes_{{\Cal A}}
s_\beta\bigr)\\
=\;&\;a^\alpha \bigl(d \rho_\alpha^\beta + \rho_\alpha^\gamma
\rho_\gamma^\beta\bigr) \otimes_{{\Cal A}} s_\beta\;.
\tag 2.21
\endalign
$$
Thus, the curvature tensor $R(\nabla)$ is completely determined by the
$N\times N$ matrix $\theta\equiv(\theta_\alpha^\beta)$ of 2-forms
given by
$$
\theta_\alpha^\beta\;=\;d \rho_\alpha^\beta\;+\;\rho_\alpha^\gamma
\;\rho_\gamma^\beta\;.
\tag 2.22
$$
The curvature matrix $\theta$ satisfies the \ub{Bianchi identity}
$$
d\theta + \rho\theta - \theta\rho\;\equiv\;\bigl( d\theta_\alpha^\beta
+ \rho_\alpha^\gamma \theta_\gamma^\beta - \theta_\alpha^\gamma
\rho_\gamma^\beta\bigr)\;=\;0\;.
\tag 2.23
$$
If one introduces a new basis
$$
\tilde s_\alpha\;=\;M_\alpha^\beta s_\beta, \; M_\alpha^\beta \in {\Cal A},
\; \alpha,\beta\;=\;1,\cdots,N,
\tag 2.24
$$
where the matrix $M\equiv (M_\alpha^\beta)$ is invertible, then the
components, $\tilde\rho$, of $\nabla$ in the new basis
$\{\tilde s_1, \cdots, \tilde s_N\}$ of $E^{(N)}$ are given by
$$
\tilde\rho\;=\;M\rho\;M^{-1}\;-\;dM\cdot M^{-1},
\tag 2.25
$$
and the components of the curvature $R(\nabla)$ transform according to
$$
\widetilde\theta\;=\; M \theta M^{-1},
\tag 2.26
$$
as one easily checks.

Given a basis $\{ s_1,\cdots, s_N\}$ of $E^{(N)}$, one may define a
\ub{Hermitian structure} $\langle\cdot,\cdot\rangle$  on $E^{(N)}$ by
setting
$$
\langle s_\alpha, s_\beta\rangle\;=\;\delta_{\alpha\beta}\;\id \;,
\tag 2.27
$$
with
$$
\langle a^\alpha s_\alpha, b^\beta s_\beta\rangle\;=\; a^\alpha
\langle s_\alpha, s_\beta\rangle (b^\beta)^*\;=\; \sum_\alpha a^\alpha
(b^\alpha)^* \in {\Cal A}\;.
\tag 2.28
$$
The definition of $\langle\cdot ,\cdot\rangle$ can be extended
canonically to $\Omega_D(E^{(N)})$, and there is then an obvious
notion of ``\ub{unitary connection}'' on $E^{(N)}$: $\nabla$ is
unitary iff
$$
d\langle s,s'\rangle\;=\;\langle\nabla s, s'\rangle\;-\;\langle s,
\nabla s'\rangle\;.
\tag 2.29
$$
This is equivalent to the condition that
$$
\rho_\alpha^\beta\;=\;\bigl( \rho_\beta^\alpha\bigr)^*,
\tag 2.30
$$
where the $\rho_\alpha^\beta$ are the components of $\nabla$ in the
orthonormal basis $\{ s_1,\cdots , s_N\}$ of $E^{(N)}$.

In the examples studied in Sections 3 through 5, we shall consider
unitary connections on trivial vector bundles, in particular on
``\ub{line bundles}'' for which $N=1$. A (unitary) connection $\nabla$
on a line bundle $E^{(1)} \simeq {\Cal A}$ is completely determined
by a (selfadjoint) 1-form $\rho \in \Omega_D^1 ({\Cal A})$.

The data $({\Cal A},\pi,H,D)$ defining a non-commutative space with
differentiable structure is also called a \ub{Fredholm module}.
Following [4], we shall say that the Fredholm module $({\Cal
A},\pi,H,D)$ is $(d,\infty)$-summable if
$$
tr\,\bigl(D^2 + \id\bigr)^{- p/2} \;<\;\infty , \quad \text{for
all}\quad p > d\;.
\tag 2.31
$$
Let $Tr_\omega (\cdot)$ denote the so-called \ub{Dixmier trace} on
$B(H)$ which is a positive, cyclic trace vanishing on trace-class
operators; see [4]. We define a notion of \ub{integration of forms},
$\int\!\!\!\!\!-$, by setting
$$
\int\!\!\!\!\!\!- \alpha\;:=\; Tr_\omega\;\bigl( \alpha\mid
D\mid^{-d}\bigr),
\tag 2.32
$$
for \ $\alpha \in \Omega ({\Cal A}) = \oplus \Omega^n ({\Cal A})$; \
(see (2.2), (2.7)). If $d=\infty$ but \ exp$\;(-\varepsilon D^2)$ is
trace class, for any $\varepsilon > 0 $ (as assumed), we set
$$
\int\!\!\!\!\!\!-
\alpha\;:=\;\displaystyle\mathop{Lim_\omega}_{\varepsilon\downarrow
0}\; \frac{tr \bigl(\alpha \exp (-\varepsilon D^2)\bigr)}{tr
\bigl(\exp (- \varepsilon D^2)\bigr)}\;,
\tag 2.33
$$
(on forms $\alpha$ which are ``
analytic elements'' for the automorphism group determined by the
dynamics \ exp$(it D^2), t\in\Bbb R$; see [12]) and $Lim_\omega$
denotes a limit defined in terms of a kind of ``Cesaro mean''
described in [4]. Then
$$
\int\!\!\!\!\!\!- \alpha\beta\;=\;\int\!\!\!\!\!\!- \beta\alpha,
\tag 2.34
$$
i.e., $\int\!\!\!\!\!-$ is cyclic; it is also a \ub{non-negative}
linear functional on $\Omega ({\Cal A})$. It can thus be used to
define a positive semi-definite inner product on $\Omega({\Cal A})$:
For $\alpha$ and $\beta$ in $\Omega({\Cal A})$, we set
$$
(\alpha,\beta)\;=\;\int\!\!\!\!\!\!- \alpha\beta^*.
\tag 2.35
$$
Then the closure of $\Omega({\Cal A})$ (mod kernel of $(\cdot,\cdot)$)
in the norm determined by $(\cdot,\cdot)$ is a Hilbert space, denoted
by $L^2\bigl(\Omega({\Cal A})\bigr)$. Given an element
$\alpha\in\Omega^n({\Cal A})$, we can now define a canonical
representative, $\alpha^\perp$, in the equivalence class $\alpha$ (mod
$Aux^n) \in \Omega_D^n({\Cal A})$ as the unique (modulo the kernel of
$(\cdot,\cdot)$) operator in $\alpha$ (mod $Aux^n)$ which is
orthogonal to $Aux^n$ in the scalar product $(\cdot,\cdot)$ given by
(2.35); $(Aux^n$ has been defined in eq. (2.3)). Then, for $\alpha$
and $\beta$ in $\Omega_D({\Cal A})$, we set
$$
(\alpha,\beta)\;:=\;(\alpha^\perp,
\beta^\perp)\;\equiv\;\int\!\!\!\!\!\!- \alpha^\perp (\beta^\perp)^*,
\tag 2.36
$$
and this defines a positive semi-definite inner product on
$\Omega_D({\Cal A})$. The closure of $\Omega_D({\Cal A})$ (mod kernel
of $(\cdot,\cdot))$ in the norm determined by $(\cdot,\cdot)$ is the
Hilbert space of ``square-integrable differential forms'', denoted by
$\Lambda_D({\Cal A})$.

In order to define the Chern-Simons forms and Chern-Simons actions in
non-commut\-ative geometry, it is useful to consider a trivial example
of the notions introduced, so far. Let $I$ denote the interval
$[0,1]\subset\Bbb R$. Let ${\Cal A}_1 = C^\infty (I)$ be the algebra
of smooth functions, $f(t)$, on the open
interval (0,1) which, together with all their derivatives in $t$, have
(finite) limits as $t$ tends to 0 or 1. Let $H_1=L^2(I)\otimes\Bbb
C^2$ denote the Hilbert space of square-integrable (with respect to
Lebesgue measure, $dt$, on $I$) two-component spinors, and $D_1 = i
\frac{\partial}{\partial t} \otimes \sigma_1$ the one-dimensional
Dirac operator (with appropriate selfadjoint boundary conditions),
where $\sigma_1,\sigma_2$ and $\sigma_3$ are the usual Pauli matrices.
A representation $\pi_1$ of ${\Cal A}_1$ on $H_1$ is defined by
setting
$$
\pi_1(a)\;=\;a \otimes \id_2,\quad a\in{\Cal A}_1.
\tag 2.37
$$
The geometry of $I$ is then coded into the space $({\Cal A}_1,
\pi_1,H_1,D_1)$. The space of 1-forms is given by
$$
\Omega_{D_1}^1({\Cal A}_1)\;=\;\bigl\{ \omega\otimes\sigma_1 : \omega
= \sum_i a^i \partial_t b^i; a^i, b^i \in {\Cal A}_1\bigr\}.
\tag 2.38
$$
The space, $\Omega_{D_1}^2 ({\Cal A}_1)$, of 2-forms is easily seen to
be \ub{trivial}, and the cohomology groups vanish. The Fredholm module
$({\Cal A}_1,\pi_1, H_1, D_1)$ is $\Bbb Z_2$-graded.
The $\Bbb Z_2$-grading, $\gamma$, is given by
$$
\gamma\;=\;\id \otimes \sigma_3,
\tag 2.39
$$
and $[\gamma, \pi_1 (a)] = 0$, for all $a\in {\Cal A}_1$, while
$$
\bigl\{ \gamma, D_1\bigr\}\;\equiv\;\gamma D_1 + D_1\gamma \;=\;0.
\tag 2.40
$$

Using this trivial example, we may introduce the notion of a \
``\ub{cylinder over a non-} \ub{commutative space}'': Let $({\Cal
A},\pi,H,D)$ be an arbitrary non-commutative space, and let $({\Cal
A}_1,\pi_1,H_1,D_1)$ be as specified in the above example. Then we
define the cylinder over $({\Cal A},\pi,H,D)$ to be given by the
non-commutative space $(\widetilde{\Cal A},\widetilde\pi,\widetilde H,
\widetilde D)$, where
$$
\widetilde H = H\otimes H_1,\quad
\widetilde\pi = \pi \otimes \pi_1,\quad
\widetilde{\Cal A} = {\Cal A}\otimes{\Cal A}_1,
\tag 2.41
$$
and
$$
\widetilde D\;=\;\id \otimes D_1\;+\;D \otimes \gamma,
\tag 2.42
$$
with $\gamma$ as in (2.39). The space $(\widetilde{\Cal
A},\widetilde\pi,\widetilde H, \widetilde D)$ is $\Bbb Z_2$-graded: We
define
$$
\Gamma\;=\;\id \otimes (\id \otimes \sigma_2),
\tag 2.43
$$
$\widetilde D_1 := \id\otimes D_1$, $\widetilde D_2 = D\otimes\gamma$. Then
$$
\{ \Gamma,\widetilde D_1\}\;=\;\{ \Gamma,\widetilde
D_2\}\;=\;\{\Gamma,\widetilde D\}\;=\; \{\widetilde D_1, \widetilde
D_2\}\;=\;0,
\tag 2.44
$$
and
$$
\bigl[ \Gamma, \widetilde\pi (\tilde a)\bigr]\;=\;0,\quad\text{for all}
\quad \tilde a\in\widetilde{\Cal A}.
\tag 2.45
$$
It is easy to show (see [12]) that arbitrary sums of operators of the
form
$$
\tilde a_0 [\widetilde D_{\varepsilon_1}, \tilde a_1] \cdots
[\widetilde D_{\varepsilon_n}, \tilde a_n],\quad \varepsilon_1,
\cdots, \varepsilon_n = 1,2,
\tag 2.46
$$
belong to $\Omega^k(\widetilde{\Cal A})$. Furthermore, if two or more
of the $\varepsilon_i$'s take the value 1 then the operator defined in
(2.46) belongs to $Aux^n$.

We define integration, $\int\!\!\!\!\!\!\sim (\cdot)$, on $(\widetilde{\Cal
A},\widetilde\pi,\widetilde H,\widetilde D)$ by setting, for any
$\alpha\in\Omega (\widetilde{\Cal A})$,
$$
\int\!\!\!\!\!\!\!\sim \alpha\;:=\;\int_0^1 dt \int\!\!\!\!\!\!- Tr_{\Bbb
C^2} \bigl(\alpha (t)\bigr),
\tag 2.47
$$
where $\alpha (t)$ is a 2$\times$2 matrix of elements of $\Omega({\Cal
A})$. The integral $\int\!\!\!\!\!\!\sim (\cdot)$ is positive semi-definite
and cyclic on the algebra $\Omega (\widetilde{\Cal A})$. We are now
prepared to define the Chern-Simons forms and Chern-Simons actions in
non-commutative geometry, (for connections on trivial vector bundles).
Let $({\Cal A},\pi,H,D)$ be a real, compact non-commutative space with
a differentiable structure determined by $D$. Let $E=E^{(N)} \simeq
{\Cal A}^N$ be a trivial vector bundle over ${\Cal A}$, and let
$\nabla$ be a connection on $E$. By (2.19), $\nabla$ is completely
determined by an $N\times N$ matrix $\rho = (\rho_\alpha^\beta)$ of
1-forms. By (2.21), the curvature of $\nabla$ is given by the $N\times
N$ matrix of 2-forms
$$
\theta\;=\;d \rho + \rho^2,
$$
where $d$ is the differential on $\Omega_D ({\Cal A})$ defined in
(2.10). Following Quillen [11], we define the Chern-Simons
(2$n$--1)-form associated with $\nabla$ as follows: Let $\nabla_0$
denote the flat connection on $E$ corresponding to an $N\times N$
matrix $\rho_0$ of 1-forms which, in an appropriate gauge, vanishes.
We set
$$
\rho_t\;=\;t \rho + (1-t)\;\rho_0\;=\; t \rho,
\tag 2.48
$$
for $\rho_0=0$, corresponding to the connection $\nabla_t = t\nabla +
(1-t) \nabla_0$. The curvature of $\nabla_t$ is given by the matrix
$\theta_t$ of 2-forms given by
$$
\theta_t\;=\;d\rho_t + \rho_t^2\;=\;t d\rho + t^2\rho^2.
$$
The Chern-Simons $(2n-1)$-form associated with $\nabla$ is then given
by
$$
\vartheta^{2n-1} (\rho)\;:=\;\frac{1}{(n-1)!} \int_0^1 dt \;\rho\;
\theta_t^{n-1}. \tag 2.49
$$
For $n=2$, we find
$$
\vartheta^3 (\rho)\;=\;\frac 1 2\;\bigl\{ \rho d \rho + \frac 2 3
\;\rho^3 \bigr\},
\tag 2.50
$$
and, for $n=3$,
$$
\vartheta^5 (\rho)\;=\;\frac 1 6 \bigl\{ \rho d\rho d\rho + \frac 3 4
\rho^3 d \rho + \frac 3 4 \rho (d\rho) \rho^2 + \frac 3 5 \rho^5
\bigr\}.
\tag 2.51
$$

In order to understand where these definitions come from and how to
define Chern-Simons actions, we extend $E$ to a trivial vector bundle
over the cylinder $(\widetilde{\Cal A},\widetilde\pi,\widetilde H,
\widetilde D)$ over $({\Cal A}, \pi, H, D)$: We set
$$
\widetilde E\;=\; E \otimes C^\infty (I) \otimes \id_2 \simeq
\widetilde{\Cal A}^N .
\tag 2.52
$$
We also extend the connection $\nabla$ on $E$ to a connection
$\widetilde\nabla$ on $\widetilde E$ by interpolating between $\nabla$
and the flat connection $\nabla_0$: By (2.39), (2.41) and (2.42), a
1-form in $\Omega_{\widetilde D}^1 (\widetilde{\Cal A})$ is given by
$$
\pmatrix
\rho (t) & \quad \phi (t) \\
\phi (t) & - \rho (t)
\endpmatrix ,
$$
where $\rho (t)\in \Omega_D^1({\Cal A})$ for all $t\in I$. Thus
$$
\tilde\rho_\alpha^\beta (t)\;:=\;t\enskip
\pmatrix
\rho_\alpha^\beta & \enskip 0 \\
0 & - \rho_\alpha^\beta
\endpmatrix , \quad
\alpha,\beta = 1, \cdots, N,
\tag 2.53
$$
is an element of $\Omega_{\widetilde D}^1 (\widetilde{\Cal A})$. We
define $\widetilde \nabla$ to be the connection on $\widetilde E$
determined by the matrix $\widetilde \rho (t) = \bigl(
\widetilde\rho_\alpha^\beta (t)\bigr)$ of 1-forms defined in (2.53).
Let $\tilde d$ be the differential on $\Omega_{\widetilde
D}(\widetilde{\Cal A})$ defined as in (2.10), (with ${\Cal A}$
replaced by $\widetilde{\Cal A}$ and $D$ replaced by $\widetilde D$).
By (2.42)
$$
\tilde d \tilde\rho \;=\;
\pmatrix 0 &- i\rho \\
i\rho &\enskip 0
\endpmatrix\;+\;t\;
\pmatrix
d\rho & 0 \\
\enskip 0 & d\rho
\endpmatrix ,
\tag 2.54
$$
with $\rho = (\rho_\alpha^\beta)$. Hence the curvature of
$\widetilde\nabla$ is
given by the matrix of 2-forms $\widetilde\theta$, with
$$
\widetilde\theta (t)\;=\;\theta_t\otimes\id_2 + \rho\otimes\sigma_2,
\tag 2.55
$$
where
$$
\theta_t\;=\;t d \rho + t^2 \rho^2.
$$
Let $\varepsilon$ be an arbitrary bounded operator on $H$ commuting
with $D$ and with all operators in $\pi ({\Cal A})$. Then,
$$
\widetilde\varepsilon\;:=\;\varepsilon\otimes\id_2
$$
commutes with $\widetilde D$ and with $\widetilde\pi (\widetilde{\Cal
A})$ and hence with $\Omega(\widetilde{\Cal A})$. It also commutes
with the $\Bbb Z_2$-grading $\Gamma =\id\otimes (\id \otimes \sigma_2)$, (as
defined in (2.43)). We now define a graded trace $\tau_\epsilon
(\cdot)$ on $\Omega(\widetilde{\Cal A})$ by setting
$$
\tau_\varepsilon (\alpha)\;:=\;\int\!\!\!\!\!\!\!\sim (\tilde\varepsilon
\Gamma \alpha), \quad \alpha \in\Omega (\widetilde{\Cal A}),
\tag 2.56
$$
where $\int\!\!\!\!\!\!\sim (\cdot)$ is given by (2.47). It is easy to show
that
$$
\tau_\varepsilon (\alpha)\;=\;0 \quad \text{if deg } \alpha\quad
\text{is \ub{odd}},
\tag 2.57
$$
and
$$
\tau_\varepsilon\bigl([\alpha,\beta]_*\bigr)\;=\;0,\quad
\text{for all } \alpha,\beta\quad\text{in } \Omega(\widetilde{\Cal
A}), \tag 2.58
$$
where
$$
[\alpha,\beta]_*\;=\;\alpha\cdot\beta - (-1)^{\deg \alpha \deg
\beta}\; \beta\cdot\alpha
$$
is the graded cummutator.

Using the Bianchi identity,
$$
d\theta^n + [\rho, \theta^n] = 0,
$$
which follows from eq.~(2.23) by induction, and the graded cyclicity
of $\tau_\varepsilon$ (see (2.57), (2.58)) one shows that
$$
\tau_\varepsilon \bigl((\widetilde\theta^n)^\perp\bigr)\;=\; n!
\tau_\varepsilon \bigl(\bigl(\tilde d \vartheta^{2n-1}
(\widetilde\rho)\bigr)^\perp\bigr),
\tag 2.59
$$
where $\alpha^\perp$ is the canonical representative in the
equivalence class $\alpha$ (mod $Aux^m) \in \Omega_{\widetilde D}^m
(\widetilde{\Cal A})$ orthogonal to $Aux^m$, for any $m=1,2,\cdots$,
(as explained after eq.~(2.35)). The calculation proving (2.59) is
indicated in [11]; (see also [12] for details). In fact, eq.~(2.59) is
a general identity valid for arbitrary connections on trivial vector
bundles over a non-commutative space and arbitrary graded traces [11].

In the case considered here, the l.h.s. of eq.~(2.59) can be rewritten
in the following interesting way:
$$
\align
\tau_\varepsilon\bigl((\widetilde\theta^n)^\perp\bigr)\;&=\;
\int_0^1 dt \int\!\!\!\!\!\!- Tr_{\Bbb C^2} \bigl(\tilde\varepsilon\; \Gamma
\bigl( \widetilde\theta^n (t)\bigr)^\perp\bigr) \\
&=\;n \int_0^1 dt \int\!\!\!\!\!\!- Tr_{\Bbb C^2}
\bigl((\varepsilon\otimes\id_2)\, \Gamma\, (\rho\otimes\sigma_2)(\theta_t^{n-1}
\otimes \id_2)\bigr).
\tag 2.60
\endalign
$$
This is shown by plugging eq.~(2.55) for $\widetilde\theta (t)$ into
the expression in the middle of (2.60) and noticing that (1) all terms
contributing to $\widetilde\theta^n(t)$ with more than one factor
proportional to $[\widetilde D_1, \tilde a]$, i.e., with more than one
factor of the form $\rho\otimes\sigma_2$, are projected out when
passing from $\widetilde\theta^n(t)$ to $\bigl(\widetilde\theta^n
(t)\bigr)^\perp$, (see the remark following eq.~(2.46)), and \ (2) \
$Tr_{\Bbb C^2} \bigl((\varepsilon\otimes\id_2)\; \Gamma$ \break
$(\theta_t^n\otimes\id_2)\bigr)=0$.

Evaluating the trace, $Tr_{\Bbb C^2}$, on the r.h.s. of (2.60) and
recalling the definition (2.49) of the Chern-Simons form, we finally
conclude that
$$
\align
\tau_\varepsilon\bigl((\widetilde\theta^n)^\perp\bigr)\;
&=\;2n \int_0^1 dt \int\!\!\!\!\!\!-
\bigl(\varepsilon\rho\theta_t^{n-1}\bigr)^\perp \\
&=\;2n! \int\!\!\!\!\!\!- \bigl(\varepsilon\bigl(\vartheta^{2n-1}
(\rho)\bigr)^\perp\bigr).
\tag 2.61
\endalign
$$
\bf\ub{Remark}\rm . The r.h.s. of (2.59) can actually be rewritten as
$$
n! \int\!\!\!\!\!\!- Tr_{\Bbb C^2} \bigl( \id \otimes \sigma_3 \bigl(
\vartheta^{2n-1} \bigl( \widetilde\rho (1)\bigr)\bigr)^\perp\bigr);
$$
see [12].

\ub{Chern-Simons actions}, $I_\varepsilon$, in non-commutative
geometry are defined by setting
$$
I_\varepsilon^{2n-1}(\rho)\;:=\;\kappa\int\!\!\!\!\!\!-
\bigl(\varepsilon\bigl( \vartheta^{2n-1} (\rho)\bigr)^\perp\bigr),
\tag 2.62
$$
where $\kappa$ is a constant. Using (2.50) and (2.51) and using the
properties of $\varepsilon$ and the cyclicity of $\int\!\!\!\!\!-
(\cdot)$, we find
$$
I_\varepsilon^3 (\rho)\;=\;\frac\kappa 2 \int\!\!\!\!\!\!-
\bigl(\varepsilon\bigl( \rho d \rho + \frac 2 3
\rho^3\bigr)^\perp\bigr),
$$
and
$$
I_\varepsilon^5 (\rho)\;=\;\frac\kappa 6 \int\!\!\!\!\!\!-
\bigl(\varepsilon \bigl( \rho d \rho d \rho + \frac 3 2 \rho^3 d \rho
+ \frac 3 5 \rho^5 \bigr)^\perp\bigr).
\tag 2.63
$$
A particularly important special case is obtained by choosing the
operator $\varepsilon$ to belong to $\Omega({\Cal A})$.
Since $\varepsilon$ commutes with $D$ and with $\pi({\Cal A})$, this
implies that $\varepsilon$ belongs to the \ub{centre} of the algebra
$\Omega({\Cal A})$. In the examples discussed in the remainder of this
paper, this property is always assumed.

One point of formula (2.61) and generalizations thereof, discussed in
[12] (and involving ``higher-dimensional cylinders''), is that it
enables us to define \ub{differences} of Chern-Simons actions even
when the underlying vector bundle is \ub{non-trivial}. If $\nabla_0$
denotes a fixed reference connection on a vector bundle $E$ over a
non-commutative space $({\Cal A},\pi,H,D)$ and $\nabla$ is an
arbitrary connection on $E$ we set
$$
\int\!\!\!\!\!\!- \bigl(\varepsilon\bigl(\vartheta^{2n-1}
(\nabla)\bigr)^\perp\bigr)\;:=\;\tau_\varepsilon \bigl(
(\widetilde\theta^n)^\perp\bigr)\;+\;\text{const.,}
\tag 2.64
$$
where $\widetilde\theta$ is the curvature of a connection
$\widetilde\nabla$ on a vector bundle $\widetilde E$ over the cylinder
$(\widetilde{\Cal A},\widetilde\pi,\widetilde H, \widetilde D)$
interpolating between $\nabla$ and $\nabla_0$, and the constant on the
r.h.s. of (2.64) is related to the choice of $\widetilde\nabla$ and of
the Chern-Simons action associated with $\nabla_0$.

Formulas (2.62) and (2.64) are helpful in understanding the
topological nature of Chern-Simons actions.

Next, we propose to discuss various concrete examples and indicate
some applications to theories of gravity.

\vskip 1.5truecm
\bf 3. \underbar{Some ``three-dimensional'' Chern-Simons actions}\rm.
\medskip

We consider a ``Euclidean space-time manifold'' $X$ which is the
Cartesian product of a Riemann surface $M_2$ and a two-point set, i.e.
$X$ consists of two copies of $M_2$. The algebra ${\Cal A}$ used in
the definition of the non-commutative space considered in this section
is given by
$$
{\Cal A}\;=\;C^\infty (M_2) \otimes {\Cal A}_0,
\tag 3.1
$$
where ${\Cal A}_0$ is a finite-dimensional, unital $^*$algebra of
$M\times M$ matrices. The Hilbert space $H$ is chosen to be
$$
H\;=\;H_0\oplus H_0,
\tag 3.2
$$
where
$$
H_0\;=\;\Bbb C^N \otimes L^2 (S) \otimes \Bbb C^M,
\tag 3.3
$$
and $L^2(S)$ is the Hilbert space of square-integrable spinors on
$M_2$ for some choices of a spin structure and of a (Riemannian)
volume form on $M_2$.

The representation $\pi$ of ${\Cal A}$ on $H$ is given by
$$
\pi (a)\;=\;
\pmatrix
\id_N \otimes a & \quad 0 \\
\quad 0 & \id_N\otimes a
\endpmatrix ,
\tag 3.4
$$
for $a \in {\Cal A}$.

We shall work locally over some coordinate chart of $M_2$, but we do
not describe how to glue together different charts (this is standard),
and we shall write ``$M_2$'' even when we mean a coordinate chart of
$M_2$. Let $g=(g_{\mu\nu})$ be some fixed, Riemannian reference metric
on $M_2$, and $(e_\mu^a)$ a section of orthonormal 2-frames, $\mu,a =
1,2$. Let $\gamma^1, \gamma^2$ denote the two-dimensional Dirac
matrices satisfying
$$
\bigl\{ \gamma^a, \gamma^b\bigr\}\;\equiv\;
\gamma^a\gamma^b + \gamma^b\gamma^a\;=\;2\delta^{ab},
\tag 3.5
$$
and
$$
\gamma^5\;=\;\gamma^1 \gamma^2.
\tag 3.6
$$
The matrices $\id_N\otimes\gamma^a$ will henceforth also be denoted by
$\gamma^a$. Let $\partial\!\!\!/$
denote the covariant Dirac operator on $\Bbb C^N \otimes L^2(S)$
corresponding to the Levi-Civita spin connection determined by
$(e_\mu^a)$ and acting trivially on $\Bbb C^N$. Let $K$ denote an
operator of the form
$$
K\;=\;k\otimes\id\otimes\id,
\tag 3.7
$$
where $k$ is some real, symmetric $N\times N$ matrix. The vector space
$\Bbb C^N$ and the matrix $k$ do not play any interesting role in the
present section but are introduced for later convenience. Let $\phi_0$
be a hermitian $M\times M$ matrix $(\neq \id_M)$. The operator $D$ on
$H$ required in the definition of a non-commutative space is chosen as
$$
D\;=\; \pmatrix
\partial\!\!\!/ \otimes \id_M\quad & i\gamma^5 K\otimes \phi_0 \\
- i\gamma^5 K\otimes\phi_0 & \partial\!\!\!/ \otimes \id_M \endpmatrix ,
\tag 3.8
$$
Then (\ub{locally} on $M_2$) the space of 1-forms, $\Omega_D^1 ({\Cal
A})$, (the ``cotangent bundle'') is a free, hermitian ${\Cal
A}$-bimodule of dimension 3, with an orthonormal basis given by
$$
\varepsilon^a\;=\;\pmatrix
\gamma^a \otimes \id_M \quad & \quad 0 \quad\\
\quad 0\quad  & \gamma^a \otimes \id_M \endpmatrix ,
\quad a\;=\;1,2,
\tag 3.9
$$
and
$$
\varepsilon^3\;=\; \pmatrix
\quad 0 \qquad & - \gamma^5 \otimes \id_M \\
\gamma^5 \otimes \id_M & \quad 0 \endpmatrix
\phantom{\quad a\;=\;1,1,\quad }
\tag 3.10
$$
and the hermitian structure is given by the normalized trace, $tr$, on
$\Bbb M_N(\Bbb C)\otimes$ Cliff. Then, for $a,b=1,2,3,$
$$
\langle \varepsilon^a,\varepsilon^b\rangle\;=\;tr\,\bigl(\varepsilon^a
(\varepsilon^b)^*\bigr)\;=\;\delta^{ab}\id_M.
$$
We define a central element $\varepsilon\in\Omega^3({\Cal A})$ by
setting
$$
\varepsilon\;=\;\varepsilon^1\varepsilon^2\varepsilon^3\;=\;\pmatrix
\quad 0\quad & \id\\
- \id & 0\endpmatrix .
\tag 3.11
$$
It is trivial to verify that $\varepsilon $ commutes with the operator
$D$ and with $\pi({\Cal A})$, and, since $\varepsilon^1,\varepsilon^2$
and $\varepsilon^3$ belong to $\Omega^1 ({\Cal A})$, $\varepsilon$
belongs to $\Omega^3({\Cal A})$.

A 1-form $\rho$ has the form
$$
\rho\;=\;\sum_j \pi (a^j) \bigl[ D, \pi (b^j)\bigr], \quad a^j,b^j\in
{\Cal A},
\tag 3.12
$$
and, without loss of generality, we may impose the constraint
$$
\sum_j a^j b^j\;=\;\id .
\tag 3.13
$$
Then
$$
\rho\;=\;\pmatrix
\quad A \qquad & i \gamma^5 K\phi \\
- i \gamma^5 K\phi & \quad A \quad \endpmatrix ,
\tag 3.14
$$
where \ $A = \sum_j a^j (\partial\!\!\!/ b^j)$, and
$\phi+\phi_0=\id_N\otimes \bigl( \sum_j a^j \phi_0 b^j\bigr)$. \
The 1-form $\rho$ given in (3.14) determines a connection, $\nabla$,
on the ``line bundle'' $E \simeq {\Cal A}$. The three-dimensional
Chern-Simons action of $\nabla$ is then given by
$$
\align
I_\varepsilon^3 (\rho)\; &=\;\frac \kappa 2 \;\int\!\!\!\!\!\!- \;
\bigl(\varepsilon (\rho\, d\rho + \frac 2 3\;\rho^3)^\perp\bigr)  \\
&=\;\frac \kappa 2\; Tr_\omega\;\bigl(\varepsilon (\rho\, d\rho\;+\;
\frac 2 3 \;\rho^3)^\perp D^{-2}\bigr),
\tag 3.15
\endalign
$$
as follows from eqs.~(2.63) and (2.32); \ (the Fredholm module \ $({\Cal
A},\pi,H,D)$ \ is \ $(2,\infty)$-summable!).

In order to proceed in our calculation, we must determine the spaces
of ``auxiliary fields'' $Aux^n$, for $n=1,2,3$. Clearly $Aux^1=0$. To
identify $Aux^2$, we consider a 1-form
$$
\rho\;=\;\sum_j\pi (a^j) \bigl[ D,\pi (b^j)\bigr]\;=\;
\pmatrix
\quad A \quad & i\gamma^5 K\phi \\
- i\gamma^5 K\phi & \quad A \quad \endpmatrix .
$$
Then
$$
\align
d\rho\;&=\;\sum_j\bigl[ D,\pi (a^j)\bigr]\bigl[D,\pi\bigr)b^j)\bigr]
\\
&=\;\pmatrix
\frac  1 2 \;\gamma^{\mu\nu}\partial_\mu A_\nu + X, \phantom{Zeichnung}&
-i\gamma^5\gamma^\mu K (\partial_\mu\phi + A_\mu \phi_0 - \phi_0
A_\mu) \\
i\gamma^5\gamma^\mu K (\partial_\mu\phi + A_\mu \phi_0-\phi_0 A_\mu), &
\frac 1 2 \; \gamma^{\mu\nu} \partial_\mu A_\nu + X
\phantom{Zeichnung}\endpmatrix ,
\endalign
$$
where $X=\id_N\otimes\bigl(\sum_j a^j \partial^\mu \partial_\mu
b^j\bigr)+\partial^\mu A_\mu$ is an arbitrary element of $\id_N\otimes
{\Cal A}$, and $\gamma^{\mu\nu} := [\gamma^\mu, \gamma^\nu]$. Hence
$$
Aux^2\;\simeq\;\pi ({\Cal A}).
\tag 3.16
$$
Next, let $\eta\in\Omega^2 ({\Cal A})$. Then one finds that
$$
d \eta\bigm|_{\eta=0} \;=\;\varepsilon\enskip
\pmatrix
\gamma^\mu X_\mu  & i\gamma^5 KX \\
- i\gamma^5 K X & \gamma^\mu X_\mu \endpmatrix ,
\tag 3.17
$$
where $X_\mu$ and $X$ are arbitrary elements of $\id_N \otimes {\Cal
A}$. Thus, in a 3-form $\vartheta$, terms proportional to $\gamma^\mu
\otimes \id_M$ in the off-diagonal elements and terms proportional to
$\gamma^5 K\otimes \id_M$ in the diagonal elements must be discarded
when evaluating $\vartheta^\perp$.

Next, we propose to check under what conditions the Chern-Simons
action $I_\varepsilon^3 (\rho)$ is gauge-invariant. In eq.~(2.25) we
have seen that, under a gauge transformation $M\in\pi({\Cal A})$,
$\rho$ transforms according to
$$
\rho\mapsto \tilde\rho\;=\; M\rho M^{-1} - (dM) M^{-1}\;=\;
g^{-1} \rho g + g^{-1} dg,
\tag 3.18
$$
with $g=M^{-1}$. From this equation and the cyclicity of
``integration'', $\int\!\!\!\!\!- (\cdot)$, we deduce that
$$
\align
I_\varepsilon^3 (\tilde \rho) \;&=\; I_\varepsilon^3 (\rho)\;+\\
&\frac \kappa 2\;\int\!\!\!\!\!\!- \bigl(\varepsilon \bigl\{ dg^{-1}
\rho dg + g^{-1} d\rho dg - \frac 1 3 \; (g^{-1} dg)^3\bigr\}^\perp
\bigr) .
\tag 3.19
\endalign
$$
The second term on the r.h.s. of (3.19) is equal to
$$
\align
\int_{M_2} d\,vol\,tr\;\bigl(\varepsilon\bigl\{[D,g^{-1}]
&\rho [D,g] + g^{-1}  d\rho [D,g] \\
&- \frac 1 3\; \bigl(g^{-1} [D,g]\bigr)^3\bigr\}^\perp\bigr).
\tag 3.20
\endalign
$$
Here, and in the following, $tr(\cdot)$ denotes a \ub{normalized
trace}, $\bigl( tr (\id) =1\bigr)$. A straightforward calculation
shows that
$$
[D,g]\;=\;\pmatrix
\partial\!\!\!/\; g& i\gamma^5 K (\phi_0 g - g\phi_0) \\
- i \gamma^5 K (\phi_0 g - g\phi_0) & \partial\!\!\!/\; g \endpmatrix ,
\tag 3.21
$$
and expression (3.20) is found to be given by
$$
\align
- i \,tr\,K \int_{M_2} \partial_\mu\,
&tr\;\bigl[ g^{-1} \phi \partial_\nu g + A_\nu \bigl( \phi_0 - g
\phi_0 g^{-1}\bigr) \\
&-\;\bigl( g \partial_\nu g^{-1} \phi_0 + g^{-1} \partial_\nu g
\phi_0\bigr)\bigr] \; dx^\mu \wedge dx^\nu,
\tag 3.22
\endalign
$$
which vanishes if $\partial M_2 = \phi$ (i.e., $M_2$ has no boundary).
\bigskip
\bf\ub{Remark}. \rm \ Had we considered a more general setting with \
${\Cal A} = C^\infty (M_2) \otimes {\Cal A}_1 \oplus C^\infty (M_2)
\otimes {\Cal A}_2$, where ${\Cal A}_1$ and ${\Cal A}_2$ are two
independent matrix algebras, and $\pi(a)=\id_N\otimes a$, for $a\in
{\Cal A}$, then, with $\varepsilon$ chosen as above,
$I_\varepsilon^3(\rho)$ would fail to be gauge-invariant.

Thus the condition for $I_\varepsilon^3(\rho)$ to be gauge-invariant
is that $\partial M_2=\phi$ and that the non-commutative space is
invariant under permuting the two copies of $M_2$ (of the space
$X_c$), i.e., the elements of $\pi ({\Cal A})$ commute with the
operator $\varepsilon$ defined in (3.11).

Under this condition one finds, after a certain amount of algebra,
that
$$
I_\varepsilon^3(\rho)\;=\;i\;\kappa \int_{M_2} tr\;(\Phi F),
\tag 3.23
$$
where
$$
\Phi\;=\;K (\phi+\phi_0),
$$
and
$$
F\;=\;\bigl(\partial_\mu A_\nu - \partial_\nu A_\mu + [A_\mu,
A_\nu]\bigr)\;
dx^\mu\wedge dx^\nu .
\tag 3.24
$$
We note that $I_\varepsilon^3(\rho)$ is obviously gauge-invariant and
topological, i.e., metric-independent. Since the ``Dirac operator''
$D$ depends on the reference metric $g$ on $M_2$, the
metric-independence of $I_\varepsilon^3$ is not, a priori, obvious
from its definition (3.15).

Let us consider the special case where
$$
{\Cal A}_o\;=\;\Bbb M_3 (\Bbb R),
\tag 3.25
$$
the algebra of \ub{real} 3$\times$3 matrices. Then
$$
I_\varepsilon^3 (\rho)\;=\;i\,\kappa \int_{M_2} \biggl( \sum_{A=1}^3
\Phi^A F_{\mu\nu}^A\biggr)\; dx^\mu \wedge dx^\nu
\tag 3.26
$$
where $F_{\mu\nu}^A=\partial_\mu A_\nu^A-\partial_\nu A_\mu^A+
\varepsilon^{ABC} A_{[\mu}\, ^B \; A_{\nu]}\,^C$,
with $A=a,3,a=1.2.$ \ Setting
$$
A_\mu^a\;=\;e_\mu^a,\quad A_\mu^3\;=\;\frac 1
2\;\omega_\mu^{ab}\;\varepsilon_{ab} \;\equiv\;\omega_\mu
\tag 3.27
$$
one observes that the action $I_\varepsilon^3$ is the one of
two-dimensional topological gravity introduced in [13]. Varying
$I_\varepsilon^3$ w.r. to $\Phi^A$ one obtains the zero-torsion and
constant-curvature conditions:
$$
\align
&\varepsilon^{\mu\nu} F_{\mu\nu}^a\;\equiv\;
 \varepsilon^{\mu\nu} T_{\mu\nu}^a\;=\;\varepsilon^{\mu\nu}
\bigl( \partial_\mu e_\nu^a+\,\frac 1 2 \sum_b \omega_\mu
\varepsilon^{ab} e_\nu^b\bigr)\;=\;0 \\
&\varepsilon^{\mu\nu} F_{\mu\nu}^3\;=\;\frac 1 2\;
 \varepsilon^{\mu\nu} R_{\mu\nu}^{ab}\, \varepsilon_{ab}\;=\;\frac 1 2
\;\varepsilon^{\mu\nu}\,
\bigl( \partial_\mu \omega_\nu + 2 \;\varepsilon_{ab}\; e_\mu^a\;
e_\nu^b\bigr)\;=\;0.
\tag 3.28
\endalign
$$
Variation of $I_\varepsilon^3$ with respect to $A_\mu^A$ implies that
$\Phi^A$ is covariantly constant, i.e.,
$$
D_\mu \Phi^A\;=\;\partial_\mu\Phi^A + \varepsilon^{ABC} A_\mu^B
\Phi^C\;=\;0.
\tag 3.29
$$
The space of solutions of (3.28) and (3.29) is characterized in [13].

These results suggest that the study of Chern-Simons actions in
non-commutative geometry is worthwhile.
\bigskip

\vskip 1.5truecm
\bf 4. \ub{Some ``five-dimensional'' Chern-Simons actions} \rm .
\medskip
In this section, we consider non-commutative space $({\Cal
A},\pi,H,D)$ with ``cotangent bundles'' $\Omega_D^1({\Cal A})$ that
are free, hermitian ${\Cal A}$-bimodules of dimension 5, and we
evaluate the ``five-dimensional'' Chern-Simons action,
$I_\varepsilon^5(\rho)$, defined in eq.~(2.63), for connections on the
``line bundle'' $E=E^{(1)} \simeq {\Cal A}$. We shall consider
algebras ${\Cal A}$ generated by matrix-valued functions on Riemann
surfaces or on four-dimensional spin manifolds. We start with the
analysis of the latter example.
\medskip

\bf (I)\rm \ We choose $X=M_4\times \{-1,1\}$, where $M_4$ is a
four-dimensional, smooth Riemannian spin manifold. The non-commutative
space $({\Cal A},\pi,H,D)$ is chosen as in Sect.~3, except that $M_2$
is replaced by $M_4$, the 2$\times$2 Dirac matrices
$\gamma^1,\gamma^2$ are replaced by the 4$\times$4 Dirac matrices
$\gamma^1,\gamma^2,\gamma^3,\gamma^4$, and
$\gamma^5=\gamma^1 \gamma^2 \gamma^3\gamma^4$. The definition of the
``Dirac operator'' $D$ is analogous to that in eq.~(3.8).

An orthonormal basis for $\Omega_D^1({\Cal A})$ is then given
(locally on $M_4$) by
$$
\varepsilon^a\;=\;\pmatrix
\gamma^a\otimes \id_M & \quad 0\quad \\
\quad 0\quad & \gamma^a\otimes\id_M \endpmatrix ,
 a=1,\cdots,4, \quad \varepsilon^5\;=\;
\pmatrix
\quad 0\quad & \gamma^5\otimes\id_M \\
- \gamma^5\otimes\id_M & \quad 0\quad \endpmatrix,
\tag 4.1
$$
and
$$
\varepsilon\;=\;\varepsilon^1\;\varepsilon^2\;\varepsilon^3\;
\varepsilon^4\; \varepsilon^5\;=\;\pmatrix
\enskip 0\quad & \id \\
-\id & 0 \endpmatrix ,
\tag 4.2
$$
similarly as in (3.11).

Again, we must determine the spaces $Aux^n$, $n=1,2,3,4,5$, of
``auxiliary fields''. The most important one is $Aux^5$. To determine
it, let us consider a vanishing element $\eta$ of $\Omega^4({\Cal
A})$ and compute $d\eta$, as given by eq.~(2.10). After a certain
amount of labouring one finds that
$$
d\eta\bigm|_{\eta=0}\;=\;\varepsilon \pmatrix
\gamma^{\mu\nu\rho} X_{\mu\nu\rho} + \gamma^\mu (K^2 X_\mu+Y_\mu),
& i \gamma^5(\gamma^{\mu\nu} K X_{\mu\nu} + K^3 X+KY) \\
- i \gamma^5 (\gamma^{\mu\nu} K X_{\mu\nu} + K^3 X+KY),
& \gamma^{\mu\nu\rho} X_{\mu\nu\rho} + \gamma^\mu (K^2 X_\mu+Y_\mu)
\endpmatrix ,
\tag 4.3
$$
where \ $X_{\mu\nu\rho}, X_{\mu\nu}, X_\mu, X, Y_\mu$ and $Y$ are
arbitrary
elements of $\id_N\otimes{\Cal A}$, and $\gamma^{\mu\nu\rho} =
\displaystyle\mathop{\Sigma}_{a,b,c}$ sig \  $\mu\nu\rho \choose a b c $ \
$\gamma^a\gamma^b\gamma^c$.
By (4.3), the passage from an element $\vartheta\in\Omega^5({\Cal A})$
to $\vartheta^\perp$ amounts to discarding all terms proportional to
$\gamma^{\mu\nu\rho} \otimes\id_M$, $K^2\gamma^\mu\otimes\id_M$ and
$\gamma^\mu\otimes\id_M$ from off-diagonal elements of $\vartheta$
and all terms proportional to $K\gamma^5\gamma^{\mu\nu}\otimes\id_M$,
$K^3\gamma^5\otimes \id_M$ and $K\gamma^5\otimes\id_M$ from the
diagonal elements of $\vartheta$. Now we start understanding the
useful role played by the matrix $K$.

It is then easy  to
evaluate $I_\varepsilon^5(\rho)$, with $\rho$ given by
$$
\rho\;=\;\pmatrix \quad A\quad & i\gamma^5 K\phi \\
- i \gamma^5 K \phi &\quad A\quad \endpmatrix ,
\quad A\;=\;\gamma^\mu A_\mu .
\tag 4.4
$$
Using eq.~(2.63), the result is
$$
I_\varepsilon^5 (\rho)\;=\;i\;\frac{3\kappa}{4}\;\int_{M_4} Tr\;
(\Phi\;F\wedge F) ,
\tag 4.5
$$
where $\Phi = K(\phi+\phi_0)$, and $F=F_{\mu\nu} dx^\mu \wedge
dx^\nu$, with $F_{\mu\nu}$ the curvature, or field strength, of
$A_\mu$. Provided that $\partial M_4=\emptyset$, $I_\varepsilon^5$ is
gauge-invariant and topological (metric-independent), as expected. The
field equation obtained by varying $I_\varepsilon^5$ w.r. to $\Phi$ is
$$
\varepsilon^{\mu\nu\rho\sigma}\;F_{\mu\nu}\;F_{\rho\sigma}\;=\;0 .
\tag 4.6
$$
Setting $\Phi$ to a constant, $I_\varepsilon^5$ turns out to be the
action of four-dimensional, topological Yang-Mills theory [16] before
gauge-fixing.
\medskip
\bf (II) \rm \ We choose $X=M_2\times \{-1,1\}$ and ${\Cal A},\pi$ and
$H$ as above, but the operator $D$ is given by
$$
D\;=\;\pmatrix
\quad\partial\!\!\!/\quad & K \gamma^\alpha \phi_{0\alpha} \\
- K \gamma^\alpha \phi_{0\alpha} &\quad\partial\!\!\!/\quad
\endpmatrix ,
\tag 4.7
$$
where $\partial\!\!\!/ = \gamma^1\partial_1+\gamma^2\partial_2$, and
$\alpha = 3,4,5$. The matrices $\gamma^1,\cdots,\gamma^4$ are
antihermitian 4$\times$4 Dirac matrices, and, \ub{only in this
paragraph}, $\gamma^5 = i\,\gamma^1\gamma^2\gamma^3\gamma^4$, so that
$\gamma^5$ is now antihermitian, too, rather than hermitian (as in the
rest of this paper). Locally on $M_2$, the cotangent bundle
$\Omega_D^1 ({\Cal A})$ is a free, hermitian ${\Cal A}$-bimodule of
dimension 5, with an orthonormal basis given by
$$
\varepsilon^a\;=\;\pmatrix
\gamma^a\otimes\id_M &\quad 0\quad\\
\quad 0\quad & \gamma^a\otimes \id_M \endpmatrix,
 a= 1,2,\quad\varepsilon^\alpha\;=\;
\pmatrix
\quad 0\quad & i\gamma^\alpha \otimes \id_M \\
- i\gamma^\alpha \otimes \id_M &\quad 0\quad \endpmatrix ,
\alpha = 3,4,5,
\tag 4.8
$$
and $\varepsilon$ is taken to be
$$
\varepsilon\;=\;\varepsilon^1 \varepsilon^2 \varepsilon^3
\varepsilon^4 \varepsilon^5\;=\; \pmatrix
\enskip 0\quad &\id \\
- \id & 0 \endpmatrix .
\tag 4.9
$$
A 1-form $\rho = \sum_j \pi (a^j) \bigl[ D,\pi (b^j)\bigr]$ has the
form
$$
\rho\;=\;\pmatrix
\quad A\quad & K \gamma^\alpha \phi_\alpha \\
- K \gamma^\alpha \phi_\alpha & \quad A\quad \endpmatrix ,
\tag 4.10
$$
with $A=\sum_j a^j \partial\!\!\!/ b^j$ and $\phi_\alpha +
\phi_{0\alpha} = \sum_j a^j \phi_{0\alpha} b^j$. Evaluating $d\rho$ as
in eq.~(2.10), one finds that
$$
d\rho\;=\;\pmatrix
\gamma^{\mu\nu} \partial_\mu A_\nu - K^2 \gamma^{\alpha\beta}
L_{\alpha\beta} + X - K^2 L_\alpha^\alpha, &
K \gamma^\mu \gamma^\alpha D_\mu^0 \phi_\alpha \\
- K \gamma^\mu \gamma^\alpha D_\mu^0 \phi_\alpha, &
\gamma^{\mu\nu} \partial_\mu A_\nu - K^2\gamma^{\alpha\beta}
L_{\alpha\beta} + X - K^2 L_\alpha^\alpha \endpmatrix ,
\tag 4.11
$$
where
$$
L_{\alpha\beta}\;=\; \phi_{0\alpha} \phi_\beta + \phi_\alpha
\phi_{0\beta} + \sum_j a^j \bigl[ b^j, \phi_{0\alpha}
\phi_{0\beta}\bigr],
$$
$$
X\;=\; -\;\sum_j a^j \;\partial\!\!\!/\,^2\;b^j\;+\;\partial^\mu A_\mu
,
$$
and
$$
D_\mu^0 \phi_\alpha\;=\; \partial_\mu \phi_\alpha\;+\;A_\mu
\phi_{0\alpha}\;-\;\phi_{0\alpha} A_\mu .
\tag 4.12
$$
For simplicity we assume that
$$
\bigl[ \phi_{0\alpha}, \phi_{0\beta}\bigr]\;=\; 0, \quad \text{and }
\phi_{0\alpha} \phi_0^\alpha\;=\; 1 .
\tag 4.13
$$
Since we may assume that $\sum a^j b^j = 1$, we then have that
$L_{\alpha\beta} = \phi_{0\alpha} \phi_\beta + \phi_\alpha
\phi_{0\beta}$, for $L_{[\alpha\beta]}$ and $L_\alpha^{\enskip\alpha}$
appearing in (4.11), which is not an auxiliary field. A tedious calculation
then yields the formula
$$
\align
I_\varepsilon^5 (\rho)\;&=\; 2\kappa \int_{M_3} \varepsilon^{\mu\nu}
\varepsilon^{\alpha\beta\gamma} tr K^3 \bigl[\bigl( L_{\alpha\beta}
\phi_\gamma + \phi_\alpha L_{\beta\gamma}\bigr)\;\partial_\mu A_\nu
\\
&-\;\phi_{0\alpha} D_\mu^0 \phi_\beta D_\nu^0 \phi_\gamma + A_\mu
L_{\alpha\beta} D_\nu^0 \phi_\gamma + A_\mu D_\nu^0 \phi_\alpha
L_{\beta\gamma} \\
&+\; \frac 3 2 \; \phi_\alpha A_\mu A_\nu L_{\beta\gamma} +\frac 3 2\;
\phi_\alpha \phi_\beta\phi_\gamma\partial_\mu A_\nu \\
&-\;\frac 3 2 \; A_\mu \phi_\alpha A_\nu L_{\beta\gamma} + \frac 3 2 \;
A_\mu A_\nu \phi_\alpha L_{\beta\gamma} \\
&-\;\frac 3 2\; \phi_\alpha A_\mu \phi_\beta D_\nu^0 \phi_\gamma
+ \frac 3 2\; \phi_\alpha \phi_\beta A_\mu D_\nu^0 \phi_\gamma \\
&+\;\frac 3 2\; A_\mu \phi_\alpha\phi_\beta D_\nu^0 + 3\; A_\mu A_\nu
\phi_\alpha\phi_\beta\phi_\gamma \\
&-\;3\;\phi_\alpha A_\mu \phi_\beta A_\nu \phi_\gamma \bigr]\; d^2 x .
\tag 4.14
\endalign
$$
If $\partial M_2=\emptyset$, and after further algebraic manipulations,
the action (4.14) can be shown to have the manifestly gauge-invariant
form
$$
\align
I_\varepsilon^5 (\rho)\;=\; 2\kappa &\int_{M_2}
\varepsilon^{\alpha\beta\gamma} tr\;\bigl[ - \Phi_\alpha \bigl( D_\mu
\Phi_\beta\bigr) \bigl(D_\nu \Phi_\gamma\bigr) \\
& +\;2\;\Phi_\alpha \Phi_\beta\Phi_\gamma\;\bigl(\partial_\mu A_\nu +
A_\mu A_\nu\bigr)\bigr]\; dx^\mu \wedge dx^\nu ,
\tag 4.15
\endalign
$$
where $\Phi_\alpha = K (\phi_\alpha + \phi_{0\alpha})$.

If the constraints (4.13) are not imposed then one must explicitly
determine $Aux^5$, in order to derive an explicit expression for
$I_\varepsilon^5$. The result is that (4.15) still holds.

It is remarkable that all the Chern-Simons actions derived in
eqs.~(3.23), (4.5) and (4.15) can be obtained from Chern-Simons
actions for connections on vector bundles over classical, commutative
manifolds by \ub{dimensional reduction}. For example, setting $M_3 =
M_2\times S^1$ and $\phi := A_3$, and assuming that $A_1, A_2$ and
$A_3$ are indpendent of the coordinate (angle) parametrizing $S^1$, we
find that
$$
\align
I^3 (A)\;&=\;i \kappa' \int_{M_3} tr\;\bigl( A \wedge dA + \frac 2 3
\; A \wedge A \wedge A \bigr) \\
&=\;i \kappa' \int_{M_2} tr\; (\phi \;F) ,
\tag 4.16
\endalign
$$
where \ $F = \bigl(\partial_1 A_2 - \partial_2 A_1 + [ A_1,
A_2]\bigr)\;dx^1\wedge dx^2.$ \ Setting $\kappa' =\kappa\; tr K$, (4.16)
reduces to (3.23). Similarly, reducing a classical, five-dimensional
Chern-Simons action to four dimensions, with $M_5 = M_4 \times S^1$,
results in
$$
\align
I^5(A)\;&=\;i \kappa' \int_{M_5} tr\;\bigl( A\wedge dA \wedge dA +\;
\frac 3 2 \; A\wedge A\wedge A\wedge dA \\
&\phantom{Zeichnung} +\; \frac 3 5\; A \wedge A\wedge A\wedge A\wedge A\bigr)
\\
&=\;i\;\frac{3\kappa'}{4}\; \int_{M_4} tr\;(\phi \;F\wedge F),
\endalign
$$
with $\phi := A_5$, and $A_1,\cdots,A_5$ independent of the angle
parametrizing $S^1$. Thus we recover (4.5). Finally, dimensionally
reducing $I^5(A)$ to a two-dimensional surface (setting $M_5 = M_2
\times S^1 \times S^1 \times S^1)$ reproduces the action (4.15).

The advantage of the non-commutative formulation is that it
automatically eliminates all excited modes corresponding to a
non-trivial dependence of the gauge potential $A$ on angular
variables.

\vfill\eject
\bf 5. \ub{Relation to four-dimensional gravity and supergravity} \rm.
\medskip
Chern-Simons actions are topological actions. In order to obtain
dynamical actions from Chern-Simons actions, one would have to
impose  constraints on the field configuration space.
In this section, we explore this possibility. As a result, we are able
to derive some action functionals of four-dimensional gravity and
supergravity theory.

We propose to impose a constraint on the scalar multiplet $\Phi$
appearing in the Chern-Simons action (4.5). The non-commutative space
$({\Cal A}, \pi, H, D)$ is chosen as in example (I) of Sect.~4; (see
also Sect.~3). Let us compute the curvature 2-form,
$$
\theta\;=\; ( d\rho + \rho^2)^\perp ,
$$
of a connection $\nabla$ on the line bundle $E\simeq A$ given by a
1-form $\rho$ as displayed in eq.~(4.4). Then
$$
\theta\;=\;\pmatrix
\frac 1 2\;\gamma^{\mu\nu} F_{\mu\nu}+(K^2)^\perp
\bigl((\phi+\phi_0)^2-\phi_0^2\bigr),
& - K i \gamma^5\gamma^\mu D_\mu (\phi +\phi_0) \\
K i \gamma^5 \gamma^\mu D_\mu (\phi +\phi_0),
&\frac 1 2\;\gamma^{\mu\nu} F_{\mu\nu} + (K^2)^\perp
\bigl((\phi+\phi_0)^2 - \phi_0^2\bigr) \endpmatrix ,
\tag 5.1
$$
where $(K^2)^\perp = K^2 - (tr\;K^2) \id$; (recall that $tr (\cdot)$
is normalized: $tr (\id) = 1)$. The appearance of $(K^2)^\perp$ is due
to the circumstance that when passing from $d\rho+\rho^2$ to $(d\rho
+\rho^2)^\perp$ terms proportional to $\id_N$ must be removed. Let
$p\,tr (\cdot)$ denote the partial trace over the Dirac-Clifford
algebra. Then
$$
p\,tr\;(\theta)\;=\;(K^2)^\perp \bigl((\phi+\phi_0)^2-\phi_0^2\bigr),
\tag 5.2
$$
and we shall impose the constraint
$$
p\,tr\;(\theta)\;=\;0 .
\tag 5.3
$$
Choosing $\phi_0$ to satisfy $\phi_0^2=\id$, and renaming
$\phi+\phi_0$ to read $\phi$, the constraint (5.3) becomes
$$
\phi^2\;=\;\id,
\tag 5.4
$$
provided $(K^2)^\perp \;\neq\; 0. $

As our matrix algebra ${\Cal A}_0$ (see eq.~(3.1)) we choose
$$
{\Cal A}_0\;=\enskip\text{real part of Cliff } \bigl(SO(4)\bigr) .
\tag 5.5
$$
We propose to show that, for this choice of ${\Cal A}_0$ and assuming
that the constraint (5.4) is satisfied, the Chern-Simons action (4.5)
is the action of the metric-independent (first-order) formulation of
four-dimensional gravity theory.

Let $\Gamma_1,\cdots,\Gamma_4$ denote the usual generators of ${\Cal
A}_0$, (i.e., 4$\times$4 Dirac matrices in a real representation), and
$\Gamma_5 = \Gamma_1\Gamma_2\Gamma_3\Gamma_4$. Then
$$
\bigl\{ \Gamma_a, \Gamma_b\bigr\}\;=\;- 2 \delta_{ab},\;
\Gamma_a^*\;=\;- \Gamma_a,\;a,b\;=\;1,\cdots,4,
$$
and $\Gamma_5^* = \Gamma_5$. A basis for ${\Cal A}_0$ is then given by
$\id_4, \Gamma_a, a=1,\cdots,4, \Gamma_5, \Gamma_{ab}, a,b
=1,\cdots,4$, and $\Gamma_a\Gamma_5$. For a 1-form $\rho$ as in
eq.~(4.4), we may expand the gauge potential $A$ and the scalar field
$\phi$ in the basis of ${\Cal A}_0$ just described:
$$
A\;=\;\gamma^\mu \;\bigl( A_\mu^0 \id + A_\mu^a \Gamma_a + A_\mu^{ab}
\Gamma_{ab} + A_\mu^5 \Gamma_5 + A_\mu^{a5} \Gamma_a \Gamma_5 \bigr) ,
\tag 5.6
$$
and
$$
\phi\;=\;\bigl( \phi^0 \id + \phi^a \Gamma_a + \phi^{ab} \Gamma_{ab} +
\phi^5 \Gamma_5 + \phi^{a5} \Gamma_a \Gamma_5\bigr) .
\tag 5.7
$$
In this section, we only consider \ub{unitary} connections on $E\equiv
E^{(1)} \simeq {\Cal A}$; see eq.~(2.29). By (2.30), this is
equivalent to \ub{hermiticity} of $\rho$. This implies that
$$
A_\mu^0=- \overline{A_\mu^0}, A_\mu^a = \overline{A_\mu^a},
A_\mu^{ab} = \overline{A_\mu^{ab}}, A_\mu^5= - \overline{A_\mu^5},
A_\mu^{a5} = - \overline{A_\mu^{a5}},
\tag 5.8
$$
and
$$
\phi^0 = \overline{\phi^0}, \phi^a = - \overline{\phi^a},
\phi^{ab} = - \overline{\phi^{ab}}, \phi^5 = \overline{\phi^5},
\phi^{a5} = \overline{\phi^{a5}},
\tag 5.9
$$
where $\bar z$ denotes the complex conjugate of $z$. Since ${\Cal
A}_0$ is chosen to be real, the coefficients of $A$ and $\phi$ should
be chosen to be real. It then follows from (5.8) and (5.9) that
$$
A\;=\;\gamma^\mu \biggl(\frac{1}{2\kappa}\;e_\mu^a \Gamma_a + \frac 1
4\; \omega_\mu^{ab} \Gamma_{ab}\biggr)
\tag 5.10
$$
and
$$
\phi\;=\;\phi^0+\phi^5\Gamma_5 + \phi^{a5} \Gamma_a \Gamma_5,
\tag 5.11
$$
where we have set $A_\mu^a =: \frac{1}{2\kappa}\;e_\mu^a$, and
$A_\mu^{ab} =: \frac 1 4 \; \omega_\mu^{ab}$, and $\kappa^{-1}$ is the
Planck scale.

Imposing the constraint that \ $tr_{{\Cal A}_0} (\varepsilon \rho) =
0$ implies that
$$
\phi^0\;=\; 0 .
\tag 5.12
$$
Constraints (5.12) and (5.4) then yield the condition
$$
(\phi^5)^2\;+\;(\phi^{a5})^2\;=\;1 .
\tag 5.13
$$
Under a gauge transformation $M\equiv g^1$, $\rho$ transforms according to
$$
\rho\;\mapsto\; M\rho M^{-1} - (dM) M^{-1},\quad M\in\pi ({\Cal A}),
$$
see (2.25), which implies the transformation law
$$
\phi \;\mapsto\; g^{-1}  \phi g, \quad g\;=\;\exp \;\frac 1 2\; \bigl(
\Lambda^a \Gamma_a + \Lambda^{ab} \Gamma_{ab}\bigr),
\tag 5.14
$$
where $\Lambda^a$ and $\Lambda^{ab}$ are smooth functions on $M_4$.
The infinitesimal form of (5.14) reads
$$
\align
\delta \phi^5\;&=\;-\;\sum_a \Lambda^a \phi^{a5} , \\
\delta \phi^{a5}\;&=\;-\;\Lambda^a \phi^5 \;-\sum_b \Lambda^{ab}
\phi^{b5} .
\tag 5.15
\endalign
$$
 From this it follows that, locally, we can choose a gauge such that
$$
\phi^{a5}\;=\;0 .
\tag 5.16
$$
In this gauge, the constraint (5.13) has the solutions
$$
\phi^5\;=\; \pm \;1 .
\tag 5.17
$$
The action (4.5) then becomes
$$
I_\varepsilon^5 (\rho)\;=\; \pm\; k \int_{M_4} tr\; (\Gamma_5 F\wedge
F) ,
\tag 5.18
$$
(with $k=i \frac{3\kappa}{4}$, in the notation of Sect.~4). Next, we
expand the field strength $F_{\mu\nu}$ in our Clifford algebra basis
which yields
$$
F_{\mu\nu}\;=\;\frac{1}{2\kappa}\; F_{\mu\nu}^a\; \Gamma_a \;+\;\frac 1
4\; F_{\mu\nu}^{ab} \;\Gamma_{ab} ,
\tag 5.19
$$
where
$$
F_{\mu\nu}^a\;=\;\partial_\mu\; e_\nu^a + \omega_{\mu\;b}^a\;
e_\nu^b\;-\;(\mu\leftrightarrow \nu) ,
\tag 5.20
$$
$$
F_{\mu\nu}^{ab}\;=\;\partial_\mu \;\omega_\nu^{ab} +
\omega_{\mu\;c}^a\;\omega_\nu^{c\;b} + \;\frac{1}{\kappa^2}\;
e_\mu^a\; e_\nu^b \;-\;(\mu\leftrightarrow \nu) ,
\tag 5.21
$$
and the indices $a,b,\cdots$ are raised and lowered with the flat
metric $\eta_{ab} = - \delta_{ab}$.

The only non-vanishing contribution to (5.18) comes from the trace \
$tr (\Gamma_5\Gamma_{ab}\Gamma_{cd}) = \varepsilon_{abcd}$, and
$I_\varepsilon^5$ is found to be given by
$$
\align
I_\varepsilon^5\;=\;\pm\;k \int_{M_4} \varepsilon_{abcd}\;\bigl(
R_{\mu\nu}^{ab} +\;\frac{2}{\kappa^2}\; e_\mu^a\;&e_\nu^b\bigr)
\bigl( R_{\rho\sigma}^{cd} + \;\frac{2}{\kappa^2}\;e_\rho^c\;
e_\sigma^d \bigr) \\
&\times\; dx^\mu \wedge dx^\nu \wedge dx^\rho \wedge dx^\sigma ,
\tag 5.22
\endalign
$$
where
$$
R_{\mu\nu}^{ab}\;=\;\partial_\mu\;\omega_\nu^{ab} +
\omega_{\mu\;c}^a\;\omega_\nu^{c\;b}\;-\; (\mu \leftrightarrow \nu) .
\tag 5.23
$$
Interpreting $\omega_\mu^{ab}$ as the components of a connection on
the spinor bundle over $M_4$, $R_{\mu\nu}^{ab}$ are the components of
its curvature, and $F_{\mu\nu}^a$ are the components of its torsion,
as is well known from the Cartan structure equations.

Setting the variation of $I_\varepsilon^5$ with respect to
$\omega_\mu^{ab}$ to zero, we find that the torsion of $\omega$
vanishes:
$$
F_{\mu\nu}^a\;=\;0,\quad\text{for all }\mu,\nu\text{ and } a.
\tag 5.24
$$
If the frame $\bigl(e_\mu^a\bigr)$ is invertible, (5.24) can be solved
for $\omega_\mu^{ab}$:
$$
\omega_\mu^{ab}\;=\;\frac 1 2 \; \bigl( \Omega_{\mu a b} -
\Omega_{ab\mu} + \Omega_{b\mu a}\bigr),
\tag 5.25
$$
where
$$
\Omega_{ab}\;^c\;=\; e_a^\mu\;e_b^\nu\;\bigl(
\partial_\mu\;e_\nu\,^c\;-\;\partial_\nu\; e_\mu\,^c \bigr).
$$
Substituting (5.25) back into (5.22) yields a functional that depends
only on the metric
$$
g_{\mu\nu}\;=\;e_\mu^a\; e_{\nu a},
\tag 5.26
$$
and is given by
$$
\align
I_\varepsilon^5\;=\;\pm\;k \int_{M_4} d^4 x\, \sqrt{g}\; \bigl[\bigl(\;
&4\,R_{\mu\nu\rho\sigma} R^{\mu\nu\rho\sigma} \;-\;4\, R_{\mu\nu} R^{\mu\nu}
+ R^2 \bigr) \\
&+\;\frac{16}{\kappa^2}\;R\;+\;\frac{96}{\kappa^4}\;\bigr]
 \tag 5.27
\endalign
$$
where $R_{\mu\nu\rho\sigma}$ is the Riemann curvature tensor,
$R_{\mu\nu}$ is the Ricci tensor, and $R$ is the scalar curvature
determined by the metric $g_{\mu\nu}$ given in (5.26). The term in
round brackets on the r.h.s. of (5.27) yields the topological
Gauss-Bonnet term for $M_4$, the second term yields the
Einstein-Hilbert action, and the last term is a cosmological constant.

Next, we show how to derive a metric-independent formulation of \
\ub{four-dimensional} \ub{supergravity} from the action $I_\varepsilon^5$
given in eq.~(4.5). For this purpose we choose the algebra ${\Cal
A}_0$ in (3.1) to be a graded algebra [18]:
$$
{\Cal A}_0\;=\;\text{real part of SU}(4\mid 1) .
\tag 5.28
$$
This algebra is generated by graded 5$\times$5 matrices preserving the
quadratic form
$$
(\vartheta^\alpha)^*\; C_{\alpha\beta} \;\vartheta^\beta\;-\;z^* \;z ,
\tag 5.29
$$
where $C_{\alpha\beta}$ is an antisymmetric matrix and
$\vartheta^\alpha$ is a Dirac spinor. At this point, one must note
that we are leaving the conventional framework of non-commutative
geometry, since, for ${\Cal A}_0$ as in (5.28), the algebra ${\Cal A}$
is not a $^*$algebra of operators. But let us try to proceed and find
out what the result is.

Let $\rho$ be a 1-form as in eq.~(4.4). Then the matrix elements
$A_\mu$ and $\phi$ of $\rho$ have the graded matrix representation
$$
\phi\;=\;\pmatrix
\Pi_\alpha^\beta & \lambda_\alpha \\
\bar\lambda^\alpha & \Pi_1 \endpmatrix ,
\tag 5.30
$$
and
$$
A_\mu\;=\; \pmatrix
M_{\mu \alpha}^{\enskip\beta} & \sqrt{\kappa}\;\psi_{\mu\alpha} \\
- \sqrt{\kappa}\;\bar\psi_\mu^\alpha &\quad B_\mu \endpmatrix .
\tag 5.31
$$
The reality conditions for $\phi$ and $A_\mu$ imply that
$\lambda_\alpha$ and $\psi_{\mu\alpha}$ are Majorana spinors:
$$
\lambda_\alpha\;=\;C_{\alpha\beta}\;\bar\lambda^\beta,
\;\psi_{\mu\alpha}\;=\;C_{\alpha\beta}\;\bar\psi_\mu^\beta .
$$
Furthermore, one finds that
$$
\align
\Pi_\alpha^\beta\;&=\;\biggl( \frac 1 4\; \Pi^0 \id \;+\; \Pi^5 \Gamma_5 \;+\;
\Pi^{a5} \Gamma_a\Gamma_5\biggr)_\alpha^\beta , \\
M_{\mu\alpha}^\beta\;&=\;\biggl(
\frac{1}{2\kappa}\;e_\mu^e\;\Gamma_a \;+\;\frac 1 4
\;\omega_\mu^{ab}\;\Gamma_{ab}\biggr)_\alpha^\beta ,
\tag 5.32
\endalign
$$
and
$$
B_\mu\;=\;0 .
$$

We shall now impose the constraints
$$
Str\;(\varepsilon\;\rho)\;=\;0 ,
\tag 5.33
$$
$$
Str\;(\theta)\;=\;0 ,
\tag 5.34
$$
and
$$
Str \;\biggl(\varepsilon\;\bigl( \rho d\rho\;+\;\frac 2
3\;\rho^3\bigr)^\perp \biggr)\;=\;0 ,
\tag 5.35
$$
along with $Str \;\phi_0^2 = 1$, and $Str\;\phi_0 = 0$. Here \ $Str
(\cdot)$ denotes the graded trace on ${\Cal A}_0$. Renaming $\phi +
\phi_0$ to read $\phi$, these constraints
imply that
$$
\Pi^0\;=\;\Pi_1 ,
\tag 5.36
$$
$$
-\;\frac 3 4 \; \Pi_1^2 + 4\bigl((\Pi^5)^2 - \sum_a (\Pi^{a5})^2\bigr)
+ \bar\lambda \lambda \;=\;1 ,
\tag 5.37
$$
and
$$
(K^3)^\perp\;S tr (\phi^3)\;=\; 0,
\tag 5.38
$$
where $(K^3)^\perp$ is defined so as to satisfy \ $tr
\bigl(K(K^3)^\perp\bigr) = 0$.

In order to determine the dynamical contents of a theory with an
action $I_\varepsilon^5$ given by (4.5), ${\Cal A}_0$ as in (5.28) and
constraints (5.36) through (5.38), it is convenient to work in a
special gauge, the unitary gauge. Consider a gauge transformation
$$
g\;=\;\exp\; \pmatrix
\bigl(\Lambda_\alpha^\beta\bigr)\quad &\sqrt{\kappa}\;\varepsilon_\alpha \\
- \sqrt{\kappa}\;\bar\varepsilon^\alpha & 0 \endpmatrix ,
\tag 5.39
$$
where $\Lambda_\alpha^\beta = \frac 1 2$ $\bigl( \Lambda^a \Gamma_a +
\Lambda^{ab} \Gamma_{ab}\bigr)_\alpha^\beta $. The transformation law of
$\phi$ is then given by $\phi \mapsto g^{-1} \phi g$. From this we
find the infinitesimal gauge transformations of the fields $\Pi$ and
$\lambda$:
$$
\align
&\delta\; \Pi_1\;=\;2\;\sqrt{\kappa}\; \bar\varepsilon \lambda, \\
&\delta\; \Pi^5\;=\;+\; \Lambda^a \Pi^{a5} + \frac{\sqrt{\kappa}}{2}\;
\bar\varepsilon \;\Gamma^5 \lambda , \\
&\delta\;\Pi^{a5}\;=\; -\; \Lambda^{ab} \Pi^{b5} - \Lambda^a \Pi^5
- \frac 1 2\;\bar\varepsilon \;\Gamma^a\Gamma^5\lambda, \\
&\delta\;\lambda_\alpha\;=\;\sqrt{\kappa}\;\bigl(- \frac 3 4\; \Pi_1 +
\Pi^5 \Gamma_5 + \Pi^{a5}
\Gamma_a\Gamma_5\bigr) \varepsilon_\alpha - \frac 1 2\;
\bigl(\Lambda^a\Gamma_a+\Lambda^{ab}\Gamma_{ab}\bigr)_\alpha^\beta\;
\lambda_\beta .
\tag 5.40
\endalign
$$
Thus, locally, we can choose the gauge
$$
\Pi^{a5}\;=\;0, \text{ and } \lambda_a\;=\; 0 .
\tag 5.41
$$
The constraints (5.37) and (5.38) then reduce to
$$
\align
-\;\frac 3 4\;\Pi_1^2 + 4\; (\Pi^5)^2\;&=\;1 , \\
\Pi_1\bigl(-\frac{5}{16}\;\Pi_1^2 + (\Pi^5)^2\bigr)\;&=\; 0 .
\tag 5.42
\endalign
$$
These equations have the solutions
$$
\Pi_1\;=\;0,\quad \Pi^5\;=\;\pm\;\frac 1 2 ,
\tag 5.43
$$
and
$$
\Pi_1\;=\;\pm\;\sqrt{2}\;, \;\Pi^5\;=\;\pm\;\sqrt{\frac{5\,}{8\,}} .
\tag 5.44
$$
We further study the first solution. Inserting it into the action
(4.5), we arrive at the expression
$$
I_\varepsilon^5\;=\;\pm\;\frac k 2 \int_{M_4} Str\;\biggl( {\Gamma_5
\quad 0 \choose \enskip 0\quad 0}\; F_{\mu\nu} F_{\rho\sigma}\biggr)\; dx^\mu
\wedge dx^\nu\wedge dx^\rho \wedge dx^\sigma ,
\tag 5.45
$$
where
$$
F_{\mu\nu}\;=\;\pmatrix
\frac 1 4 \; F_{\mu\nu}^{ab}\;\Gamma_{ab} +
\frac{1}{2\kappa}\;F_{\mu\nu}^a\;\Gamma_a, &
\sqrt{\kappa}\;\psi_{\mu\nu} \\
-\;\sqrt{\kappa}\;\bar\psi_{\mu\nu}\phantom{ZeichnungZ}, &\quad 0
\endpmatrix ,
\tag 5.46
$$
with
$$
\align
F_{\mu\nu}^a\;&=\;\partial_\mu\;e_\nu^a + \omega_{\mu\; b}^a\;e_\nu^b
-\;\frac{\kappa^2}{2}\;\bar\psi_\mu \Gamma^a\psi_\nu -
(\mu\leftrightarrow \nu) , \\
F_{\mu\nu}^{ab}\;&=\;R_{\mu\nu}^{ab} + \frac{1}{\kappa^2}\; \bigl(
e_\mu^a\;e_\nu^b - e_\nu^a \; e_\mu^b\bigr) + \kappa \bar\psi_\mu
\Gamma^{ab} \psi_\nu , \\
\psi_{\mu\nu\alpha}\;&=\;\partial_\mu\;\psi_{\nu\alpha} + \frac 1 4\;
\omega_\mu^{ab} \;\bigl( \Gamma_{ab}\psi_\nu\bigr)_\alpha +
\frac{1}{2\kappa}\; e_\mu^a \bigl( \Gamma_a\psi_\nu\bigr)_\alpha \\
&\phantom{ZeichnungZeichnung} -\;(\mu\leftrightarrow \nu) .
\tag 5.47
\endalign
$$
After some further manipulations and evaluating all the traces, one
obtains the elegant result that the action reduces to that proposed in
[17], namely
$$
\align
I_\varepsilon^5\;=\;\pm\;k\int_{M_4} \biggl[ \frac 1 4\;
\varepsilon_{abcd} F_{\mu\nu}^{ab} &F_{\rho\sigma}^{cd} + \alpha \kappa
\; \bar\psi_{\mu\nu} \Gamma_5 \psi_{\rho\sigma}\biggr] \\
& \times\; dx^\mu \wedge dx^\nu \wedge dx^\rho \wedge dx^\sigma ,
\tag 5.48 \
\endalign
$$
where $\alpha$ is some constant introduced for later convenience, but
here $\alpha=1$. Substituting eqs.~(5.47) into (5.48), one obtains
that
$$
\align
I_\varepsilon^5\;=\;&\pm\;k \biggl\{ \int_{M_4} \varepsilon_{abcd}\;
\frac 1 4 \;\bigl[ R_{\mu\nu}^{ab} R_{\rho\sigma}^{cd} + 2\kappa\;
R_{\mu\nu}^{ab} \bigl( \bar\psi_\rho \Gamma^{cd} \psi_\sigma\bigr)\\
&\phantom{Zeich} + \kappa^2 \bigl(\bar\psi_\mu \Gamma^{ab}
\psi_\nu\bigr) \bigl(\bar\psi_\rho \Gamma^{cd}
\psi_\sigma\bigr)\bigr]\; dx^\mu\wedge dx^\nu\wedge dx^\rho\wedge
dx^\sigma \\
&+\;\frac{4}{\kappa^2}\;\int_{M_4} d^4x\; \sqrt{g} \;e_a^\mu e_b^\nu
\bigl(R_{\mu\nu}^{ab} + \kappa \bigl(\bar\psi_\mu \Gamma^{ab} \psi_\nu
\bigr)\bigr) \\
&+\;4 \alpha \kappa \int_{M_4} \bigl( D_\mu \bar\psi_\nu\bigr)
\Gamma^5\; \bigl(D_\rho \psi_\sigma\bigr) \; dx^\mu \wedge \cdots
\wedge dx^\sigma \\
&+\; 4\alpha \int_{M_4} \bigl( \bar\psi_\mu \Gamma_\nu \Gamma^5 D_\rho
\psi_\sigma\bigr) \; dx^\mu \wedge \cdots \wedge dx^\sigma \\
&+\;\frac{2\alpha}{\kappa}\;\int_{M_4} d^4x\; \sqrt{g}\; \bar\psi_\mu
\Gamma^{\mu\nu} \psi_\nu + \frac{24}{\kappa^4}\;\int_{M_4} d^4 x\,
\sqrt{g} \biggr\} ,
\tag 5.49
\endalign
$$
where
$$
D_\mu \;\psi_\nu\;=\;\partial_\mu\; \psi_\nu \;+\;\frac 1 4 \;
\omega_\mu^{ab}\; \Gamma_{ab}\; \psi_\nu .
$$
After Fierz reshuffling, the term quartic in the gravitino field
$\psi_\mu$ disappears. The remaining terms describe massive
supergravity with a Gauss-Bonnet term. It is an interesting fact that
the action (5.48), with $\alpha = 2$ (!), is invariant under
the same supersymmetry transformation obtained form the variation of
$\Pi(\rho)$, except for $\delta \omega_\mu^{ab}$ which is chosen to
preserve the constraint [18]:
$$
F_{\mu\nu}^{\enskip a}\;=\; 0 .
\tag 5.50
$$
The supersymmetry transformations can be read by substituting (5.10)
into eq.~(3.18):
$$
\align
&\delta\;e_\mu^a\;=\;\kappa\;\bar\varepsilon\;\Gamma^a \psi_\mu,\\
&\delta\;\psi_\mu\;=\;\bigl( \partial_\mu + \frac 1 4\; \omega_\mu^{ab}
\Gamma_{ab} + \frac{1}{2\kappa}\;e_\mu^a \Gamma_a\bigr) \varepsilon,
\tag 5.51
\endalign
$$
and, for $F_{\mu\nu}^{ab}$ and $\psi_{\mu\nu}$, they are
$$
\align
&\delta\;F_{\mu\nu}^{ab}\;=\;\kappa\;\bar\varepsilon \,\Gamma^{ab}
\psi_{\mu\nu}, \\
&\delta\;\psi_{\mu\nu}\;=\;-\;\frac 1 4\;
F_{\mu\nu}^{ab}\;(\Gamma_{ab}\; \varepsilon ).
\tag 5.52
\endalign
$$
When $\alpha=2$ the action (5.48) becomes invariant under the
transformations (5.51) with the constraint (5.50), and the action
corresponds to de Sitter supergravity where the cosmological constant
and the gravitino mass-like term are fixed with respect to each other.
In this case the action (5.48) simplifies to
$$
\align
I_{sg}\;=\;&-\;\biggl[\int_{M_4} d^4x\; \varepsilon^{\mu\nu\rho\sigma}
\,\bigl(\frac 1 4\;\varepsilon_{abcd} R_{\mu\nu}^{\enskip ab}
R_{\rho\sigma}^{\enskip cd} + 8\, \bar\psi_\mu \Gamma_\nu \Gamma_5
D_\rho \psi_\sigma\bigr) \\
&+\;4 \int d^4x\,e\;\bigl(e_a^\mu\, e_b^\nu\,R_{\mu\nu}^{\enskip ab} +
\frac 2 \kappa\;\bar\psi_\mu\; \Gamma^{\mu\nu} \psi_\nu +
\frac{6}{\kappa^4}\bigr)\biggr] .
\tag 5.52
\endalign
$$
The first term in (5.52) is a topological invariant and can be removed
from the action without affecting its invariance. After rescaling
$$
\align
e_\mu^a\;&\to\;r\;e_\mu^a \\
\psi_{\mu\alpha}\;&\to\;\sqrt{r} \; \psi_{\mu\alpha} \\
I_{sg}\;&\to\;8\,r^2\;I_{sg}
\tag 5.53
\endalign
$$
and taking the limit $r\to0$ the action (5.52) reduces to that of
$N$=1 supergravity [19]:
$$
I_{sg}\;=\;-\;\frac{1}{2\kappa^2} \int_{M_4} d^4x\,
e\;e_a^\mu\,e_b^\nu\,R_{\mu\nu}^{\;ab} - \int_{M_4}
d^4x\,\varepsilon^{\mu\nu\rho\sigma} \,\bar\psi_\mu \Gamma_5
\Gamma_\nu D_\rho \psi_\sigma .
\tag 5.54
$$
The significance of the constraint (5.50) and the choice $\alpha=2$ in
the non-commutative construction is not clear to us. It would be
helpful to better understand  this point.

If we had worked instead with the solution (5.44), then additional
terms which are dynamically trivial will be present. We shall not
present the details for this case.

\vskip 1.5truecm

\bf 6. \ub{Conclusions and outlook} \rm .
\medskip
In this paper, we have shown how to construct Chern-Simons forms and
Chern-Simons actions in real, non-commutative geometry; (more detailed
results will appear in [12]). We have illustrated the general,
mathematical results of Sect.~2 by discussing a number of examples.
These examples involve non-commutative spaces described by
$^*$algebras of matrix-valued functions over even-dimensional spin
manifolds. As expected, the Chern-Simons actions associated with these
spaces are manifestly topological (metric-independent). By imposing
constraints on the field configurations on which these action
functionals depend (and choosing convenient gauges) we have been able
to derive the metric-independent, first-order formulation of
four-dimensional gravity theory from a Chern-Simons action over a
``five-dimensional'' non-commutative space. By extending the
mathematical framework, formally, to allow for graded algebras, we
have also recovered an action functional for supergravity.

It would appear to be of interest to study Chern-Simons actions for
more general non-commutative spaces, e.g. those considered in [8], and
to derive from them theories of interest to physics. In this regard,
one should recall that a rather profound theory has the form of a
Chern-Simons theory: Witten's open string field theory [15]. We are
presently attempting to formulate that theory within Connes'
mathematical framework of non-commutative geometry, using a variant of
the formalism developed in Sect. 2.

On the mathematical side, it appears to be of interest to better
understand the topological nature of Chern-Simons actions over general
non-commutative spaces, to understand the connection between the
material presented in Sect.~2 and the theory of characteristic classes
in non-commutative geometry and cyclic cohomology, see [3,10], and, most
importantly, to learn how to quantize Chern-Simons theories in
non-commutative geometry, in order to construct new topological field
theories.

\vskip 1.5truecm
\bf \ub{References}\rm\;:
\medskip

\item{[1]} E. Witten, ``Quantum field theory and the Jones
polynomial'', Comm. Math. Phys. \ub{121}, (1989) 351.

\item{[2]} J. Fr\"ohlich ``Statistics of fields, the Yang-Baxter
equation, and the theory of knots and links'', 1987 Carg\`ese
lectures, in: ``Non-Perturbative Quantum Field Theory'', eds.: G. 't~Hooft
et al., (Plenum Press, New York, 1988).

\item{[3]} M. Dubois-Violette, C.R. Acad. Sc. Paris, \ub{307}, I, (1988)
403.
\item{} J. Fr\"ohlich and C. King, Comm. Math. Phys. \ub{126} (1989)
187.
\item{} E. Guadagnini, M. Martellini and M. Mintchev, Nucl. Phys.
B\ub{330} (1990) 575.

\item{[4]} A. Connes, Publ. Math. I.H.E.S. \ub{62} (1985) 41-144,
``Non-Commutative Geometry'', Academic Press, to appear (1994).

\item{[5]} A. Connes and J. Lott, Nucl. Phys. B (Proc.~Supp.) \ub{18}B
(1990) 29; in Proc. 1991 Summer Carg\`ese Conference, eds.: J.
Fr\"ohlich et al., (Plenum Press, New York 1992).

\item{[6]} R. Coquereaux, G. Esposito-Far\`ese, G. Vaillant, Nucl.
Phys. B\ub{353} (1991) 689;
\item{} M. Dubois-Violette, R. Kerner, J. Madore, J. Math. Phys.
\ub{31} (1990) 316;
\item{} B. Balakrishna, F. G\"ursey and K.C. Wali, Phys. Lett.
\ub{254}B (1991) 430; Phys. Rev. D\ub{46} (1991) 6498.
\item{} R. Coquereaux, G. Esposito-Far\`ese and F. Scheck,
Int.~J.~Mod.~Phys.~A\ub{7} (1992) 6555.

\item{[7]} D. Kastler, ``A detailed account of Alain Connes' version
of the standard model in non-commutative geometry'' I, II and III, to
appear in Rev. Math. Phys.;
\item{} D. Kastler and M. Mebkhout, ``Lectures on non-commutative
differential geometry and the standard model'', World Scientific, to
be published;
\item{} D. Kastler and T. Sch\"ucker, Theor. Math. Phys. \ub{92}
(1992) 522.

\item{[8]} A.H. Chamseddine, G. Felder and J. Fr\"ohlich, Phys. Lett.
B\ub{296} (1992) 109; Nucl. Phys. B\ub{395} (1993) 672.

\item{[9]} A.H. Chamseddine, G. Felder and J. Fr\"ohlich, Commun. Math.
Phys. \ub{155} (1993) 205.

\item{[10]} A.H. Chamseddine and J. Fr\"ohlich ``Some elements of
Connes' non-commutative geometry and space-time geometry'', to appear
in Yang-Festschrift.

\item{[11]} D. Quillen, ``Chern-Simons forms and cyclic cohomology'',
in: ``The Interface of Mathematics and Particle Physics'', D. Quillen,
G. Segal and S. Tsou (eds.), Oxford University Press, Oxford 1990.

\item{[12]} A.H. Chamseddine, J. Fr\"ohlich and O. Grandjean, in
preparation.

\item{[13]} A.H. Chamseddine and D. Wyler, Phys. Lett. B\ub{228} (1989)
75; Nucl. Phys. B\ub{340} (1990) 595;
\item{} E. Witten, ``Surprises with topological field theories'' in
Proc. ``Strings 90'', eds R. Arnowitt et al.

\item{[14]} E. Witten, Nucl. Phys. B\ub{311} (1988) 96; B\ub{323}
(1989) 113;
\item{} A.H. Chamseddine, Nucl. Phys. B\ub{346} (1990) 213.

\item{[15]} E. Witten, Nucl. Phys. B\ub{268} (1986) 253; Nucl. Phys.
B\ub{276} (1986) 291.

\item{[16]} E. Witten, Commun. Math. Phys. \ub{117} (1988) 353.

\item{[17]} A.H. Chamseddine, Ann. Phys. \ub{113} (1978) 219; Nucl.
Phys. B\ub{131} (1977) 494;
\item{} K. Stelle and P. West, J. Phys. A\ub{12} (1979) 1205.

\item{[18]} P. van Nieuwenhuizen, Phys. Rep. \ub{68} (1981) 189.

\item{[19]} D. Freedman, P. van Nieuwenhuizen and S. Ferrara, Phys.
Rev. D\ub{13} (1976) 3214;
\item{} S. Deser and B. Zumino, Phys. Lett. B\ub{62} (1976) 335.

\bye

{}From schultze Tue May 31 13:43:35 1994
Received: from isis.itp by itp.ethz.ch; Tue, 31 May 94 13:43:33 +0200
{}From: schultze (Annetraut Schultze)
Date: Tue, 31 May 94 13:43:33 +0200
Message-Id: <9405311143.AA11330@itp.ethz.ch>
To: chams
Status: R

\input vanilla.sty
\input mymacros
\overfullrule=0pt
\magnification=\magstep1
\vsize=23.5truecm
\hsize=16.5truecm
\baselineskip=.6truecm

\def\ub#1{{\underbar{#1}}}

\font\ti=cmbx10 scaled\magstep1
\TagsOnRight
\loadmsam
\loadmsbm
\UseAMSsymbols

$\;$

\pageno=0

\rightline{\bf ETH-TH/94-11\rm\quad}
\vskip 1truecm
\centerline{\ti{THE \ CHERN-SIMONS \ ACTION IN}}

\centerline{\ti{NON-COMMUTATIVE GEOMETRY}}\rm
\vskip 3truecm
\bf
\centerline{A.H. Chamseddine \ and \ J. Fr\"ohlich}

\centerline{Theoretical Physics}

\centerline{ETH-H\"onggerberg}

\centerline{CH-8093 \ Z\"urich}

\vskip 5truecm

\noindent Abstract\rm . \ A general definition of Chern-Simons actions
in non-commutative geometry is proposed and illustrated in several
examples. These examples are based on ``space-times'' which are
products of even-dimensional, Riemannian spin manifolds by a discrete
(two-point) set. If the $^*$algebras of operators describing the
non-commutative spaces are generated by functions over such
``space-times'' with values in certain Clifford algebras the
Chern-Simons actions turn out to be the actions of topological gravity
on the even-dimensional spin manifolds. By constraining the space of
field configurations in these examples in an appropriate manner one is
able to extract dynamical actions from Chern-Simons actions.

\vskip 1truecm

\noindent PACS No. \ A0210 / A0240 / A0350K / A0420C / A0420F
\vfill\eject

\pageno=1

\bf
\noindent 1. \ub{Introduction}\rm

During the past several years, topological field theories have been
the subject of a lot of interesting work. For example, deep connections
between three-dimensional, topological Chern-Simons theories [1], or,
equivalently, two-dimensional, chiral conformal field theories [2], on
one hand, and a large family of invariants of links, including the
famous Jones polynomial, and of three-manifolds [1], on the other
hand, have been discovered. Other topological field theories have been
invented to analyze e.g. the moduli space of flat connections on
vector bundles over Riemann surfaces or to elucidate the Donaldson
invariants of four-manifolds.
These topological field theories are formulated as theories over some
classical (topological or differentiable) manifolds.

Connes has proposed notions of non-commutative spaces generalizing,
for example, the notion of a classical differentiable manifold [4].
His theory is known under the name of ``non-commutative geometry''.
Dubois-Violette [3] and Connes have proposed to study field theories
over non-commutative spaces. In joint work with J. Lott [4], Connes
has found a construction of the classical action of the standard
model, using tools of non-commutative geometry, which yields a
geometrical interpretation of the scalar Higgs field responsible for
the ``spontaneous breaking of the electroweak gauge symmetry''. In
fact, the Higgs field appears as a component of a generalized gauge
field (connection 1-form) associated with the gauge group,
SU(2)$_w$\;$\times$\;U(1)$_{em}$,  of electroweak interactions. This is
accomplished by formulating gauge theory on a generalized space
consisting of two copies of standard Euclidian space-time the
``distance'' between which is determined by the weak scale. Although
the space-time model underlying the Connes-Lott construction is a
commutative space, it is not a classical manifold, and analysis on
space-times of the Connes-Lott type requires some of the tools of
non-commutative geometry.

The results of Connes and Lott have been reformulated and refined in
[6,7] and extended to grand-unified theories in [8]. In [9], G. Felder
and the authors have proposed some form of non-commutative Riemannian
geometry and applied it to derive an analogue of the Einstein-Hilbert
action in non-commutative geometry.

Our aim in this article is to attempt to do some steps towards a
synthesis between the different developments just described. Some of
our results have been described in our review paper [10]. We start by
presenting a general definition of the Chern-Simons action in
non-commutative geometry, (Section 2). Our definition is motivated by
some results of Quillen [11] and is based on joint work with O.
Grandjean [12]. In Section 3, we discuss a first family of examples.
In these examples, the non-commutative space is described in terms of
a $^*$algebra of matrix-valued functions over a Connes-Lott type
``space-time'', i.e., over a commutative space consisting of two
copies of an even-dimensional, differentiable spin manifold. The
Chern-Simons actions on such non-commutative spaces turn out to be
actions of topological gauge- and gravity theories, as studied in
[13,14]. In Section 3, the dimension of the continuous, differentiable
spin manifold is two, i.e., we consider products of Riemann surfaces
by discrete sets, and our Chern-Simons action is based on the
Chern-Simons 3-form.

In Section 4, we consider two- and four-dimensional topological
theories derived from a Chern-Simons action based on the Chern-Simons
5-form.

In Section 5, we describe connections of the theories found in Section
4 with four-dimensional gravity and supergravity theories.

In Section 6, we suggest applications of our ideas to string field
theory [15], and we draw some conclusions.


\vskip 1.5truecm

\bf
\ub{Acknowledgements}\rm\;. \ We thank G. Felder, K. Gaw\c edzki and
D. Kastler for their stimulating interest and O. Grandjean for very
helpful discussions on the definition of the Chern-Simons action in
non-commutative geometry and for collaboration on related matters, [12].

\vskip 1.5truecm

\bf
2. \ub{Elements of non-commutative geometry}.\rm
\medskip
This section is based on Connes' theory of non-commutative geometry,
as described in [4], and on results in [9,10,11,12].

We start by recalling the definition of a special case of Connes'
general definition of non-commutative spaces. A \ub{real, compact
non-commutative space} is defined by the data $({\Cal A}, \pi, H,
D)$, where ${\Cal A}$ is a $^*$algebra of bounded operators containing
an identity element, $\pi$ is a $^*$representation of ${\Cal A}$ on
$H$, where $H$ is a separable Hilbert space, and $D$ is a
\ub{selfadjoint} operator on $H$, with the following properties:
\medskip
\item{(i)} $\bigl[D,\pi(a)\bigr]$ is a bounded operator on $H$, for
all $a\in{\Cal A}$. [This condition determines the analogue of a
differentiable structure on the non-commutative space described by
${\Cal A}$.]

\item{} In the following, we shall usually identify the
algebra ${\Cal A}$ with the $^*$subalgebra $\pi({\Cal A})$ of the
algebra $B(H)$ of all bounded operators on $H$; (we shall thus assume
that the kernel of the representation $\pi$ in ${\Cal A}$ is trivial).
We shall often write ``$a$'' for both, the element $a$ of ${\Cal A}$
and the operator $\pi(a)$ on $H$.

\item{(ii)} $(D^2+\id)^{-1}$ is a compact operator on $H$. More
precisely, $\exp (-\varepsilon D^2)$ is trace-class, for any
$\varepsilon > 0$.
\medskip
Given a real, compact non-commutative space $({\Cal A}, \pi, H, D)$,
one defines a differential algebra, $\Omega_D({\Cal A})$, of forms as
follows: 0-forms (``scalars'') form a $^*$algebra with identity,
$\Omega_D^0({\Cal A})$, given by $\pi({\Cal A})$; $n$-forms form a
linear space, $\Omega_D^n({\Cal A})$, spanned by equivalence classes
of operators on $H$,
$$
\Omega_D^n({\Cal A})\;:=\;\Omega^n({\Cal A}) / Aux^n\;,\tag 2.1
$$
where the linear space $\Omega^n({\Cal A})$ is spanned by the
operators
$$
\bigl\{ \sum_i a_0^i [D,a_1^i]\cdots[D,a_n^i]\;:\; a_j^i \in\;
{\Cal A}\;\equiv \pi({\Cal A}),\;\forall i,j\bigr\},
\tag 2.2
$$
and $Aux^n$, the space of ``auxiliary fields'' [5], is spanned by
operators of the form
$$
\align
Aux^n\;:=\;\bigl\{\;&\sum_i [D,a_0^i][D,a_1^i]\cdots[D,a_n^i]\;:\\
&\sum_i a_0^i [D,a_1^i]\cdots [D,a_n^i]\;=\;0,\; a_j^i\in {\Cal
A}\bigr\}\;. \tag 2.3
\endalign
$$
Using the Leibniz rule
$$
[D,ab]\;=\;[D,a]b\;+\;a[D,b],\; a,b\in {\Cal A},
\tag 2.4
$$
and
$$
[D,a]^*\;=\;-\;[D,a^*],\quad a\in {\Cal A},
\tag 2.5
$$
we see that the spaces $\Omega^n({\Cal A})$ are ${\Cal A}$-bimodules
closed under the involution $^*$ and that
$$
Aux\;:=\;\oplus \;Aux^n
\tag 2.6
$$
is a two-sided ideal in
$$
\Omega ({\Cal A})\;:=\;\oplus\;\Omega^n ({\Cal A}),
\tag 2.7
$$
closed under the operation $^*$. Thus, for each $n$, $\Omega_D^n
({\Cal A})$ is an ${\Cal A}$-bimodule closed under $^*$. It follows
that
$$
\Omega_D ({\Cal A})\;:=\;\oplus\;\Omega_D^n({\Cal A})
\tag 2.8
$$
is a $^*$algebra of equivalence classes of bounded operators on $H$,
with multiplication defined as the multiplication of operators on $H$.
Since ${\Cal A} = \Omega^0({\Cal A})=\Omega_D^0({\Cal A})$ is a
$^*$subalgebra of $\Omega_D({\Cal A})$ containing an identity element,
$\Omega_D({\Cal A})$ is a unital $^*$algebra of equivalence classes \
(mod $Aux)$ of bounded operators on $H$ which is an ${\Cal
A}$-bimodule.

The degree of a form $\alpha \in \Omega_D^n({\Cal A})$ is defined by
$$
deg (\alpha)\;=\;n,\quad n\;=\;0,1,2,\cdots .
\tag 2.9
$$
Clearly, $deg (\alpha^*) = deg (\alpha)$, by (2.4), (2.5). With this
definition of $deg$, $\Omega_D({\Cal A})$ is $\Bbb Z$-graded. If
$\alpha$ is given by
$$
\alpha\;=\;\sum_i a_0^i [D,a_1^i]\cdots [D,a_n^i] (\text{mod} Aux^n) \in
\Omega_D^n ({\Cal A})
$$
we set
$$
d\alpha\;:=\;\sum_i [D,a_0^i] [D,a_1^i]\cdots [D,a_n^i] \in
\Omega_D^{n+1} ({\Cal A})\;. \tag 2.10
$$
The map
$$
d\;:\;\Omega_D^n ({\Cal A}) \;\rightarrow\;\Omega_D^{n+1} ({\Cal
A}), \quad \alpha\;\mapsto\; d\alpha
\tag 2.11
$$
is a $\Bbb C$-linear map from $\Omega_D({\Cal A})$ to itself which
increases the degree of a form by one and satisfies
$$
d(\alpha\cdot\beta)\;=\;(d\alpha)\cdot\beta\;+\;(-1)^{deg\,\alpha}\;
\alpha\cdot(d\beta),
\tag 2.12
$$
for any homogeneous element $\alpha$ of $\Omega_D({\Cal A})$ (Leibniz
rule) and
$$
d^2\;=\;0\;.
\tag 2.13
$$
Hence $\Omega_D({\Cal A})$ is a differential algebra which is a $\Bbb
Z$-graded complex.

These notions are described in detail (and in a more general setting)
in [4].

In non-commutative geometry,\  \ub{vector bundles} \ over a
non-commutative space described by a $^*$algebra ${\Cal A}$ are
defined as \ub{finitely generated}, \ub{projective left} ${\Cal
A}$-\ub{modules}. Let $E$ denote (the ``space of sections'' of) a
vector bundle over ${\Cal A}$. A connection $\nabla$ on $E$ is a
$\Bbb C$-linear map
$$
\nabla\;:\;E\;\rightarrow\;\Omega_D^1 ({\Cal A}) \;\otimes_{{\Cal
A}}\;E
\tag 2.14
$$
with the property that (with $da = [D,a]$, for all $a\in{\Cal A}$)
$$
\nabla (as)\;=\;da\;\otimes_{{\Cal A}}\;s + a\;\nabla s,
\tag 2.15
$$
for arbitrary $a\in{\Cal A}$, $s\in E$. The definition of $\nabla$
can be extended to the space
$$
\Omega_D(E)\;=\;\Omega_D({\Cal A})\;\otimes_{{\Cal A}}\;E
\tag 2.16
$$
in a canonical way, and, for $s\in \Omega_D(E)$ and a homogeneous form
$\alpha\in\Omega_D({\Cal A})$,
$$
\nabla(\alpha s)\;=\;(d\alpha) s\;+\;(-1)^{deg\,\alpha}\;\alpha \;
\nabla s .\tag 2.17
$$
Thanks to (2.14) - (2.17), it makes sense to define the
\ub{curvature}, $R(\nabla)$, of the connection $\nabla$ as the $\Bbb
C$-linear map
$$
R(\nabla)\;:=\;- \nabla^2
\tag 2.18
$$
from $\Omega_D(E)$ to $\Omega_D(E)$. Actually, it is easy to check
that $R(\nabla)$ is ${\Cal A}$-\ub{linear}, i.e. $R(\nabla)$ is a
\ub{tensor}.

A \ub{trivial} vector bundle, $E^{(N)}$, corresponds to a finitely
generated, \ub{free} left ${\Cal A}$-module, i.e., one that has a
basis $\{ s_1,\cdots, s_N\}$, for some finite $N$ called its
dimension. Then
$$
E^{(N)}\; \simeq\; {\Cal A} \oplus \cdots \oplus {\Cal A}\;\equiv\;
{\Cal A}^n,
$$
(with $N$ summands). The affine space of connections on $E^{(N)}$ can
be characterized as follows: Given a basis $\{ s_1,\cdots, s_N\}$ of
$E^{(N)}$, there are $N^2$ 1-forms $\rho_\alpha^\beta \in \Omega_D^1
({\Cal A})$, the \ub{components} of the connection $\nabla$, such that
$$
\nabla s_\alpha\;=\;-\;\rho_\alpha^\beta \otimes_{{\Cal A}} s_\beta,
\tag 2.19
$$
(where, here and in the following, we are using the summation
convention). Then
$$
\nabla (a^\alpha s_\alpha)\;=\;da^\alpha \otimes_{{\Cal A}} s_\alpha -
a^\alpha \rho_\alpha^\beta \otimes_{{\Cal A}} s_\beta,
\tag 2.20
$$
by (2.15). Furthermore, by (2.18) and (2.20),
$$
\align
R(\nabla)(a^\alpha s_\alpha)\;=\;
&-\;\nabla \bigl( da^\alpha \otimes_{{\Cal A}} s_\alpha - a^\alpha
\rho_\alpha^\beta \otimes_{{\Cal A}} s_\beta\bigr) \\
=\;&-\;\bigl(d^2 a^\alpha \otimes_{{\Cal A}} s_\alpha + da^\alpha
\rho_\alpha^\beta \otimes_{{\Cal A}} s_\beta \\
&-\;da^\alpha \rho_\alpha^\beta \otimes_{{\Cal A}} s_\beta - a^\alpha
d \rho_\alpha^\beta \otimes_{{\Cal A}} s_\beta \\
&-\; a^\alpha \rho_\alpha^\gamma \rho_\gamma^\beta \otimes_{{\Cal A}}
s_\beta\bigr)\\
=\;&\;a^\alpha \bigl(d \rho_\alpha^\beta + \rho_\alpha^\gamma
\rho_\gamma^\beta\bigr) \otimes_{{\Cal A}} s_\beta\;.
\tag 2.21
\endalign
$$
Thus, the curvature tensor $R(\nabla)$ is completely determined by the
$N\times N$ matrix $\theta\equiv(\theta_\alpha^\beta)$ of 2-forms
given by
$$
\theta_\alpha^\beta\;=\;d \rho_\alpha^\beta\;+\;\rho_\alpha^\gamma
\;\rho_\gamma^\beta\;.
\tag 2.22
$$
The curvature matrix $\theta$ satisfies the \ub{Bianchi identity}
$$
d\theta + \rho\theta - \theta\rho\;\equiv\;\bigl( d\theta_\alpha^\beta
+ \rho_\alpha^\gamma \theta_\gamma^\beta - \theta_\alpha^\gamma
\rho_\gamma^\beta\bigr)\;=\;0\;.
\tag 2.23
$$
If one introduces a new basis
$$
\tilde s_\alpha\;=\;M_\alpha^\beta s_\beta, \; M_\alpha^\beta \in {\Cal A},
\; \alpha,\beta\;=\;1,\cdots,N,
\tag 2.24
$$
where the matrix $M\equiv (M_\alpha^\beta)$ is invertible, then the
components, $\tilde\rho$, of $\nabla$ in the new basis
$\{\tilde s_1, \cdots, \tilde s_N\}$ of $E^{(N)}$ are given by
$$
\tilde\rho\;=\;M\rho\;M^{-1}\;-\;dM\cdot M^{-1},
\tag 2.25
$$
and the components of the curvature $R(\nabla)$ transform according to
$$
\widetilde\theta\;=\; M \theta M^{-1},
\tag 2.26
$$
as one easily checks.

Given a basis $\{ s_1,\cdots, s_N\}$ of $E^{(N)}$, one may define a
\ub{Hermitian structure} $\langle\cdot,\cdot\rangle$  on $E^{(N)}$ by
setting
$$
\langle s_\alpha, s_\beta\rangle\;=\;\delta_{\alpha\beta}\;\id \;,
\tag 2.27
$$
with
$$
\langle a^\alpha s_\alpha, b^\beta s_\beta\rangle\;=\; a^\alpha
\langle s_\alpha, s_\beta\rangle (b^\beta)^*\;=\; \sum_\alpha a^\alpha
(b^\alpha)^* \in {\Cal A}\;.
\tag 2.28
$$
The definition of $\langle\cdot ,\cdot\rangle$ can be extended
canonically to $\Omega_D(E^{(N)})$, and there is then an obvious
notion of ``\ub{unitary connection}'' on $E^{(N)}$: $\nabla$ is
unitary iff
$$
d\langle s,s'\rangle\;=\;\langle\nabla s, s'\rangle\;-\;\langle s,
\nabla s'\rangle\;.
\tag 2.29
$$
This is equivalent to the condition that
$$
\rho_\alpha^\beta\;=\;\bigl( \rho_\beta^\alpha\bigr)^*,
\tag 2.30
$$
where the $\rho_\alpha^\beta$ are the components of $\nabla$ in the
orthonormal basis $\{ s_1,\cdots , s_N\}$ of $E^{(N)}$.

In the examples studied in Sections 3 through 5, we shall consider
unitary connections on trivial vector bundles, in particular on
``\ub{line bundles}'' for which $N=1$. A (unitary) connection $\nabla$
on a line bundle $E^{(1)} \simeq {\Cal A}$ is completely determined
by a (selfadjoint) 1-form $\rho \in \Omega_D^1 ({\Cal A})$.

The data $({\Cal A},\pi,H,D)$ defining a non-commutative space with
differentiable structure is also called a \ub{Fredholm module}.
Following [4], we shall say that the Fredholm module $({\Cal
A},\pi,H,D)$ is $(d,\infty)$-summable if
$$
tr\,\bigl(D^2 + \id\bigr)^{- p/2} \;<\;\infty , \quad \text{for
all}\quad p > d\;.
\tag 2.31
$$
Let $Tr_\omega (\cdot)$ denote the so-called \ub{Dixmier trace} on
$B(H)$ which is a positive, cyclic trace vanishing on trace-class
operators; see [4]. We define a notion of \ub{integration of forms},
$\int\!\!\!\!\!-$, by setting
$$
\int\!\!\!\!\!\!- \alpha\;:=\; Tr_\omega\;\bigl( \alpha\mid
D\mid^{-d}\bigr),
\tag 2.32
$$
for \ $\alpha \in \Omega ({\Cal A}) = \oplus \Omega^n ({\Cal A})$; \
(see (2.2), (2.7)). If $d=\infty$ but \ exp$\;(-\varepsilon D^2)$ is
trace class, for any $\varepsilon > 0 $ (as assumed), we set
$$
\int\!\!\!\!\!\!-
\alpha\;:=\;\displaystyle\mathop{Lim_\omega}_{\varepsilon\downarrow
0}\; \frac{tr \bigl(\alpha \exp (-\varepsilon D^2)\bigr)}{tr
\bigl(\exp (- \varepsilon D^2)\bigr)}\;,
\tag 2.33
$$
(on forms $\alpha$ which are ``
analytic elements'' for the automorphism group determined by the
dynamics \ exp$(it D^2), t\in\Bbb R$; see [12]) and $Lim_\omega$
denotes a limit defined in terms of a kind of ``Cesaro mean''
described in [4]. Then
$$
\int\!\!\!\!\!\!- \alpha\beta\;=\;\int\!\!\!\!\!\!- \beta\alpha,
\tag 2.34
$$
i.e., $\int\!\!\!\!\!-$ is cyclic; it is also a \ub{non-negative}
linear functional on $\Omega ({\Cal A})$. It can thus be used to
define a positive semi-definite inner product on $\Omega({\Cal A})$:
For $\alpha$ and $\beta$ in $\Omega({\Cal A})$, we set
$$
(\alpha,\beta)\;=\;\int\!\!\!\!\!\!- \alpha\beta^*.
\tag 2.35
$$
Then the closure of $\Omega({\Cal A})$ (mod kernel of $(\cdot,\cdot)$)
in the norm determined by $(\cdot,\cdot)$ is a Hilbert space, denoted
by $L^2\bigl(\Omega({\Cal A})\bigr)$. Given an element
$\alpha\in\Omega^n({\Cal A})$, we can now define a canonical
representative, $\alpha^\perp$, in the equivalence class $\alpha$ (mod
$Aux^n) \in \Omega_D^n({\Cal A})$ as the unique (modulo the kernel of
$(\cdot,\cdot)$) operator in $\alpha$ (mod $Aux^n)$ which is
orthogonal to $Aux^n$ in the scalar product $(\cdot,\cdot)$ given by
(2.35); $(Aux^n$ has been defined in eq. (2.3)). Then, for $\alpha$
and $\beta$ in $\Omega_D({\Cal A})$, we set
$$
(\alpha,\beta)\;:=\;(\alpha^\perp,
\beta^\perp)\;\equiv\;\int\!\!\!\!\!\!- \alpha^\perp (\beta^\perp)^*,
\tag 2.36
$$
and this defines a positive semi-definite inner product on
$\Omega_D({\Cal A})$. The closure of $\Omega_D({\Cal A})$ (mod kernel
of $(\cdot,\cdot))$ in the norm determined by $(\cdot,\cdot)$ is the
Hilbert space of ``square-integrable differential forms'', denoted by
$\Lambda_D({\Cal A})$.

In order to define the Chern-Simons forms and Chern-Simons actions in
non-commut\-ative geometry, it is useful to consider a trivial example
of the notions introduced, so far. Let $I$ denote the interval
$[0,1]\subset\Bbb R$. Let ${\Cal A}_1 = C^\infty (I)$ be the algebra
of smooth functions, $f(t)$, on the open
interval (0,1) which, together with all their derivatives in $t$, have
(finite) limits as $t$ tends to 0 or 1. Let $H_1=L^2(I)\otimes\Bbb
C^2$ denote the Hilbert space of square-integrable (with respect to
Lebesgue measure, $dt$, on $I$) two-component spinors, and $D_1 = i
\frac{\partial}{\partial t} \otimes \sigma_1$ the one-dimensional
Dirac operator (with appropriate selfadjoint boundary conditions),
where $\sigma_1,\sigma_2$ and $\sigma_3$ are the usual Pauli matrices.
A representation $\pi_1$ of ${\Cal A}_1$ on $H_1$ is defined by
setting
$$
\pi_1(a)\;=\;a \otimes \id_2,\quad a\in{\Cal A}_1.
\tag 2.37
$$
The geometry of $I$ is then coded into the space $({\Cal A}_1,
\pi_1,H_1,D_1)$. The space of 1-forms is given by
$$
\Omega_{D_1}^1({\Cal A}_1)\;=\;\bigl\{ \omega\otimes\sigma_1 : \omega
= \sum_i a^i \partial_t b^i; a^i, b^i \in {\Cal A}_1\bigr\}.
\tag 2.38
$$
The space, $\Omega_{D_1}^2 ({\Cal A}_1)$, of 2-forms is easily seen to
be \ub{trivial}, and the cohomology groups vanish. The Fredholm module
$({\Cal A}_1,\pi_1, H_1, D_1)$ is $\Bbb Z_2$-graded.
The $\Bbb Z_2$-grading, $\gamma$, is given by
$$
\gamma\;=\;\id \otimes \sigma_3,
\tag 2.39
$$
and $[\gamma, \pi_1 (a)] = 0$, for all $a\in {\Cal A}_1$, while
$$
\bigl\{ \gamma, D_1\bigr\}\;\equiv\;\gamma D_1 + D_1\gamma \;=\;0.
\tag 2.40
$$

Using this trivial example, we may introduce the notion of a \
``\ub{cylinder over a non-} \ub{commutative space}'': Let $({\Cal
A},\pi,H,D)$ be an arbitrary non-commutative space, and let $({\Cal
A}_1,\pi_1,H_1,D_1)$ be as specified in the above example. Then we
define the cylinder over $({\Cal A},\pi,H,D)$ to be given by the
non-commutative space $(\widetilde{\Cal A},\widetilde\pi,\widetilde H,
\widetilde D)$, where
$$
\widetilde H = H\otimes H_1,\quad
\widetilde\pi = \pi \otimes \pi_1,\quad
\widetilde{\Cal A} = {\Cal A}\otimes{\Cal A}_1,
\tag 2.41
$$
and
$$
\widetilde D\;=\;\id \otimes D_1\;+\;D \otimes \gamma,
\tag 2.42
$$
with $\gamma$ as in (2.39). The space $(\widetilde{\Cal
A},\widetilde\pi,\widetilde H, \widetilde D)$ is $\Bbb Z_2$-graded: We
define
$$
\Gamma\;=\;\id \otimes (\id \otimes \sigma_2),
\tag 2.43
$$
$\widetilde D_1 := \id\otimes D_1$, $\widetilde D_2 = D\otimes\gamma$. Then
$$
\{ \Gamma,\widetilde D_1\}\;=\;\{ \Gamma,\widetilde
D_2\}\;=\;\{\Gamma,\widetilde D\}\;=\; \{\widetilde D_1, \widetilde
D_2\}\;=\;0,
\tag 2.44
$$
and
$$
\bigl[ \Gamma, \widetilde\pi (\tilde a)\bigr]\;=\;0,\quad\text{for all}
\quad \tilde a\in\widetilde{\Cal A}.
\tag 2.45
$$
It is easy to show (see [12]) that arbitrary sums of operators of the
form
$$
\tilde a_0 [\widetilde D_{\varepsilon_1}, \tilde a_1] \cdots
[\widetilde D_{\varepsilon_n}, \tilde a_n],\quad \varepsilon_1,
\cdots, \varepsilon_n = 1,2,
\tag 2.46
$$
belong to $\Omega^k(\widetilde{\Cal A})$. Furthermore, if two or more
of the $\varepsilon_i$'s take the value 1 then the operator defined in
(2.46) belongs to $Aux^n$.

We define integration, $\int\!\!\!\!\!\!\sim (\cdot)$, on $(\widetilde{\Cal
A},\widetilde\pi,\widetilde H,\widetilde D)$ by setting, for any
$\alpha\in\Omega (\widetilde{\Cal A})$,
$$
\int\!\!\!\!\!\!\!\sim \alpha\;:=\;\int_0^1 dt \int\!\!\!\!\!\!- Tr_{\Bbb
C^2} \bigl(\alpha (t)\bigr),
\tag 2.47
$$
where $\alpha (t)$ is a 2$\times$2 matrix of elements of $\Omega({\Cal
A})$. The integral $\int\!\!\!\!\!\!\sim (\cdot)$ is positive semi-definite
and cyclic on the algebra $\Omega (\widetilde{\Cal A})$. We are now
prepared to define the Chern-Simons forms and Chern-Simons actions in
non-commutative geometry, (for connections on trivial vector bundles).
Let $({\Cal A},\pi,H,D)$ be a real, compact non-commutative space with
a differentiable structure determined by $D$. Let $E=E^{(N)} \simeq
{\Cal A}^N$ be a trivial vector bundle over ${\Cal A}$, and let
$\nabla$ be a connection on $E$. By (2.19), $\nabla$ is completely
determined by an $N\times N$ matrix $\rho = (\rho_\alpha^\beta)$ of
1-forms. By (2.21), the curvature of $\nabla$ is given by the $N\times
N$ matrix of 2-forms
$$
\theta\;=\;d \rho + \rho^2,
$$
where $d$ is the differential on $\Omega_D ({\Cal A})$ defined in
(2.10). Following Quillen [11], we define the Chern-Simons
(2$n$--1)-form associated with $\nabla$ as follows: Let $\nabla_0$
denote the flat connection on $E$ corresponding to an $N\times N$
matrix $\rho_0$ of 1-forms which, in an appropriate gauge, vanishes.
We set
$$
\rho_t\;=\;t \rho + (1-t)\;\rho_0\;=\; t \rho,
\tag 2.48
$$
for $\rho_0=0$, corresponding to the connection $\nabla_t = t\nabla +
(1-t) \nabla_0$. The curvature of $\nabla_t$ is given by the matrix
$\theta_t$ of 2-forms given by
$$
\theta_t\;=\;d\rho_t + \rho_t^2\;=\;t d\rho + t^2\rho^2.
$$
The Chern-Simons $(2n-1)$-form associated with $\nabla$ is then given
by
$$
\vartheta^{2n-1} (\rho)\;:=\;\frac{1}{(n-1)!} \int_0^1 dt \;\rho\;
\theta_t^{n-1}. \tag 2.49
$$
For $n=2$, we find
$$
\vartheta^3 (\rho)\;=\;\frac 1 2\;\bigl\{ \rho d \rho + \frac 2 3
\;\rho^3 \bigr\},
\tag 2.50
$$
and, for $n=3$,
$$
\vartheta^5 (\rho)\;=\;\frac 1 6 \bigl\{ \rho d\rho d\rho + \frac 3 4
\rho^3 d \rho + \frac 3 4 \rho (d\rho) \rho^2 + \frac 3 5 \rho^5
\bigr\}.
\tag 2.51
$$

In order to understand where these definitions come from and how to
define Chern-Simons actions, we extend $E$ to a trivial vector bundle
over the cylinder $(\widetilde{\Cal A},\widetilde\pi,\widetilde H,
\widetilde D)$ over $({\Cal A}, \pi, H, D)$: We set
$$
\widetilde E\;=\; E \otimes C^\infty (I) \otimes \id_2 \simeq
\widetilde{\Cal A}^N .
\tag 2.52
$$
We also extend the connection $\nabla$ on $E$ to a connection
$\widetilde\nabla$ on $\widetilde E$ by interpolating between $\nabla$
and the flat connection $\nabla_0$: By (2.39), (2.41) and (2.42), a
1-form in $\Omega_{\widetilde D}^1 (\widetilde{\Cal A})$ is given by
$$
\pmatrix
\rho (t) & \quad \phi (t) \\
\phi (t) & - \rho (t)
\endpmatrix ,
$$
where $\rho (t)\in \Omega_D^1({\Cal A})$ for all $t\in I$. Thus
$$
\tilde\rho_\alpha^\beta (t)\;:=\;t\enskip
\pmatrix
\rho_\alpha^\beta & \enskip 0 \\
0 & - \rho_\alpha^\beta
\endpmatrix , \quad
\alpha,\beta = 1, \cdots, N,
\tag 2.53
$$
is an element of $\Omega_{\widetilde D}^1 (\widetilde{\Cal A})$. We
define $\widetilde \nabla$ to be the connection on $\widetilde E$
determined by the matrix $\widetilde \rho (t) = \bigl(
\widetilde\rho_\alpha^\beta (t)\bigr)$ of 1-forms defined in (2.53).
Let $\tilde d$ be the differential on $\Omega_{\widetilde
D}(\widetilde{\Cal A})$ defined as in (2.10), (with ${\Cal A}$
replaced by $\widetilde{\Cal A}$ and $D$ replaced by $\widetilde D$).
By (2.42)
$$
\tilde d \tilde\rho \;=\;
\pmatrix 0 &- i\rho \\
i\rho &\enskip 0
\endpmatrix\;+\;t\;
\pmatrix
d\rho & 0 \\
\enskip 0 & d\rho
\endpmatrix ,
\tag 2.54
$$
with $\rho = (\rho_\alpha^\beta)$. Hence the curvature of
$\widetilde\nabla$ is
given by the matrix of 2-forms $\widetilde\theta$, with
$$
\widetilde\theta (t)\;=\;\theta_t\otimes\id_2 + \rho\otimes\sigma_2,
\tag 2.55
$$
where
$$
\theta_t\;=\;t d \rho + t^2 \rho^2.
$$
Let $\varepsilon$ be an arbitrary bounded operator on $H$ commuting
with $D$ and with all operators in $\pi ({\Cal A})$. Then,
$$
\widetilde\varepsilon\;:=\;\varepsilon\otimes\id_2
$$
commutes with $\widetilde D$ and with $\widetilde\pi (\widetilde{\Cal
A})$ and hence with $\Omega(\widetilde{\Cal A})$. It also commutes
with the $\Bbb Z_2$-grading $\Gamma =\id\otimes (\id \otimes \sigma_2)$, (as
defined in (2.43)). We now define a graded trace $\tau_\epsilon
(\cdot)$ on $\Omega(\widetilde{\Cal A})$ by setting
$$
\tau_\varepsilon (\alpha)\;:=\;\int\!\!\!\!\!\!\!\sim (\tilde\varepsilon
\Gamma \alpha), \quad \alpha \in\Omega (\widetilde{\Cal A}),
\tag 2.56
$$
where $\int\!\!\!\!\!\!\sim (\cdot)$ is given by (2.47). It is easy to show
that
$$
\tau_\varepsilon (\alpha)\;=\;0 \quad \text{if deg } \alpha\quad
\text{is \ub{odd}},
\tag 2.57
$$
and
$$
\tau_\varepsilon\bigl([\alpha,\beta]_*\bigr)\;=\;0,\quad
\text{for all } \alpha,\beta\quad\text{in } \Omega(\widetilde{\Cal
A}), \tag 2.58
$$
where
$$
[\alpha,\beta]_*\;=\;\alpha\cdot\beta - (-1)^{\deg \alpha \deg
\beta}\; \beta\cdot\alpha
$$
is the graded cummutator.

Using the Bianchi identity,
$$
d\theta^n + [\rho, \theta^n] = 0,
$$
which follows from eq.~(2.23) by induction, and the graded cyclicity
of $\tau_\varepsilon$ (see (2.57), (2.58)) one shows that
$$
\tau_\varepsilon \bigl((\widetilde\theta^n)^\perp\bigr)\;=\; n!
\tau_\varepsilon \bigl(\bigl(\tilde d \vartheta^{2n-1}
(\widetilde\rho)\bigr)^\perp\bigr),
\tag 2.59
$$
where $\alpha^\perp$ is the canonical representative in the
equivalence class $\alpha$ (mod $Aux^m) \in \Omega_{\widetilde D}^m
(\widetilde{\Cal A})$ orthogonal to $Aux^m$, for any $m=1,2,\cdots$,
(as explained after eq.~(2.35)). The calculation proving (2.59) is
indicated in [11]; (see also [12] for details). In fact, eq.~(2.59) is
a general identity valid for arbitrary connections on trivial vector
bundles over a non-commutative space and arbitrary graded traces [11].

In the case considered here, the l.h.s. of eq.~(2.59) can be rewritten
in the following interesting way:
$$
\align
\tau_\varepsilon\bigl((\widetilde\theta^n)^\perp\bigr)\;&=\;
\int_0^1 dt \int\!\!\!\!\!\!- Tr_{\Bbb C^2} \bigl(\tilde\varepsilon\; \Gamma
\bigl( \widetilde\theta^n (t)\bigr)^\perp\bigr) \\
&=\;n \int_0^1 dt \int\!\!\!\!\!\!- Tr_{\Bbb C^2}
\bigl((\varepsilon\otimes\id_2)\, \Gamma\, (\rho\otimes\sigma_2)(\theta_t^{n-1}
\otimes \id_2)\bigr).
\tag 2.60
\endalign
$$
This is shown by plugging eq.~(2.55) for $\widetilde\theta (t)$ into
the expression in the middle of (2.60) and noticing that (1) all terms
contributing to $\widetilde\theta^n(t)$ with more than one factor
proportional to $[\widetilde D_1, \tilde a]$, i.e., with more than one
factor of the form $\rho\otimes\sigma_2$, are projected out when
passing from $\widetilde\theta^n(t)$ to $\bigl(\widetilde\theta^n
(t)\bigr)^\perp$, (see the remark following eq.~(2.46)), and \ (2) \
$Tr_{\Bbb C^2} \bigl((\varepsilon\otimes\id_2)\; \Gamma$ \break
$(\theta_t^n\otimes\id_2)\bigr)=0$.

Evaluating the trace, $Tr_{\Bbb C^2}$, on the r.h.s. of (2.60) and
recalling the definition (2.49) of the Chern-Simons form, we finally
conclude that
$$
\align
\tau_\varepsilon\bigl((\widetilde\theta^n)^\perp\bigr)\;
&=\;2n \int_0^1 dt \int\!\!\!\!\!\!-
\bigl(\varepsilon\rho\theta_t^{n-1}\bigr)^\perp \\
&=\;2n! \int\!\!\!\!\!\!- \bigl(\varepsilon\bigl(\vartheta^{2n-1}
(\rho)\bigr)^\perp\bigr).
\tag 2.61
\endalign
$$
\bf\ub{Remark}\rm . The r.h.s. of (2.59) can actually be rewritten as
$$
n! \int\!\!\!\!\!\!- Tr_{\Bbb C^2} \bigl( \id \otimes \sigma_3 \bigl(
\vartheta^{2n-1} \bigl( \widetilde\rho (1)\bigr)\bigr)^\perp\bigr);
$$
see [12].

\ub{Chern-Simons actions}, $I_\varepsilon$, in non-commutative
geometry are defined by setting
$$
I_\varepsilon^{2n-1}(\rho)\;:=\;\kappa\int\!\!\!\!\!\!-
\bigl(\varepsilon\bigl( \vartheta^{2n-1} (\rho)\bigr)^\perp\bigr),
\tag 2.62
$$
where $\kappa$ is a constant. Using (2.50) and (2.51) and using the
properties of $\varepsilon$ and the cyclicity of $\int\!\!\!\!\!-
(\cdot)$, we find
$$
I_\varepsilon^3 (\rho)\;=\;\frac\kappa 2 \int\!\!\!\!\!\!-
\bigl(\varepsilon\bigl( \rho d \rho + \frac 2 3
\rho^3\bigr)^\perp\bigr),
$$
and
$$
I_\varepsilon^5 (\rho)\;=\;\frac\kappa 6 \int\!\!\!\!\!\!-
\bigl(\varepsilon \bigl( \rho d \rho d \rho + \frac 3 2 \rho^3 d \rho
+ \frac 3 5 \rho^5 \bigr)^\perp\bigr).
\tag 2.63
$$
A particularly important special case is obtained by choosing the
operator $\varepsilon$ to belong to $\Omega({\Cal A})$.
Since $\varepsilon$ commutes with $D$ and with $\pi({\Cal A})$, this
implies that $\varepsilon$ belongs to the \ub{centre} of the algebra
$\Omega({\Cal A})$. In the examples discussed in the remainder of this
paper, this property is always assumed.

One point of formula (2.61) and generalizations thereof, discussed in
[12] (and involving ``higher-dimensional cylinders''), is that it
enables us to define \ub{differences} of Chern-Simons actions even
when the underlying vector bundle is \ub{non-trivial}. If $\nabla_0$
denotes a fixed reference connection on a vector bundle $E$ over a
non-commutative space $({\Cal A},\pi,H,D)$ and $\nabla$ is an
arbitrary connection on $E$ we set
$$
\int\!\!\!\!\!\!- \bigl(\varepsilon\bigl(\vartheta^{2n-1}
(\nabla)\bigr)^\perp\bigr)\;:=\;\tau_\varepsilon \bigl(
(\widetilde\theta^n)^\perp\bigr)\;+\;\text{const.,}
\tag 2.64
$$
where $\widetilde\theta$ is the curvature of a connection
$\widetilde\nabla$ on a vector bundle $\widetilde E$ over the cylinder
$(\widetilde{\Cal A},\widetilde\pi,\widetilde H, \widetilde D)$
interpolating between $\nabla$ and $\nabla_0$, and the constant on the
r.h.s. of (2.64) is related to the choice of $\widetilde\nabla$ and of
the Chern-Simons action associated with $\nabla_0$.

Formulas (2.62) and (2.64) are helpful in understanding the
topological nature of Chern-Simons actions.

Next, we propose to discuss various concrete examples and indicate
some applications to theories of gravity.

\vskip 1.5truecm
\bf 3. \underbar{Some ``three-dimensional'' Chern-Simons actions}\rm.
\medskip

We consider a ``Euclidean space-time manifold'' $X$ which is the
Cartesian product of a Riemann surface $M_2$ and a two-point set, i.e.
$X$ consists of two copies of $M_2$. The algebra ${\Cal A}$ used in
the definition of the non-commutative space considered in this section
is given by
$$
{\Cal A}\;=\;C^\infty (M_2) \otimes {\Cal A}_0,
\tag 3.1
$$
where ${\Cal A}_0$ is a finite-dimensional, unital $^*$algebra of
$M\times M$ matrices. The Hilbert space $H$ is chosen to be
$$
H\;=\;H_0\oplus H_0,
\tag 3.2
$$
where
$$
H_0\;=\;\Bbb C^N \otimes L^2 (S) \otimes \Bbb C^M,
\tag 3.3
$$
and $L^2(S)$ is the Hilbert space of square-integrable spinors on
$M_2$ for some choices of a spin structure and of a (Riemannian)
volume form on $M_2$.

The representation $\pi$ of ${\Cal A}$ on $H$ is given by
$$
\pi (a)\;=\;
\pmatrix
\id_N \otimes a & \quad 0 \\
\quad 0 & \id_N\otimes a
\endpmatrix ,
\tag 3.4
$$
for $a \in {\Cal A}$.

We shall work locally over some coordinate chart of $M_2$, but we do
not describe how to glue together different charts (this is standard),
and we shall write ``$M_2$'' even when we mean a coordinate chart of
$M_2$. Let $g=(g_{\mu\nu})$ be some fixed, Riemannian reference metric
on $M_2$, and $(e_\mu^a)$ a section of orthonormal 2-frames, $\mu,a =
1,2$. Let $\gamma^1, \gamma^2$ denote the two-dimensional Dirac
matrices satisfying
$$
\bigl\{ \gamma^a, \gamma^b\bigr\}\;\equiv\;
\gamma^a\gamma^b + \gamma^b\gamma^a\;=\;2\delta^{ab},
\tag 3.5
$$
and
$$
\gamma^5\;=\;\gamma^1 \gamma^2.
\tag 3.6
$$
The matrices $\id_N\otimes\gamma^a$ will henceforth also be denoted by
$\gamma^a$. Let $\partial\!\!\!/$
denote the covariant Dirac operator on $\Bbb C^N \otimes L^2(S)$
corresponding to the Levi-Civita spin connection determined by
$(e_\mu^a)$ and acting trivially on $\Bbb C^N$. Let $K$ denote an
operator of the form
$$
K\;=\;k\otimes\id\otimes\id,
\tag 3.7
$$
where $k$ is some real, symmetric $N\times N$ matrix. The vector space
$\Bbb C^N$ and the matrix $k$ do not play any interesting role in the
present section but are introduced for later convenience. Let $\phi_0$
be a hermitian $M\times M$ matrix $(\neq \id_M)$. The operator $D$ on
$H$ required in the definition of a non-commutative space is chosen as
$$
D\;=\; \pmatrix
\partial\!\!\!/ \otimes \id_M\quad & i\gamma^5 K\otimes \phi_0 \\
- i\gamma^5 K\otimes\phi_0 & \partial\!\!\!/ \otimes \id_M \endpmatrix ,
\tag 3.8
$$
Then (\ub{locally} on $M_2$) the space of 1-forms, $\Omega_D^1 ({\Cal
A})$, (the ``cotangent bundle'') is a free, hermitian ${\Cal
A}$-bimodule of dimension 3, with an orthonormal basis given by
$$
\varepsilon^a\;=\;\pmatrix
\gamma^a \otimes \id_M \quad & \quad 0 \quad\\
\quad 0\quad  & \gamma^a \otimes \id_M \endpmatrix ,
\quad a\;=\;1,2,
\tag 3.9
$$
and
$$
\varepsilon^3\;=\; \pmatrix
\quad 0 \qquad & - \gamma^5 \otimes \id_M \\
\gamma^5 \otimes \id_M & \quad 0 \endpmatrix
\phantom{\quad a\;=\;1,1,\quad }
\tag 3.10
$$
and the hermitian structure is given by the normalized trace, $tr$, on
$\Bbb M_N(\Bbb C)\otimes$ Cliff. Then, for $a,b=1,2,3,$
$$
\langle \varepsilon^a,\varepsilon^b\rangle\;=\;tr\,\bigl(\varepsilon^a
(\varepsilon^b)^*\bigr)\;=\;\delta^{ab}\id_M.
$$
We define a central element $\varepsilon\in\Omega^3({\Cal A})$ by
setting
$$
\varepsilon\;=\;\varepsilon^1\varepsilon^2\varepsilon^3\;=\;\pmatrix
\quad 0\quad & \id\\
- \id & 0\endpmatrix .
\tag 3.11
$$
It is trivial to verify that $\varepsilon $ commutes with the operator
$D$ and with $\pi({\Cal A})$, and, since $\varepsilon^1,\varepsilon^2$
and $\varepsilon^3$ belong to $\Omega^1 ({\Cal A})$, $\varepsilon$
belongs to $\Omega^3({\Cal A})$.

A 1-form $\rho$ has the form
$$
\rho\;=\;\sum_j \pi (a^j) \bigl[ D, \pi (b^j)\bigr], \quad a^j,b^j\in
{\Cal A},
\tag 3.12
$$
and, without loss of generality, we may impose the constraint
$$
\sum_j a^j b^j\;=\;\id .
\tag 3.13
$$
Then
$$
\rho\;=\;\pmatrix
\quad A \qquad & i \gamma^5 K\phi \\
- i \gamma^5 K\phi & \quad A \quad \endpmatrix ,
\tag 3.14
$$
where \ $A = \sum_j a^j (\partial\!\!\!/ b^j)$, and
$\phi+\phi_0=\id_N\otimes \bigl( \sum_j a^j \phi_0 b^j\bigr)$. \
The 1-form $\rho$ given in (3.14) determines a connection, $\nabla$,
on the ``line bundle'' $E \simeq {\Cal A}$. The three-dimensional
Chern-Simons action of $\nabla$ is then given by
$$
\align
I_\varepsilon^3 (\rho)\; &=\;\frac \kappa 2 \;\int\!\!\!\!\!\!- \;
\bigl(\varepsilon (\rho\, d\rho + \frac 2 3\;\rho^3)^\perp\bigr)  \\
&=\;\frac \kappa 2\; Tr_\omega\;\bigl(\varepsilon (\rho\, d\rho\;+\;
\frac 2 3 \;\rho^3)^\perp D^{-2}\bigr),
\tag 3.15
\endalign
$$
as follows from eqs.~(2.63) and (2.32); \ (the Fredholm module \ $({\Cal
A},\pi,H,D)$ \ is \ $(2,\infty)$-summable!).

In order to proceed in our calculation, we must determine the spaces
of ``auxiliary fields'' $Aux^n$, for $n=1,2,3$. Clearly $Aux^1=0$. To
identify $Aux^2$, we consider a 1-form
$$
\rho\;=\;\sum_j\pi (a^j) \bigl[ D,\pi (b^j)\bigr]\;=\;
\pmatrix
\quad A \quad & i\gamma^5 K\phi \\
- i\gamma^5 K\phi & \quad A \quad \endpmatrix .
$$
Then
$$
\align
d\rho\;&=\;\sum_j\bigl[ D,\pi (a^j)\bigr]\bigl[D,\pi\bigr)b^j)\bigr]
\\
&=\;\pmatrix
\frac  1 2 \;\gamma^{\mu\nu}\partial_\mu A_\nu + X, \phantom{Zeichnung}&
-i\gamma^5\gamma^\mu K (\partial_\mu\phi + A_\mu \phi_0 - \phi_0
A_\mu) \\
i\gamma^5\gamma^\mu K (\partial_\mu\phi + A_\mu \phi_0-\phi_0 A_\mu), &
\frac 1 2 \; \gamma^{\mu\nu} \partial_\mu A_\nu + X
\phantom{Zeichnung}\endpmatrix ,
\endalign
$$
where $X=\id_N\otimes\bigl(\sum_j a^j \partial^\mu \partial_\mu
b^j\bigr)+\partial^\mu A_\mu$ is an arbitrary element of $\id_N\otimes
{\Cal A}$, and $\gamma^{\mu\nu} := [\gamma^\mu, \gamma^\nu]$. Hence
$$
Aux^2\;\simeq\;\pi ({\Cal A}).
\tag 3.16
$$
Next, let $\eta\in\Omega^2 ({\Cal A})$. Then one finds that
$$
d \eta\bigm|_{\eta=0} \;=\;\varepsilon\enskip
\pmatrix
\gamma^\mu X_\mu  & i\gamma^5 KX \\
- i\gamma^5 K X & \gamma^\mu X_\mu \endpmatrix ,
\tag 3.17
$$
where $X_\mu$ and $X$ are arbitrary elements of $\id_N \otimes {\Cal
A}$. Thus, in a 3-form $\vartheta$, terms proportional to $\gamma^\mu
\otimes \id_M$ in the off-diagonal elements and terms proportional to
$\gamma^5 K\otimes \id_M$ in the diagonal elements must be discarded
when evaluating $\vartheta^\perp$.

Next, we propose to check under what conditions the Chern-Simons
action $I_\varepsilon^3 (\rho)$ is gauge-invariant. In eq.~(2.25) we
have seen that, under a gauge transformation $M\in\pi({\Cal A})$,
$\rho$ transforms according to
$$
\rho\mapsto \tilde\rho\;=\; M\rho M^{-1} - (dM) M^{-1}\;=\;
g^{-1} \rho g + g^{-1} dg,
\tag 3.18
$$
with $g=M^{-1}$. From this equation and the cyclicity of
``integration'', $\int\!\!\!\!\!- (\cdot)$, we deduce that
$$
\align
I_\varepsilon^3 (\tilde \rho) \;&=\; I_\varepsilon^3 (\rho)\;+\\
&\frac \kappa 2\;\int\!\!\!\!\!\!- \bigl(\varepsilon \bigl\{ dg^{-1}
\rho dg + g^{-1} d\rho dg - \frac 1 3 \; (g^{-1} dg)^3\bigr\}^\perp
\bigr) .
\tag 3.19
\endalign
$$
The second term on the r.h.s. of (3.19) is equal to
$$
\align
\int_{M_2} d\,vol\,tr\;\bigl(\varepsilon\bigl\{[D,g^{-1}]
&\rho [D,g] + g^{-1}  d\rho [D,g] \\
&- \frac 1 3\; \bigl(g^{-1} [D,g]\bigr)^3\bigr\}^\perp\bigr).
\tag 3.20
\endalign
$$
Here, and in the following, $tr(\cdot)$ denotes a \ub{normalized
trace}, $\bigl( tr (\id) =1\bigr)$. A straightforward calculation
shows that
$$
[D,g]\;=\;\pmatrix
\partial\!\!\!/\; g& i\gamma^5 K (\phi_0 g - g\phi_0) \\
- i \gamma^5 K (\phi_0 g - g\phi_0) & \partial\!\!\!/\; g \endpmatrix ,
\tag 3.21
$$
and expression (3.20) is found to be given by
$$
\align
- i \,tr\,K \int_{M_2} \partial_\mu\,
&tr\;\bigl[ g^{-1} \phi \partial_\nu g + A_\nu \bigl( \phi_0 - g
\phi_0 g^{-1}\bigr) \\
&-\;\bigl( g \partial_\nu g^{-1} \phi_0 + g^{-1} \partial_\nu g
\phi_0\bigr)\bigr] \; dx^\mu \wedge dx^\nu,
\tag 3.22
\endalign
$$
which vanishes if $\partial M_2 = \phi$ (i.e., $M_2$ has no boundary).
\bigskip
\bf\ub{Remark}. \rm \ Had we considered a more general setting with \
${\Cal A} = C^\infty (M_2) \otimes {\Cal A}_1 \oplus C^\infty (M_2)
\otimes {\Cal A}_2$, where ${\Cal A}_1$ and ${\Cal A}_2$ are two
independent matrix algebras, and $\pi(a)=\id_N\otimes a$, for $a\in
{\Cal A}$, then, with $\varepsilon$ chosen as above,
$I_\varepsilon^3(\rho)$ would fail to be gauge-invariant.

Thus the condition for $I_\varepsilon^3(\rho)$ to be gauge-invariant
is that $\partial M_2=\phi$ and that the non-commutative space is
invariant under permuting the two copies of $M_2$ (of the space
$X_c$), i.e., the elements of $\pi ({\Cal A})$ commute with the
operator $\varepsilon$ defined in (3.11).

Under this condition one finds, after a certain amount of algebra,
that
$$
I_\varepsilon^3(\rho)\;=\;i\;\kappa \int_{M_2} tr\;(\Phi F),
\tag 3.23
$$
where
$$
\Phi\;=\;K (\phi+\phi_0),
$$
and
$$
F\;=\;\bigl(\partial_\mu A_\nu - \partial_\nu A_\mu + [A_\mu,
A_\nu]\bigr)\;
dx^\mu\wedge dx^\nu .
\tag 3.24
$$
We note that $I_\varepsilon^3(\rho)$ is obviously gauge-invariant and
topological, i.e., metric-independent. Since the ``Dirac operator''
$D$ depends on the reference metric $g$ on $M_2$, the
metric-independence of $I_\varepsilon^3$ is not, a priori, obvious
from its definition (3.15).

Let us consider the special case where
$$
{\Cal A}_o\;=\;\Bbb M_3 (\Bbb R),
\tag 3.25
$$
the algebra of \ub{real} 3$\times$3 matrices. Then
$$
I_\varepsilon^3 (\rho)\;=\;i\,\kappa \int_{M_2} \biggl( \sum_{A=1}^3
\Phi^A F_{\mu\nu}^A\biggr)\; dx^\mu \wedge dx^\nu
\tag 3.26
$$
where $F_{\mu\nu}^A=\partial_\mu A_\nu^A-\partial_\nu A_\mu^A+
\varepsilon^{ABC} A_{[\mu}\, ^B \; A_{\nu]}\,^C$,
with $A=a,3,a=1.2.$ \ Setting
$$
A_\mu^a\;=\;e_\mu^a,\quad A_\mu^3\;=\;\frac 1
2\;\omega_\mu^{ab}\;\varepsilon_{ab} \;\equiv\;\omega_\mu
\tag 3.27
$$
one observes that the action $I_\varepsilon^3$ is the one of
two-dimensional topological gravity introduced in [13]. Varying
$I_\varepsilon^3$ w.r. to $\Phi^A$ one obtains the zero-torsion and
constant-curvature conditions:
$$
\align
&\varepsilon^{\mu\nu} F_{\mu\nu}^a\;\equiv\;
 \varepsilon^{\mu\nu} T_{\mu\nu}^a\;=\;\varepsilon^{\mu\nu}
\bigl( \partial_\mu e_\nu^a+\,\frac 1 2 \sum_b \omega_\mu
\varepsilon^{ab} e_\nu^b\bigr)\;=\;0 \\
&\varepsilon^{\mu\nu} F_{\mu\nu}^3\;=\;\frac 1 2\;
 \varepsilon^{\mu\nu} R_{\mu\nu}^{ab}\, \varepsilon_{ab}\;=\;\frac 1 2
\;\varepsilon^{\mu\nu}\,
\bigl( \partial_\mu \omega_\nu + 2 \;\varepsilon_{ab}\; e_\mu^a\;
e_\nu^b\bigr)\;=\;0.
\tag 3.28
\endalign
$$
Variation of $I_\varepsilon^3$ with respect to $A_\mu^A$ implies that
$\Phi^A$ is covariantly constant, i.e.,
$$
D_\mu \Phi^A\;=\;\partial_\mu\Phi^A + \varepsilon^{ABC} A_\mu^B
\Phi^C\;=\;0.
\tag 3.29
$$
The space of solutions of (3.28) and (3.29) is characterized in [13].

These results suggest that the study of Chern-Simons actions in
non-commutative geometry is worthwhile.
\bigskip

\vskip 1.5truecm
\bf 4. \ub{Some ``five-dimensional'' Chern-Simons actions} \rm .
\medskip
In this section, we consider non-commutative space $({\Cal
A},\pi,H,D)$ with ``cotangent bundles'' $\Omega_D^1({\Cal A})$ that
are free, hermitian ${\Cal A}$-bimodules of dimension 5, and we
evaluate the ``five-dimensional'' Chern-Simons action,
$I_\varepsilon^5(\rho)$, defined in eq.~(2.63), for connections on the
``line bundle'' $E=E^{(1)} \simeq {\Cal A}$. We shall consider
algebras ${\Cal A}$ generated by matrix-valued functions on Riemann
surfaces or on four-dimensional spin manifolds. We start with the
analysis of the latter example.
\medskip

\bf (I)\rm \ We choose $X=M_4\times \{-1,1\}$, where $M_4$ is a
four-dimensional, smooth Riemannian spin manifold. The non-commutative
space $({\Cal A},\pi,H,D)$ is chosen as in Sect.~3, except that $M_2$
is replaced by $M_4$, the 2$\times$2 Dirac matrices
$\gamma^1,\gamma^2$ are replaced by the 4$\times$4 Dirac matrices
$\gamma^1,\gamma^2,\gamma^3,\gamma^4$, and
$\gamma^5=\gamma^1 \gamma^2 \gamma^3\gamma^4$. The definition of the
``Dirac operator'' $D$ is analogous to that in eq.~(3.8).

An orthonormal basis for $\Omega_D^1({\Cal A})$ is then given
(locally on $M_4$) by
$$
\varepsilon^a\;=\;\pmatrix
\gamma^a\otimes \id_M & \quad 0\quad \\
\quad 0\quad & \gamma^a\otimes\id_M \endpmatrix ,
 a=1,\cdots,4, \quad \varepsilon^5\;=\;
\pmatrix
\quad 0\quad & \gamma^5\otimes\id_M \\
- \gamma^5\otimes\id_M & \quad 0\quad \endpmatrix,
\tag 4.1
$$
and
$$
\varepsilon\;=\;\varepsilon^1\;\varepsilon^2\;\varepsilon^3\;
\varepsilon^4\; \varepsilon^5\;=\;\pmatrix
\enskip 0\quad & \id \\
-\id & 0 \endpmatrix ,
\tag 4.2
$$
similarly as in (3.11).

Again, we must determine the spaces $Aux^n$, $n=1,2,3,4,5$, of
``auxiliary fields''. The most important one is $Aux^5$. To determine
it, let us consider a vanishing element $\eta$ of $\Omega^4({\Cal
A})$ and compute $d\eta$, as given by eq.~(2.10). After a certain
amount of labouring one finds that
$$
d\eta\bigm|_{\eta=0}\;=\;\varepsilon \pmatrix
\gamma^{\mu\nu\rho} X_{\mu\nu\rho} + \gamma^\mu (K^2 X_\mu+Y_\mu),
& i \gamma^5(\gamma^{\mu\nu} K X_{\mu\nu} + K^3 X+KY) \\
- i \gamma^5 (\gamma^{\mu\nu} K X_{\mu\nu} + K^3 X+KY),
& \gamma^{\mu\nu\rho} X_{\mu\nu\rho} + \gamma^\mu (K^2 X_\mu+Y_\mu)
\endpmatrix ,
\tag 4.3
$$
where \ $X_{\mu\nu\rho}, X_{\mu\nu}, X_\mu, X, Y_\mu$ and $Y$ are
arbitrary
elements of $\id_N\otimes{\Cal A}$, and $\gamma^{\mu\nu\rho} =
\displaystyle\mathop{\Sigma}_{a,b,c}$ sig \  $\mu\nu\rho \choose a b c $ \
$\gamma^a\gamma^b\gamma^c$.
By (4.3), the passage from an element $\vartheta\in\Omega^5({\Cal A})$
to $\vartheta^\perp$ amounts to discarding all terms proportional to
$\gamma^{\mu\nu\rho} \otimes\id_M$, $K^2\gamma^\mu\otimes\id_M$ and
$\gamma^\mu\otimes\id_M$ from off-diagonal elements of $\vartheta$
and all terms proportional to $K\gamma^5\gamma^{\mu\nu}\otimes\id_M$,
$K^3\gamma^5\otimes \id_M$ and $K\gamma^5\otimes\id_M$ from the
diagonal elements of $\vartheta$. Now we start understanding the
useful role played by the matrix $K$.

It is then easy  to
evaluate $I_\varepsilon^5(\rho)$, with $\rho$ given by
$$
\rho\;=\;\pmatrix \quad A\quad & i\gamma^5 K\phi \\
- i \gamma^5 K \phi &\quad A\quad \endpmatrix ,
\quad A\;=\;\gamma^\mu A_\mu .
\tag 4.4
$$
Using eq.~(2.63), the result is
$$
I_\varepsilon^5 (\rho)\;=\;i\;\frac{3\kappa}{4}\;\int_{M_4} Tr\;
(\Phi\;F\wedge F) ,
\tag 4.5
$$
where $\Phi = K(\phi+\phi_0)$, and $F=F_{\mu\nu} dx^\mu \wedge
dx^\nu$, with $F_{\mu\nu}$ the curvature, or field strength, of
$A_\mu$. Provided that $\partial M_4=\emptyset$, $I_\varepsilon^5$ is
gauge-invariant and topological (metric-independent), as expected. The
field equation obtained by varying $I_\varepsilon^5$ w.r. to $\Phi$ is
$$
\varepsilon^{\mu\nu\rho\sigma}\;F_{\mu\nu}\;F_{\rho\sigma}\;=\;0 .
\tag 4.6
$$
Setting $\Phi$ to a constant, $I_\varepsilon^5$ turns out to be the
action of four-dimensional, topological Yang-Mills theory [16] before
gauge-fixing.
\medskip
\bf (II) \rm \ We choose $X=M_2\times \{-1,1\}$ and ${\Cal A},\pi$ and
$H$ as above, but the operator $D$ is given by
$$
D\;=\;\pmatrix
\quad\partial\!\!\!/\quad & K \gamma^\alpha \phi_{0\alpha} \\
- K \gamma^\alpha \phi_{0\alpha} &\quad\partial\!\!\!/\quad
\endpmatrix ,
\tag 4.7
$$
where $\partial\!\!\!/ = \gamma^1\partial_1+\gamma^2\partial_2$, and
$\alpha = 3,4,5$. The matrices $\gamma^1,\cdots,\gamma^4$ are
antihermitian 4$\times$4 Dirac matrices, and, \ub{only in this
paragraph}, $\gamma^5 = i\,\gamma^1\gamma^2\gamma^3\gamma^4$, so that
$\gamma^5$ is now antihermitian, too, rather than hermitian (as in the
rest of this paper). Locally on $M_2$, the cotangent bundle
$\Omega_D^1 ({\Cal A})$ is a free, hermitian ${\Cal A}$-bimodule of
dimension 5, with an orthonormal basis given by
$$
\varepsilon^a\;=\;\pmatrix
\gamma^a\otimes\id_M &\quad 0\quad\\
\quad 0\quad & \gamma^a\otimes \id_M \endpmatrix,
 a= 1,2,\quad\varepsilon^\alpha\;=\;
\pmatrix
\quad 0\quad & i\gamma^\alpha \otimes \id_M \\
- i\gamma^\alpha \otimes \id_M &\quad 0\quad \endpmatrix ,
\alpha = 3,4,5,
\tag 4.8
$$
and $\varepsilon$ is taken to be
$$
\varepsilon\;=\;\varepsilon^1 \varepsilon^2 \varepsilon^3
\varepsilon^4 \varepsilon^5\;=\; \pmatrix
\enskip 0\quad &\id \\
- \id & 0 \endpmatrix .
\tag 4.9
$$
A 1-form $\rho = \sum_j \pi (a^j) \bigl[ D,\pi (b^j)\bigr]$ has the
form
$$
\rho\;=\;\pmatrix
\quad A\quad & K \gamma^\alpha \phi_\alpha \\
- K \gamma^\alpha \phi_\alpha & \quad A\quad \endpmatrix ,
\tag 4.10
$$
with $A=\sum_j a^j \partial\!\!\!/ b^j$ and $\phi_\alpha +
\phi_{0\alpha} = \sum_j a^j \phi_{0\alpha} b^j$. Evaluating $d\rho$ as
in eq.~(2.10), one finds that
$$
d\rho\;=\;\pmatrix
\gamma^{\mu\nu} \partial_\mu A_\nu - K^2 \gamma^{\alpha\beta}
L_{\alpha\beta} + X - K^2 L_\alpha^\alpha, &
K \gamma^\mu \gamma^\alpha D_\mu^0 \phi_\alpha \\
- K \gamma^\mu \gamma^\alpha D_\mu^0 \phi_\alpha, &
\gamma^{\mu\nu} \partial_\mu A_\nu - K^2\gamma^{\alpha\beta}
L_{\alpha\beta} + X - K^2 L_\alpha^\alpha \endpmatrix ,
\tag 4.11
$$
where
$$
L_{\alpha\beta}\;=\; \phi_{0\alpha} \phi_\beta + \phi_\alpha
\phi_{0\beta} + \sum_j a^j \bigl[ b^j, \phi_{0\alpha}
\phi_{0\beta}\bigr],
$$
$$
X\;=\; -\;\sum_j a^j \;\partial\!\!\!/\,^2\;b^j\;+\;\partial^\mu A_\mu
,
$$
and
$$
D_\mu^0 \phi_\alpha\;=\; \partial_\mu \phi_\alpha\;+\;A_\mu
\phi_{0\alpha}\;-\;\phi_{0\alpha} A_\mu .
\tag 4.12
$$
For simplicity we assume that
$$
\bigl[ \phi_{0\alpha}, \phi_{0\beta}\bigr]\;=\; 0, \quad \text{and }
\phi_{0\alpha} \phi_0^\alpha\;=\; 1 .
\tag 4.13
$$
Since we may assume that $\sum a^j b^j = 1$, we then have that
$L_{\alpha\beta} = \phi_{0\alpha} \phi_\beta + \phi_\alpha
\phi_{0\beta}$, for $L_{[\alpha\beta]}$ and $L_\alpha^{\enskip\alpha}$
appearing in (4.11), which is not an auxiliary field. A tedious calculation
then yields the formula
$$
\align
I_\varepsilon^5 (\rho)\;&=\; 2\kappa \int_{M_3} \varepsilon^{\mu\nu}
\varepsilon^{\alpha\beta\gamma} tr K^3 \bigl[\bigl( L_{\alpha\beta}
\phi_\gamma + \phi_\alpha L_{\beta\gamma}\bigr)\;\partial_\mu A_\nu
\\
&-\;\phi_{0\alpha} D_\mu^0 \phi_\beta D_\nu^0 \phi_\gamma + A_\mu
L_{\alpha\beta} D_\nu^0 \phi_\gamma + A_\mu D_\nu^0 \phi_\alpha
L_{\beta\gamma} \\
&+\; \frac 3 2 \; \phi_\alpha A_\mu A_\nu L_{\beta\gamma} +\frac 3 2\;
\phi_\alpha \phi_\beta\phi_\gamma\partial_\mu A_\nu \\
&-\;\frac 3 2 \; A_\mu \phi_\alpha A_\nu L_{\beta\gamma} + \frac 3 2 \;
A_\mu A_\nu \phi_\alpha L_{\beta\gamma} \\
&-\;\frac 3 2\; \phi_\alpha A_\mu \phi_\beta D_\nu^0 \phi_\gamma
+ \frac 3 2\; \phi_\alpha \phi_\beta A_\mu D_\nu^0 \phi_\gamma \\
&+\;\frac 3 2\; A_\mu \phi_\alpha\phi_\beta D_\nu^0 + 3\; A_\mu A_\nu
\phi_\alpha\phi_\beta\phi_\gamma \\
&-\;3\;\phi_\alpha A_\mu \phi_\beta A_\nu \phi_\gamma \bigr]\; d^2 x .
\tag 4.14
\endalign
$$
If $\partial M_2=\emptyset$, and after further algebraic manipulations,
the action (4.14) can be shown to have the manifestly gauge-invariant
form
$$
\align
I_\varepsilon^5 (\rho)\;=\; 2\kappa &\int_{M_2}
\varepsilon^{\alpha\beta\gamma} tr\;\bigl[ - \Phi_\alpha \bigl( D_\mu
\Phi_\beta\bigr) \bigl(D_\nu \Phi_\gamma\bigr) \\
& +\;2\;\Phi_\alpha \Phi_\beta\Phi_\gamma\;\bigl(\partial_\mu A_\nu +
A_\mu A_\nu\bigr)\bigr]\; dx^\mu \wedge dx^\nu ,
\tag 4.15
\endalign
$$
where $\Phi_\alpha = K (\phi_\alpha + \phi_{0\alpha})$.

If the constraints (4.13) are not imposed then one must explicitly
determine $Aux^5$, in order to derive an explicit expression for
$I_\varepsilon^5$. The result is that (4.15) still holds.

It is remarkable that all the Chern-Simons actions derived in
eqs.~(3.23), (4.5) and (4.15) can be obtained from Chern-Simons
actions for connections on vector bundles over classical, commutative
manifolds by \ub{dimensional reduction}. For example, setting $M_3 =
M_2\times S^1$ and $\phi := A_3$, and assuming that $A_1, A_2$ and
$A_3$ are indpendent of the coordinate (angle) parametrizing $S^1$, we
find that
$$
\align
I^3 (A)\;&=\;i \kappa' \int_{M_3} tr\;\bigl( A \wedge dA + \frac 2 3
\; A \wedge A \wedge A \bigr) \\
&=\;i \kappa' \int_{M_2} tr\; (\phi \;F) ,
\tag 4.16
\endalign
$$
where \ $F = \bigl(\partial_1 A_2 - \partial_2 A_1 + [ A_1,
A_2]\bigr)\;dx^1\wedge dx^2.$ \ Setting $\kappa' =\kappa\; tr K$, (4.16)
reduces to (3.23). Similarly, reducing a classical, five-dimensional
Chern-Simons action to four dimensions, with $M_5 = M_4 \times S^1$,
results in
$$
\align
I^5(A)\;&=\;i \kappa' \int_{M_5} tr\;\bigl( A\wedge dA \wedge dA +\;
\frac 3 2 \; A\wedge A\wedge A\wedge dA \\
&\phantom{Zeichnung} +\; \frac 3 5\; A \wedge A\wedge A\wedge A\wedge A\bigr)
\\
&=\;i\;\frac{3\kappa'}{4}\; \int_{M_4} tr\;(\phi \;F\wedge F),
\endalign
$$
with $\phi := A_5$, and $A_1,\cdots,A_5$ independent of the angle
parametrizing $S^1$. Thus we recover (4.5). Finally, dimensionally
reducing $I^5(A)$ to a two-dimensional surface (setting $M_5 = M_2
\times S^1 \times S^1 \times S^1)$ reproduces the action (4.15).

The advantage of the non-commutative formulation is that it
automatically eliminates all excited modes corresponding to a
non-trivial dependence of the gauge potential $A$ on angular
variables.

\vfill\eject
\bf 5. \ub{Relation to four-dimensional gravity and supergravity} \rm.
\medskip
Chern-Simons actions are topological actions. In order to obtain
dynamical actions from Chern-Simons actions, one would have to
impose  constraints on the field configuration space.
In this section, we explore this possibility. As a result, we are able
to derive some action functionals of four-dimensional gravity and
supergravity theory.

We propose to impose a constraint on the scalar multiplet $\Phi$
appearing in the Chern-Simons action (4.5). The non-commutative space
$({\Cal A}, \pi, H, D)$ is chosen as in example (I) of Sect.~4; (see
also Sect.~3). Let us compute the curvature 2-form,
$$
\theta\;=\; ( d\rho + \rho^2)^\perp ,
$$
of a connection $\nabla$ on the line bundle $E\simeq A$ given by a
1-form $\rho$ as displayed in eq.~(4.4). Then
$$
\theta\;=\;\pmatrix
\frac 1 2\;\gamma^{\mu\nu} F_{\mu\nu}+(K^2)^\perp
\bigl((\phi+\phi_0)^2-\phi_0^2\bigr),
& - K i \gamma^5\gamma^\mu D_\mu (\phi +\phi_0) \\
K i \gamma^5 \gamma^\mu D_\mu (\phi +\phi_0),
&\frac 1 2\;\gamma^{\mu\nu} F_{\mu\nu} + (K^2)^\perp
\bigl((\phi+\phi_0)^2 - \phi_0^2\bigr) \endpmatrix ,
\tag 5.1
$$
where $(K^2)^\perp = K^2 - (tr\;K^2) \id$; (recall that $tr (\cdot)$
is normalized: $tr (\id) = 1)$. The appearance of $(K^2)^\perp$ is due
to the circumstance that when passing from $d\rho+\rho^2$ to $(d\rho
+\rho^2)^\perp$ terms proportional to $\id_N$ must be removed. Let
$p\,tr (\cdot)$ denote the partial trace over the Dirac-Clifford
algebra. Then
$$
p\,tr\;(\theta)\;=\;(K^2)^\perp \bigl((\phi+\phi_0)^2-\phi_0^2\bigr),
\tag 5.2
$$
and we shall impose the constraint
$$
p\,tr\;(\theta)\;=\;0 .
\tag 5.3
$$
Choosing $\phi_0$ to satisfy $\phi_0^2=\id$, and renaming
$\phi+\phi_0$ to read $\phi$, the constraint (5.3) becomes
$$
\phi^2\;=\;\id,
\tag 5.4
$$
provided $(K^2)^\perp \;\neq\; 0. $

As our matrix algebra ${\Cal A}_0$ (see eq.~(3.1)) we choose
$$
{\Cal A}_0\;=\enskip\text{real part of Cliff } \bigl(SO(4)\bigr) .
\tag 5.5
$$
We propose to show that, for this choice of ${\Cal A}_0$ and assuming
that the constraint (5.4) is satisfied, the Chern-Simons action (4.5)
is the action of the metric-independent (first-order) formulation of
four-dimensional gravity theory.

Let $\Gamma_1,\cdots,\Gamma_4$ denote the usual generators of ${\Cal
A}_0$, (i.e., 4$\times$4 Dirac matrices in a real representation), and
$\Gamma_5 = \Gamma_1\Gamma_2\Gamma_3\Gamma_4$. Then
$$
\bigl\{ \Gamma_a, \Gamma_b\bigr\}\;=\;- 2 \delta_{ab},\;
\Gamma_a^*\;=\;- \Gamma_a,\;a,b\;=\;1,\cdots,4,
$$
and $\Gamma_5^* = \Gamma_5$. A basis for ${\Cal A}_0$ is then given by
$\id_4, \Gamma_a, a=1,\cdots,4, \Gamma_5, \Gamma_{ab}, a,b
=1,\cdots,4$, and $\Gamma_a\Gamma_5$. For a 1-form $\rho$ as in
eq.~(4.4), we may expand the gauge potential $A$ and the scalar field
$\phi$ in the basis of ${\Cal A}_0$ just described:
$$
A\;=\;\gamma^\mu \;\bigl( A_\mu^0 \id + A_\mu^a \Gamma_a + A_\mu^{ab}
\Gamma_{ab} + A_\mu^5 \Gamma_5 + A_\mu^{a5} \Gamma_a \Gamma_5 \bigr) ,
\tag 5.6
$$
and
$$
\phi\;=\;\bigl( \phi^0 \id + \phi^a \Gamma_a + \phi^{ab} \Gamma_{ab} +
\phi^5 \Gamma_5 + \phi^{a5} \Gamma_a \Gamma_5\bigr) .
\tag 5.7
$$
In this section, we only consider \ub{unitary} connections on $E\equiv
E^{(1)} \simeq {\Cal A}$; see eq.~(2.29). By (2.30), this is
equivalent to \ub{hermiticity} of $\rho$. This implies that
$$
A_\mu^0=- \overline{A_\mu^0}, A_\mu^a = \overline{A_\mu^a},
A_\mu^{ab} = \overline{A_\mu^{ab}}, A_\mu^5= - \overline{A_\mu^5},
A_\mu^{a5} = - \overline{A_\mu^{a5}},
\tag 5.8
$$
and
$$
\phi^0 = \overline{\phi^0}, \phi^a = - \overline{\phi^a},
\phi^{ab} = - \overline{\phi^{ab}}, \phi^5 = \overline{\phi^5},
\phi^{a5} = \overline{\phi^{a5}},
\tag 5.9
$$
where $\bar z$ denotes the complex conjugate of $z$. Since ${\Cal
A}_0$ is chosen to be real, the coefficients of $A$ and $\phi$ should
be chosen to be real. It then follows from (5.8) and (5.9) that
$$
A\;=\;\gamma^\mu \biggl(\frac{1}{2\kappa}\;e_\mu^a \Gamma_a + \frac 1
4\; \omega_\mu^{ab} \Gamma_{ab}\biggr)
\tag 5.10
$$
and
$$
\phi\;=\;\phi^0+\phi^5\Gamma_5 + \phi^{a5} \Gamma_a \Gamma_5,
\tag 5.11
$$
where we have set $A_\mu^a =: \frac{1}{2\kappa}\;e_\mu^a$, and
$A_\mu^{ab} =: \frac 1 4 \; \omega_\mu^{ab}$, and $\kappa^{-1}$ is the
Planck scale.

Imposing the constraint that \ $tr_{{\Cal A}_0} (\varepsilon \rho) =
0$ implies that
$$
\phi^0\;=\; 0 .
\tag 5.12
$$
Constraints (5.12) and (5.4) then yield the condition
$$
(\phi^5)^2\;+\;(\phi^{a5})^2\;=\;1 .
\tag 5.13
$$
Under a gauge transformation $M\equiv g^1$, $\rho$ transforms according to
$$
\rho\;\mapsto\; M\rho M^{-1} - (dM) M^{-1},\quad M\in\pi ({\Cal A}),
$$
see (2.25), which implies the transformation law
$$
\phi \;\mapsto\; g^{-1}  \phi g, \quad g\;=\;\exp \;\frac 1 2\; \bigl(
\Lambda^a \Gamma_a + \Lambda^{ab} \Gamma_{ab}\bigr),
\tag 5.14
$$
where $\Lambda^a$ and $\Lambda^{ab}$ are smooth functions on $M_4$.
The infinitesimal form of (5.14) reads
$$
\align
\delta \phi^5\;&=\;-\;\sum_a \Lambda^a \phi^{a5} , \\
\delta \phi^{a5}\;&=\;-\;\Lambda^a \phi^5 \;-\sum_b \Lambda^{ab}
\phi^{b5} .
\tag 5.15
\endalign
$$
 From this it follows that, locally, we can choose a gauge such that
$$
\phi^{a5}\;=\;0 .
\tag 5.16
$$
In this gauge, the constraint (5.13) has the solutions
$$
\phi^5\;=\; \pm \;1 .
\tag 5.17
$$
The action (4.5) then becomes
$$
I_\varepsilon^5 (\rho)\;=\; \pm\; k \int_{M_4} tr\; (\Gamma_5 F\wedge
F) ,
\tag 5.18
$$
(with $k=i \frac{3\kappa}{4}$, in the notation of Sect.~4). Next, we
expand the field strength $F_{\mu\nu}$ in our Clifford algebra basis
which yields
$$
F_{\mu\nu}\;=\;\frac{1}{2\kappa}\; F_{\mu\nu}^a\; \Gamma_a \;+\;\frac 1
4\; F_{\mu\nu}^{ab} \;\Gamma_{ab} ,
\tag 5.19
$$
where
$$
F_{\mu\nu}^a\;=\;\partial_\mu\; e_\nu^a + \omega_{\mu\;b}^a\;
e_\nu^b\;-\;(\mu\leftrightarrow \nu) ,
\tag 5.20
$$
$$
F_{\mu\nu}^{ab}\;=\;\partial_\mu \;\omega_\nu^{ab} +
\omega_{\mu\;c}^a\;\omega_\nu^{c\;b} + \;\frac{1}{\kappa^2}\;
e_\mu^a\; e_\nu^b \;-\;(\mu\leftrightarrow \nu) ,
\tag 5.21
$$
and the indices $a,b,\cdots$ are raised and lowered with the flat
metric $\eta_{ab} = - \delta_{ab}$.

The only non-vanishing contribution to (5.18) comes from the trace \
$tr (\Gamma_5\Gamma_{ab}\Gamma_{cd}) = \varepsilon_{abcd}$, and
$I_\varepsilon^5$ is found to be given by
$$
\align
I_\varepsilon^5\;=\;\pm\;k \int_{M_4} \varepsilon_{abcd}\;\bigl(
R_{\mu\nu}^{ab} +\;\frac{2}{\kappa^2}\; e_\mu^a\;&e_\nu^b\bigr)
\bigl( R_{\rho\sigma}^{cd} + \;\frac{2}{\kappa^2}\;e_\rho^c\;
e_\sigma^d \bigr) \\
&\times\; dx^\mu \wedge dx^\nu \wedge dx^\rho \wedge dx^\sigma ,
\tag 5.22
\endalign
$$
where
$$
R_{\mu\nu}^{ab}\;=\;\partial_\mu\;\omega_\nu^{ab} +
\omega_{\mu\;c}^a\;\omega_\nu^{c\;b}\;-\; (\mu \leftrightarrow \nu) .
\tag 5.23
$$
Interpreting $\omega_\mu^{ab}$ as the components of a connection on
the spinor bundle over $M_4$, $R_{\mu\nu}^{ab}$ are the components of
its curvature, and $F_{\mu\nu}^a$ are the components of its torsion,
as is well known from the Cartan structure equations.

Setting the variation of $I_\varepsilon^5$ with respect to
$\omega_\mu^{ab}$ to zero, we find that the torsion of $\omega$
vanishes:
$$
F_{\mu\nu}^a\;=\;0,\quad\text{for all }\mu,\nu\text{ and } a.
\tag 5.24
$$
If the frame $\bigl(e_\mu^a\bigr)$ is invertible, (5.24) can be solved
for $\omega_\mu^{ab}$:
$$
\omega_\mu^{ab}\;=\;\frac 1 2 \; \bigl( \Omega_{\mu a b} -
\Omega_{ab\mu} + \Omega_{b\mu a}\bigr),
\tag 5.25
$$
where
$$
\Omega_{ab}\;^c\;=\; e_a^\mu\;e_b^\nu\;\bigl(
\partial_\mu\;e_\nu\,^c\;-\;\partial_\nu\; e_\mu\,^c \bigr).
$$
Substituting (5.25) back into (5.22) yields a functional that depends
only on the metric
$$
g_{\mu\nu}\;=\;e_\mu^a\; e_{\nu a},
\tag 5.26
$$
and is given by
$$
\align
I_\varepsilon^5\;=\;\pm\;k \int_{M_4} d^4 x\, \sqrt{g}\; \bigl[\bigl(\;
&4\,R_{\mu\nu\rho\sigma} R^{\mu\nu\rho\sigma} \;-\;4\, R_{\mu\nu} R^{\mu\nu}
+ R^2 \bigr) \\
&+\;\frac{16}{\kappa^2}\;R\;+\;\frac{96}{\kappa^4}\;\bigr]
 \tag 5.27
\endalign
$$
where $R_{\mu\nu\rho\sigma}$ is the Riemann curvature tensor,
$R_{\mu\nu}$ is the Ricci tensor, and $R$ is the scalar curvature
determined by the metric $g_{\mu\nu}$ given in (5.26). The term in
round brackets on the r.h.s. of (5.27) yields the topological
Gauss-Bonnet term for $M_4$, the second term yields the
Einstein-Hilbert action, and the last term is a cosmological constant.

Next, we show how to derive a metric-independent formulation of \
\ub{four-dimensional} \ub{supergravity} from the action $I_\varepsilon^5$
given in eq.~(4.5). For this purpose we choose the algebra ${\Cal
A}_0$ in (3.1) to be a graded algebra [18]:
$$
{\Cal A}_0\;=\;\text{real part of SU}(4\mid 1) .
\tag 5.28
$$
This algebra is generated by graded 5$\times$5 matrices preserving the
quadratic form
$$
(\vartheta^\alpha)^*\; C_{\alpha\beta} \;\vartheta^\beta\;-\;z^* \;z ,
\tag 5.29
$$
where $C_{\alpha\beta}$ is an antisymmetric matrix and
$\vartheta^\alpha$ is a Dirac spinor. At this point, one must note
that we are leaving the conventional framework of non-commutative
geometry, since, for ${\Cal A}_0$ as in (5.28), the algebra ${\Cal A}$
is not a $^*$algebra of operators. But let us try to proceed and find
out what the result is.

Let $\rho$ be a 1-form as in eq.~(4.4). Then the matrix elements
$A_\mu$ and $\phi$ of $\rho$ have the graded matrix representation
$$
\phi\;=\;\pmatrix
\Pi_\alpha^\beta & \lambda_\alpha \\
\bar\lambda^\alpha & \Pi_1 \endpmatrix ,
\tag 5.30
$$
and
$$
A_\mu\;=\; \pmatrix
M_{\mu \alpha}^{\enskip\beta} & \sqrt{\kappa}\;\psi_{\mu\alpha} \\
- \sqrt{\kappa}\;\bar\psi_\mu^\alpha &\quad B_\mu \endpmatrix .
\tag 5.31
$$
The reality conditions for $\phi$ and $A_\mu$ imply that
$\lambda_\alpha$ and $\psi_{\mu\alpha}$ are Majorana spinors:
$$
\lambda_\alpha\;=\;C_{\alpha\beta}\;\bar\lambda^\beta,
\;\psi_{\mu\alpha}\;=\;C_{\alpha\beta}\;\bar\psi_\mu^\beta .
$$
Furthermore, one finds that
$$
\align
\Pi_\alpha^\beta\;&=\;\biggl( \frac 1 4\; \Pi^0 \id \;+\; \Pi^5 \Gamma_5 \;+\;
\Pi^{a5} \Gamma_a\Gamma_5\biggr)_\alpha^\beta , \\
M_{\mu\alpha}^\beta\;&=\;\biggl(
\frac{1}{2\kappa}\;e_\mu^e\;\Gamma_a \;+\;\frac 1 4
\;\omega_\mu^{ab}\;\Gamma_{ab}\biggr)_\alpha^\beta ,
\tag 5.32
\endalign
$$
and
$$
B_\mu\;=\;0 .
$$

We shall now impose the constraints
$$
Str\;(\varepsilon\;\rho)\;=\;0 ,
\tag 5.33
$$
$$
Str\;(\theta)\;=\;0 ,
\tag 5.34
$$
and
$$
Str \;\biggl(\varepsilon\;\bigl( \rho d\rho\;+\;\frac 2
3\;\rho^3\bigr)^\perp \biggr)\;=\;0 ,
\tag 5.35
$$
along with $Str \;\phi_0^2 = 1$, and $Str\;\phi_0 = 0$. Here \ $Str
(\cdot)$ denotes the graded trace on ${\Cal A}_0$. Renaming $\phi +
\phi_0$ to read $\phi$, these constraints
imply that
$$
\Pi^0\;=\;\Pi_1 ,
\tag 5.36
$$
$$
-\;\frac 3 4 \; \Pi_1^2 + 4\bigl((\Pi^5)^2 - \sum_a (\Pi^{a5})^2\bigr)
+ \bar\lambda \lambda \;=\;1 ,
\tag 5.37
$$
and
$$
(K^3)^\perp\;S tr (\phi^3)\;=\; 0,
\tag 5.38
$$
where $(K^3)^\perp$ is defined so as to satisfy \ $tr
\bigl(K(K^3)^\perp\bigr) = 0$.

In order to determine the dynamical contents of a theory with an
action $I_\varepsilon^5$ given by (4.5), ${\Cal A}_0$ as in (5.28) and
constraints (5.36) through (5.38), it is convenient to work in a
special gauge, the unitary gauge. Consider a gauge transformation
$$
g\;=\;\exp\; \pmatrix
\bigl(\Lambda_\alpha^\beta\bigr)\quad &\sqrt{\kappa}\;\varepsilon_\alpha \\
- \sqrt{\kappa}\;\bar\varepsilon^\alpha & 0 \endpmatrix ,
\tag 5.39
$$
where $\Lambda_\alpha^\beta = \frac 1 2$ $\bigl( \Lambda^a \Gamma_a +
\Lambda^{ab} \Gamma_{ab}\bigr)_\alpha^\beta $. The transformation law of
$\phi$ is then given by $\phi \mapsto g^{-1} \phi g$. From this we
find the infinitesimal gauge transformations of the fields $\Pi$ and
$\lambda$:
$$
\align
&\delta\; \Pi_1\;=\;2\;\sqrt{\kappa}\; \bar\varepsilon \lambda, \\
&\delta\; \Pi^5\;=\;+\; \Lambda^a \Pi^{a5} + \frac{\sqrt{\kappa}}{2}\;
\bar\varepsilon \;\Gamma^5 \lambda , \\
&\delta\;\Pi^{a5}\;=\; -\; \Lambda^{ab} \Pi^{b5} - \Lambda^a \Pi^5
- \frac 1 2\;\bar\varepsilon \;\Gamma^a\Gamma^5\lambda, \\
&\delta\;\lambda_\alpha\;=\;\sqrt{\kappa}\;\bigl(- \frac 3 4\; \Pi_1 +
\Pi^5 \Gamma_5 + \Pi^{a5}
\Gamma_a\Gamma_5\bigr) \varepsilon_\alpha - \frac 1 2\;
\bigl(\Lambda^a\Gamma_a+\Lambda^{ab}\Gamma_{ab}\bigr)_\alpha^\beta\;
\lambda_\beta .
\tag 5.40
\endalign
$$
Thus, locally, we can choose the gauge
$$
\Pi^{a5}\;=\;0, \text{ and } \lambda_a\;=\; 0 .
\tag 5.41
$$
The constraints (5.37) and (5.38) then reduce to
$$
\align
-\;\frac 3 4\;\Pi_1^2 + 4\; (\Pi^5)^2\;&=\;1 , \\
\Pi_1\bigl(-\frac{5}{16}\;\Pi_1^2 + (\Pi^5)^2\bigr)\;&=\; 0 .
\tag 5.42
\endalign
$$
These equations have the solutions
$$
\Pi_1\;=\;0,\quad \Pi^5\;=\;\pm\;\frac 1 2 ,
\tag 5.43
$$
and
$$
\Pi_1\;=\;\pm\;\sqrt{2}\;, \;\Pi^5\;=\;\pm\;\sqrt{\frac{5\,}{8\,}} .
\tag 5.44
$$
We further study the first solution. Inserting it into the action
(4.5), we arrive at the expression
$$
I_\varepsilon^5\;=\;\pm\;\frac k 2 \int_{M_4} Str\;\biggl( {\Gamma_5
\quad 0 \choose \enskip 0\quad 0}\; F_{\mu\nu} F_{\rho\sigma}\biggr)\; dx^\mu
\wedge dx^\nu\wedge dx^\rho \wedge dx^\sigma ,
\tag 5.45
$$
where
$$
F_{\mu\nu}\;=\;\pmatrix
\frac 1 4 \; F_{\mu\nu}^{ab}\;\Gamma_{ab} +
\frac{1}{2\kappa}\;F_{\mu\nu}^a\;\Gamma_a, &
\sqrt{\kappa}\;\psi_{\mu\nu} \\
-\;\sqrt{\kappa}\;\bar\psi_{\mu\nu}\phantom{ZeichnungZ}, &\quad 0
\endpmatrix ,
\tag 5.46
$$
with
$$
\align
F_{\mu\nu}^a\;&=\;\partial_\mu\;e_\nu^a + \omega_{\mu\; b}^a\;e_\nu^b
-\;\frac{\kappa^2}{2}\;\bar\psi_\mu \Gamma^a\psi_\nu -
(\mu\leftrightarrow \nu) , \\
F_{\mu\nu}^{ab}\;&=\;R_{\mu\nu}^{ab} + \frac{1}{\kappa^2}\; \bigl(
e_\mu^a\;e_\nu^b - e_\nu^a \; e_\mu^b\bigr) + \kappa \bar\psi_\mu
\Gamma^{ab} \psi_\nu , \\
\psi_{\mu\nu\alpha}\;&=\;\partial_\mu\;\psi_{\nu\alpha} + \frac 1 4\;
\omega_\mu^{ab} \;\bigl( \Gamma_{ab}\psi_\nu\bigr)_\alpha +
\frac{1}{2\kappa}\; e_\mu^a \bigl( \Gamma_a\psi_\nu\bigr)_\alpha \\
&\phantom{ZeichnungZeichnung} -\;(\mu\leftrightarrow \nu) .
\tag 5.47
\endalign
$$
After some further manipulations and evaluating all the traces, one
obtains the elegant result that the action reduces to that proposed in
[17], namely
$$
\align
I_\varepsilon^5\;=\;\pm\;k\int_{M_4} \biggl[ \frac 1 4\;
\varepsilon_{abcd} F_{\mu\nu}^{ab} &F_{\rho\sigma}^{cd} + \alpha \kappa
\; \bar\psi_{\mu\nu} \Gamma_5 \psi_{\rho\sigma}\biggr] \\
& \times\; dx^\mu \wedge dx^\nu \wedge dx^\rho \wedge dx^\sigma ,
\tag 5.48 \
\endalign
$$
where $\alpha$ is some constant introduced for later convenience, but
here $\alpha=1$. Substituting eqs.~(5.47) into (5.48), one obtains
that
$$
\align
I_\varepsilon^5\;=\;&\pm\;k \biggl\{ \int_{M_4} \varepsilon_{abcd}\;
\frac 1 4 \;\bigl[ R_{\mu\nu}^{ab} R_{\rho\sigma}^{cd} + 2\kappa\;
R_{\mu\nu}^{ab} \bigl( \bar\psi_\rho \Gamma^{cd} \psi_\sigma\bigr)\\
&\phantom{Zeich} + \kappa^2 \bigl(\bar\psi_\mu \Gamma^{ab}
\psi_\nu\bigr) \bigl(\bar\psi_\rho \Gamma^{cd}
\psi_\sigma\bigr)\bigr]\; dx^\mu\wedge dx^\nu\wedge dx^\rho\wedge
dx^\sigma \\
&+\;\frac{4}{\kappa^2}\;\int_{M_4} d^4x\; \sqrt{g} \;e_a^\mu e_b^\nu
\bigl(R_{\mu\nu}^{ab} + \kappa \bigl(\bar\psi_\mu \Gamma^{ab} \psi_\nu
\bigr)\bigr) \\
&+\;4 \alpha \kappa \int_{M_4} \bigl( D_\mu \bar\psi_\nu\bigr)
\Gamma^5\; \bigl(D_\rho \psi_\sigma\bigr) \; dx^\mu \wedge \cdots
\wedge dx^\sigma \\
&+\; 4\alpha \int_{M_4} \bigl( \bar\psi_\mu \Gamma_\nu \Gamma^5 D_\rho
\psi_\sigma\bigr) \; dx^\mu \wedge \cdots \wedge dx^\sigma \\
&+\;\frac{2\alpha}{\kappa}\;\int_{M_4} d^4x\; \sqrt{g}\; \bar\psi_\mu
\Gamma^{\mu\nu} \psi_\nu + \frac{24}{\kappa^4}\;\int_{M_4} d^4 x\,
\sqrt{g} \biggr\} ,
\tag 5.49
\endalign
$$
where
$$
D_\mu \;\psi_\nu\;=\;\partial_\mu\; \psi_\nu \;+\;\frac 1 4 \;
\omega_\mu^{ab}\; \Gamma_{ab}\; \psi_\nu .
$$
After Fierz reshuffling, the term quartic in the gravitino field
$\psi_\mu$ disappears. The remaining terms describe massive
supergravity with a Gauss-Bonnet term. It is an interesting fact that
the action (5.48), with $\alpha = 2$ (!), is invariant under
the same supersymmetry transformation obtained form the variation of
$\Pi(\rho)$, except for $\delta \omega_\mu^{ab}$ which is chosen to
preserve the constraint [18]:
$$
F_{\mu\nu}^{\enskip a}\;=\; 0 .
\tag 5.50
$$
The supersymmetry transformations can be read by substituting (5.10)
into eq.~(3.18):
$$
\align
&\delta\;e_\mu^a\;=\;\kappa\;\bar\varepsilon\;\Gamma^a \psi_\mu,\\
&\delta\;\psi_\mu\;=\;\bigl( \partial_\mu + \frac 1 4\; \omega_\mu^{ab}
\Gamma_{ab} + \frac{1}{2\kappa}\;e_\mu^a \Gamma_a\bigr) \varepsilon,
\tag 5.51
\endalign
$$
and, for $F_{\mu\nu}^{ab}$ and $\psi_{\mu\nu}$, they are
$$
\align
&\delta\;F_{\mu\nu}^{ab}\;=\;\kappa\;\bar\varepsilon \,\Gamma^{ab}
\psi_{\mu\nu}, \\
&\delta\;\psi_{\mu\nu}\;=\;-\;\frac 1 4\;
F_{\mu\nu}^{ab}\;(\Gamma_{ab}\; \varepsilon ).
\tag 5.52
\endalign
$$
When $\alpha=2$ the action (5.48) becomes invariant under the
transformations (5.51) with the constraint (5.50), and the action
corresponds to de Sitter supergravity where the cosmological constant
and the gravitino mass-like term are fixed with respect to each other.
In this case the action (5.48) simplifies to
$$
\align
I_{sg}\;=\;&-\;\biggl[\int_{M_4} d^4x\; \varepsilon^{\mu\nu\rho\sigma}
\,\bigl(\frac 1 4\;\varepsilon_{abcd} R_{\mu\nu}^{\enskip ab}
R_{\rho\sigma}^{\enskip cd} + 8\, \bar\psi_\mu \Gamma_\nu \Gamma_5
D_\rho \psi_\sigma\bigr) \\
&+\;4 \int d^4x\,e\;\bigl(e_a^\mu\, e_b^\nu\,R_{\mu\nu}^{\enskip ab} +
\frac 2 \kappa\;\bar\psi_\mu\; \Gamma^{\mu\nu} \psi_\nu +
\frac{6}{\kappa^4}\bigr)\biggr] .
\tag 5.52
\endalign
$$
The first term in (5.52) is a topological invariant and can be removed
from the action without affecting its invariance. After rescaling
$$
\align
e_\mu^a\;&\to\;r\;e_\mu^a \\
\psi_{\mu\alpha}\;&\to\;\sqrt{r} \; \psi_{\mu\alpha} \\
I_{sg}\;&\to\;8\,r^2\;I_{sg}
\tag 5.53
\endalign
$$
and taking the limit $r\to0$ the action (5.52) reduces to that of
$N$=1 supergravity [19]:
$$
I_{sg}\;=\;-\;\frac{1}{2\kappa^2} \int_{M_4} d^4x\,
e\;e_a^\mu\,e_b^\nu\,R_{\mu\nu}^{\;ab} - \int_{M_4}
d^4x\,\varepsilon^{\mu\nu\rho\sigma} \,\bar\psi_\mu \Gamma_5
\Gamma_\nu D_\rho \psi_\sigma .
\tag 5.54
$$
The significance of the constraint (5.50) and the choice $\alpha=2$ in
the non-commutative construction is not clear to us. It would be
helpful to better understand  this point.

If we had worked instead with the solution (5.44), then additional
terms which are dynamically trivial will be present. We shall not
present the details for this case.

\vskip 1.5truecm

\bf 6. \ub{Conclusions and outlook} \rm .
\medskip
In this paper, we have shown how to construct Chern-Simons forms and
Chern-Simons actions in real, non-commutative geometry; (more detailed
results will appear in [12]). We have illustrated the general,
mathematical results of Sect.~2 by discussing a number of examples.
These examples involve non-commutative spaces described by
$^*$algebras of matrix-valued functions over even-dimensional spin
manifolds. As expected, the Chern-Simons actions associated with these
spaces are manifestly topological (metric-independent). By imposing
constraints on the field configurations on which these action
functionals depend (and choosing convenient gauges) we have been able
to derive the metric-independent, first-order formulation of
four-dimensional gravity theory from a Chern-Simons action over a
``five-dimensional'' non-commutative space. By extending the
mathematical framework, formally, to allow for graded algebras, we
have also recovered an action functional for supergravity.

It would appear to be of interest to study Chern-Simons actions for
more general non-commutative spaces, e.g. those considered in [8], and
to derive from them theories of interest to physics. In this regard,
one should recall that a rather profound theory has the form of a
Chern-Simons theory: Witten's open string field theory [15]. We are
presently attempting to formulate that theory within Connes'
mathematical framework of non-commutative geometry, using a variant of
the formalism developed in Sect. 2.

On the mathematical side, it appears to be of interest to better
understand the topological nature of Chern-Simons actions over general
non-commutative spaces, to understand the connection between the
material presented in Sect.~2 and the theory of characteristic classes
in non-commutative geometry and cyclic cohomology, see [3,10], and, most
importantly, to learn how to quantize Chern-Simons theories in
non-commutative geometry, in order to construct new topological field
theories.

\vskip 1.5truecm
\bf \ub{References}\rm\;:
\medskip

\item{[1]} E. Witten, ``Quantum field theory and the Jones
polynomial'', Comm. Math. Phys. \ub{121}, (1989) 351.

\item{[2]} J. Fr\"ohlich ``Statistics of fields, the Yang-Baxter
equation, and the theory of knots and links'', 1987 Carg\`ese
lectures, in: ``Non-Perturbative Quantum Field Theory'', eds.: G. 't~Hooft
et al., (Plenum Press, New York, 1988).

\item{[3]} M. Dubois-Violette, C.R. Acad. Sc. Paris, \ub{307}, I, (1988)
403.
\item{} J. Fr\"ohlich and C. King, Comm. Math. Phys. \ub{126} (1989)
187.
\item{} E. Guadagnini, M. Martellini and M. Mintchev, Nucl. Phys.
B\ub{330} (1990) 575.

\item{[4]} A. Connes, Publ. Math. I.H.E.S. \ub{62} (1985) 41-144,
``Non-Commutative Geometry'', Academic Press, to appear (1994).

\item{[5]} A. Connes and J. Lott, Nucl. Phys. B (Proc.~Supp.) \ub{18}B
(1990) 29; in Proc. 1991 Summer Carg\`ese Conference, eds.: J.
Fr\"ohlich et al., (Plenum Press, New York 1992).

\item{[6]} R. Coquereaux, G. Esposito-Far\`ese, G. Vaillant, Nucl.
Phys. B\ub{353} (1991) 689;
\item{} M. Dubois-Violette, R. Kerner, J. Madore, J. Math. Phys.
\ub{31} (1990) 316;
\item{} B. Balakrishna, F. G\"ursey and K.C. Wali, Phys. Lett.
\ub{254}B (1991) 430; Phys. Rev. D\ub{46} (1991) 6498.
\item{} R. Coquereaux, G. Esposito-Far\`ese and F. Scheck,
Int.~J.~Mod.~Phys.~A\ub{7} (1992) 6555.

\item{[7]} D. Kastler, ``A detailed account of Alain Connes' version
of the standard model in non-commutative geometry'' I, II and III, to
appear in Rev. Math. Phys.;
\item{} D. Kastler and M. Mebkhout, ``Lectures on non-commutative
differential geometry and the standard model'', World Scientific, to
be published;
\item{} D. Kastler and T. Sch\"ucker, Theor. Math. Phys. \ub{92}
(1992) 522.

\item{[8]} A.H. Chamseddine, G. Felder and J. Fr\"ohlich, Phys. Lett.
B\ub{296} (1992) 109; Nucl. Phys. B\ub{395} (1993) 672.

\item{[9]} A.H. Chamseddine, G. Felder and J. Fr\"ohlich, Commun. Math.
Phys. \ub{155} (1993) 205.

\item{[10]} A.H. Chamseddine and J. Fr\"ohlich ``Some elements of
Connes' non-commutative geometry and space-time geometry'', to appear
in Yang-Festschrift.

\item{[11]} D. Quillen, ``Chern-Simons forms and cyclic cohomology'',
in: ``The Interface of Mathematics and Particle Physics'', D. Quillen,
G. Segal and S. Tsou (eds.), Oxford University Press, Oxford 1990.

\item{[12]} A.H. Chamseddine, J. Fr\"ohlich and O. Grandjean, in
preparation.

\item{[13]} A.H. Chamseddine and D. Wyler, Phys. Lett. B\ub{228} (1989)
75; Nucl. Phys. B\ub{340} (1990) 595;
\item{} E. Witten, ``Surprises with topological field theories'' in
Proc. ``Strings 90'', eds R. Arnowitt et al.

\item{[14]} E. Witten, Nucl. Phys. B\ub{311} (1988) 96; B\ub{323}
(1989) 113;
\item{} A.H. Chamseddine, Nucl. Phys. B\ub{346} (1990) 213.

\item{[15]} E. Witten, Nucl. Phys. B\ub{268} (1986) 253; Nucl. Phys.
B\ub{276} (1986) 291.

\item{[16]} E. Witten, Commun. Math. Phys. \ub{117} (1988) 353.

\item{[17]} A.H. Chamseddine, Ann. Phys. \ub{113} (1978) 219; Nucl.
Phys. B\ub{131} (1977) 494;
\item{} K. Stelle and P. West, J. Phys. A\ub{12} (1979) 1205.

\item{[18]} P. van Nieuwenhuizen, Phys. Rep. \ub{68} (1981) 189.

\item{[19]} D. Freedman, P. van Nieuwenhuizen and S. Ferrara, Phys.
Rev. D\ub{13} (1976) 3214;
\item{} S. Deser and B. Zumino, Phys. Lett. B\ub{62} (1976) 335.

\bye

{}From schultze Tue May 31 13:43:35 1994
Received: from isis.itp by itp.ethz.ch; Tue, 31 May 94 13:43:33 +0200
{}From: schultze (Annetraut Schultze)
Date: Tue, 31 May 94 13:43:33 +0200
Message-Id: <9405311143.AA11330@itp.ethz.ch>
To: chams
Status: R

\input vanilla.sty
\input mymacros
\overfullrule=0pt
\magnification=\magstep1
\vsize=23.5truecm
\hsize=16.5truecm
\baselineskip=.6truecm

\def\ub#1{{\underbar{#1}}}

\font\ti=cmbx10 scaled\magstep1
\TagsOnRight
\loadmsam
\loadmsbm
\UseAMSsymbols

$\;$

\pageno=0

\rightline{\bf ETH-TH/94-11\rm\quad}
\vskip 1truecm
\centerline{\ti{THE \ CHERN-SIMONS \ ACTION IN}}

\centerline{\ti{NON-COMMUTATIVE GEOMETRY}}\rm
\vskip 3truecm
\bf
\centerline{A.H. Chamseddine \ and \ J. Fr\"ohlich}

\centerline{Theoretical Physics}

\centerline{ETH-H\"onggerberg}

\centerline{CH-8093 \ Z\"urich}

\vskip 5truecm

\noindent Abstract\rm . \ A general definition of Chern-Simons actions
in non-commutative geometry is proposed and illustrated in several
examples. These examples are based on ``space-times'' which are
products of even-dimensional, Riemannian spin manifolds by a discrete
(two-point) set. If the $^*$algebras of operators describing the
non-commutative spaces are generated by functions over such
``space-times'' with values in certain Clifford algebras the
Chern-Simons actions turn out to be the actions of topological gravity
on the even-dimensional spin manifolds. By constraining the space of
field configurations in these examples in an appropriate manner one is
able to extract dynamical actions from Chern-Simons actions.

\vskip 1truecm

\noindent PACS No. \ A0210 / A0240 / A0350K / A0420C / A0420F
\vfill\eject

\pageno=1

\bf
\noindent 1. \ub{Introduction}\rm

During the past several years, topological field theories have been
the subject of a lot of interesting work. For example, deep connections
between three-dimensional, topological Chern-Simons theories [1], or,
equivalently, two-dimensional, chiral conformal field theories [2], on
one hand, and a large family of invariants of links, including the
famous Jones polynomial, and of three-manifolds [1], on the other
hand, have been discovered. Other topological field theories have been
invented to analyze e.g. the moduli space of flat connections on
vector bundles over Riemann surfaces or to elucidate the Donaldson
invariants of four-manifolds.
These topological field theories are formulated as theories over some
classical (topological or differentiable) manifolds.

Connes has proposed notions of non-commutative spaces generalizing,
for example, the notion of a classical differentiable manifold [4].
His theory is known under the name of ``non-commutative geometry''.
Dubois-Violette [3] and Connes have proposed to study field theories
over non-commutative spaces. In joint work with J. Lott [4], Connes
has found a construction of the classical action of the standard
model, using tools of non-commutative geometry, which yields a
geometrical interpretation of the scalar Higgs field responsible for
the ``spontaneous breaking of the electroweak gauge symmetry''. In
fact, the Higgs field appears as a component of a generalized gauge
field (connection 1-form) associated with the gauge group,
SU(2)$_w$\;$\times$\;U(1)$_{em}$,  of electroweak interactions. This is
accomplished by formulating gauge theory on a generalized space
consisting of two copies of standard Euclidian space-time the
``distance'' between which is determined by the weak scale. Although
the space-time model underlying the Connes-Lott construction is a
commutative space, it is not a classical manifold, and analysis on
space-times of the Connes-Lott type requires some of the tools of
non-commutative geometry.

The results of Connes and Lott have been reformulated and refined in
[6,7] and extended to grand-unified theories in [8]. In [9], G. Felder
and the authors have proposed some form of non-commutative Riemannian
geometry and applied it to derive an analogue of the Einstein-Hilbert
action in non-commutative geometry.

Our aim in this article is to attempt to do some steps towards a
synthesis between the different developments just described. Some of
our results have been described in our review paper [10]. We start by
presenting a general definition of the Chern-Simons action in
non-commutative geometry, (Section 2). Our definition is motivated by
some results of Quillen [11] and is based on joint work with O.
Grandjean [12]. In Section 3, we discuss a first family of examples.
In these examples, the non-commutative space is described in terms of
a $^*$algebra of matrix-valued functions over a Connes-Lott type
``space-time'', i.e., over a commutative space consisting of two
copies of an even-dimensional, differentiable spin manifold. The
Chern-Simons actions on such non-commutative spaces turn out to be
actions of topological gauge- and gravity theories, as studied in
[13,14]. In Section 3, the dimension of the continuous, differentiable
spin manifold is two, i.e., we consider products of Riemann surfaces
by discrete sets, and our Chern-Simons action is based on the
Chern-Simons 3-form.

In Section 4, we consider two- and four-dimensional topological
theories derived from a Chern-Simons action based on the Chern-Simons
5-form.

In Section 5, we describe connections of the theories found in Section
4 with four-dimensional gravity and supergravity theories.

In Section 6, we suggest applications of our ideas to string field
theory [15], and we draw some conclusions.


\vskip 1.5truecm

\bf
\ub{Acknowledgements}\rm\;. \ We thank G. Felder, K. Gaw\c edzki and
D. Kastler for their stimulating interest and O. Grandjean for very
helpful discussions on the definition of the Chern-Simons action in
non-commutative geometry and for collaboration on related matters, [12].

\vskip 1.5truecm

\bf
2. \ub{Elements of non-commutative geometry}.\rm
\medskip
This section is based on Connes' theory of non-commutative geometry,
as described in [4], and on results in [9,10,11,12].

We start by recalling the definition of a special case of Connes'
general definition of non-commutative spaces. A \ub{real, compact
non-commutative space} is defined by the data $({\Cal A}, \pi, H,
D)$, where ${\Cal A}$ is a $^*$algebra of bounded operators containing
an identity element, $\pi$ is a $^*$representation of ${\Cal A}$ on
$H$, where $H$ is a separable Hilbert space, and $D$ is a
\ub{selfadjoint} operator on $H$, with the following properties:
\medskip
\item{(i)} $\bigl[D,\pi(a)\bigr]$ is a bounded operator on $H$, for
all $a\in{\Cal A}$. [This condition determines the analogue of a
differentiable structure on the non-commutative space described by
${\Cal A}$.]

\item{} In the following, we shall usually identify the
algebra ${\Cal A}$ with the $^*$subalgebra $\pi({\Cal A})$ of the
algebra $B(H)$ of all bounded operators on $H$; (we shall thus assume
that the kernel of the representation $\pi$ in ${\Cal A}$ is trivial).
We shall often write ``$a$'' for both, the element $a$ of ${\Cal A}$
and the operator $\pi(a)$ on $H$.

\item{(ii)} $(D^2+\id)^{-1}$ is a compact operator on $H$. More
precisely, $\exp (-\varepsilon D^2)$ is trace-class, for any
$\varepsilon > 0$.
\medskip
Given a real, compact non-commutative space $({\Cal A}, \pi, H, D)$,
one defines a differential algebra, $\Omega_D({\Cal A})$, of forms as
follows: 0-forms (``scalars'') form a $^*$algebra with identity,
$\Omega_D^0({\Cal A})$, given by $\pi({\Cal A})$; $n$-forms form a
linear space, $\Omega_D^n({\Cal A})$, spanned by equivalence classes
of operators on $H$,
$$
\Omega_D^n({\Cal A})\;:=\;\Omega^n({\Cal A}) / Aux^n\;,\tag 2.1
$$
where the linear space $\Omega^n({\Cal A})$ is spanned by the
operators
$$
\bigl\{ \sum_i a_0^i [D,a_1^i]\cdots[D,a_n^i]\;:\; a_j^i \in\;
{\Cal A}\;\equiv \pi({\Cal A}),\;\forall i,j\bigr\},
\tag 2.2
$$
and $Aux^n$, the space of ``auxiliary fields'' [5], is spanned by
operators of the form
$$
\align
Aux^n\;:=\;\bigl\{\;&\sum_i [D,a_0^i][D,a_1^i]\cdots[D,a_n^i]\;:\\
&\sum_i a_0^i [D,a_1^i]\cdots [D,a_n^i]\;=\;0,\; a_j^i\in {\Cal
A}\bigr\}\;. \tag 2.3
\endalign
$$
Using the Leibniz rule
$$
[D,ab]\;=\;[D,a]b\;+\;a[D,b],\; a,b\in {\Cal A},
\tag 2.4
$$
and
$$
[D,a]^*\;=\;-\;[D,a^*],\quad a\in {\Cal A},
\tag 2.5
$$
we see that the spaces $\Omega^n({\Cal A})$ are ${\Cal A}$-bimodules
closed under the involution $^*$ and that
$$
Aux\;:=\;\oplus \;Aux^n
\tag 2.6
$$
is a two-sided ideal in
$$
\Omega ({\Cal A})\;:=\;\oplus\;\Omega^n ({\Cal A}),
\tag 2.7
$$
closed under the operation $^*$. Thus, for each $n$, $\Omega_D^n
({\Cal A})$ is an ${\Cal A}$-bimodule closed under $^*$. It follows
that
$$
\Omega_D ({\Cal A})\;:=\;\oplus\;\Omega_D^n({\Cal A})
\tag 2.8
$$
is a $^*$algebra of equivalence classes of bounded operators on $H$,
with multiplication defined as the multiplication of operators on $H$.
Since ${\Cal A} = \Omega^0({\Cal A})=\Omega_D^0({\Cal A})$ is a
$^*$subalgebra of $\Omega_D({\Cal A})$ containing an identity element,
$\Omega_D({\Cal A})$ is a unital $^*$algebra of equivalence classes \
(mod $Aux)$ of bounded operators on $H$ which is an ${\Cal
A}$-bimodule.

The degree of a form $\alpha \in \Omega_D^n({\Cal A})$ is defined by
$$
deg (\alpha)\;=\;n,\quad n\;=\;0,1,2,\cdots .
\tag 2.9
$$
Clearly, $deg (\alpha^*) = deg (\alpha)$, by (2.4), (2.5). With this
definition of $deg$, $\Omega_D({\Cal A})$ is $\Bbb Z$-graded. If
$\alpha$ is given by
$$
\alpha\;=\;\sum_i a_0^i [D,a_1^i]\cdots [D,a_n^i] (\text{mod} Aux^n) \in
\Omega_D^n ({\Cal A})
$$
we set
$$
d\alpha\;:=\;\sum_i [D,a_0^i] [D,a_1^i]\cdots [D,a_n^i] \in
\Omega_D^{n+1} ({\Cal A})\;. \tag 2.10
$$
The map
$$
d\;:\;\Omega_D^n ({\Cal A}) \;\rightarrow\;\Omega_D^{n+1} ({\Cal
A}), \quad \alpha\;\mapsto\; d\alpha
\tag 2.11
$$
is a $\Bbb C$-linear map from $\Omega_D({\Cal A})$ to itself which
increases the degree of a form by one and satisfies
$$
d(\alpha\cdot\beta)\;=\;(d\alpha)\cdot\beta\;+\;(-1)^{deg\,\alpha}\;
\alpha\cdot(d\beta),
\tag 2.12
$$
for any homogeneous element $\alpha$ of $\Omega_D({\Cal A})$ (Leibniz
rule) and
$$
d^2\;=\;0\;.
\tag 2.13
$$
Hence $\Omega_D({\Cal A})$ is a differential algebra which is a $\Bbb
Z$-graded complex.

These notions are described in detail (and in a more general setting)
in [4].

In non-commutative geometry,\  \ub{vector bundles} \ over a
non-commutative space described by a $^*$algebra ${\Cal A}$ are
defined as \ub{finitely generated}, \ub{projective left} ${\Cal
A}$-\ub{modules}. Let $E$ denote (the ``space of sections'' of) a
vector bundle over ${\Cal A}$. A connection $\nabla$ on $E$ is a
$\Bbb C$-linear map
$$
\nabla\;:\;E\;\rightarrow\;\Omega_D^1 ({\Cal A}) \;\otimes_{{\Cal
A}}\;E
\tag 2.14
$$
with the property that (with $da = [D,a]$, for all $a\in{\Cal A}$)
$$
\nabla (as)\;=\;da\;\otimes_{{\Cal A}}\;s + a\;\nabla s,
\tag 2.15
$$
for arbitrary $a\in{\Cal A}$, $s\in E$. The definition of $\nabla$
can be extended to the space
$$
\Omega_D(E)\;=\;\Omega_D({\Cal A})\;\otimes_{{\Cal A}}\;E
\tag 2.16
$$
in a canonical way, and, for $s\in \Omega_D(E)$ and a homogeneous form
$\alpha\in\Omega_D({\Cal A})$,
$$
\nabla(\alpha s)\;=\;(d\alpha) s\;+\;(-1)^{deg\,\alpha}\;\alpha \;
\nabla s .\tag 2.17
$$
Thanks to (2.14) - (2.17), it makes sense to define the
\ub{curvature}, $R(\nabla)$, of the connection $\nabla$ as the $\Bbb
C$-linear map
$$
R(\nabla)\;:=\;- \nabla^2
\tag 2.18
$$
from $\Omega_D(E)$ to $\Omega_D(E)$. Actually, it is easy to check
that $R(\nabla)$ is ${\Cal A}$-\ub{linear}, i.e. $R(\nabla)$ is a
\ub{tensor}.

A \ub{trivial} vector bundle, $E^{(N)}$, corresponds to a finitely
generated, \ub{free} left ${\Cal A}$-module, i.e., one that has a
basis $\{ s_1,\cdots, s_N\}$, for some finite $N$ called its
dimension. Then
$$
E^{(N)}\; \simeq\; {\Cal A} \oplus \cdots \oplus {\Cal A}\;\equiv\;
{\Cal A}^n,
$$
(with $N$ summands). The affine space of connections on $E^{(N)}$ can
be characterized as follows: Given a basis $\{ s_1,\cdots, s_N\}$ of
$E^{(N)}$, there are $N^2$ 1-forms $\rho_\alpha^\beta \in \Omega_D^1
({\Cal A})$, the \ub{components} of the connection $\nabla$, such that
$$
\nabla s_\alpha\;=\;-\;\rho_\alpha^\beta \otimes_{{\Cal A}} s_\beta,
\tag 2.19
$$
(where, here and in the following, we are using the summation
convention). Then
$$
\nabla (a^\alpha s_\alpha)\;=\;da^\alpha \otimes_{{\Cal A}} s_\alpha -
a^\alpha \rho_\alpha^\beta \otimes_{{\Cal A}} s_\beta,
\tag 2.20
$$
by (2.15). Furthermore, by (2.18) and (2.20),
$$
\align
R(\nabla)(a^\alpha s_\alpha)\;=\;
&-\;\nabla \bigl( da^\alpha \otimes_{{\Cal A}} s_\alpha - a^\alpha
\rho_\alpha^\beta \otimes_{{\Cal A}} s_\beta\bigr) \\
=\;&-\;\bigl(d^2 a^\alpha \otimes_{{\Cal A}} s_\alpha + da^\alpha
\rho_\alpha^\beta \otimes_{{\Cal A}} s_\beta \\
&-\;da^\alpha \rho_\alpha^\beta \otimes_{{\Cal A}} s_\beta - a^\alpha
d \rho_\alpha^\beta \otimes_{{\Cal A}} s_\beta \\
&-\; a^\alpha \rho_\alpha^\gamma \rho_\gamma^\beta \otimes_{{\Cal A}}
s_\beta\bigr)\\
=\;&\;a^\alpha \bigl(d \rho_\alpha^\beta + \rho_\alpha^\gamma
\rho_\gamma^\beta\bigr) \otimes_{{\Cal A}} s_\beta\;.
\tag 2.21
\endalign
$$
Thus, the curvature tensor $R(\nabla)$ is completely determined by the
$N\times N$ matrix $\theta\equiv(\theta_\alpha^\beta)$ of 2-forms
given by
$$
\theta_\alpha^\beta\;=\;d \rho_\alpha^\beta\;+\;\rho_\alpha^\gamma
\;\rho_\gamma^\beta\;.
\tag 2.22
$$
The curvature matrix $\theta$ satisfies the \ub{Bianchi identity}
$$
d\theta + \rho\theta - \theta\rho\;\equiv\;\bigl( d\theta_\alpha^\beta
+ \rho_\alpha^\gamma \theta_\gamma^\beta - \theta_\alpha^\gamma
\rho_\gamma^\beta\bigr)\;=\;0\;.
\tag 2.23
$$
If one introduces a new basis
$$
\tilde s_\alpha\;=\;M_\alpha^\beta s_\beta, \; M_\alpha^\beta \in {\Cal A},
\; \alpha,\beta\;=\;1,\cdots,N,
\tag 2.24
$$
where the matrix $M\equiv (M_\alpha^\beta)$ is invertible, then the
components, $\tilde\rho$, of $\nabla$ in the new basis
$\{\tilde s_1, \cdots, \tilde s_N\}$ of $E^{(N)}$ are given by
$$
\tilde\rho\;=\;M\rho\;M^{-1}\;-\;dM\cdot M^{-1},
\tag 2.25
$$
and the components of the curvature $R(\nabla)$ transform according to
$$
\widetilde\theta\;=\; M \theta M^{-1},
\tag 2.26
$$
as one easily checks.

Given a basis $\{ s_1,\cdots, s_N\}$ of $E^{(N)}$, one may define a
\ub{Hermitian structure} $\langle\cdot,\cdot\rangle$  on $E^{(N)}$ by
setting
$$
\langle s_\alpha, s_\beta\rangle\;=\;\delta_{\alpha\beta}\;\id \;,
\tag 2.27
$$
with
$$
\langle a^\alpha s_\alpha, b^\beta s_\beta\rangle\;=\; a^\alpha
\langle s_\alpha, s_\beta\rangle (b^\beta)^*\;=\; \sum_\alpha a^\alpha
(b^\alpha)^* \in {\Cal A}\;.
\tag 2.28
$$
The definition of $\langle\cdot ,\cdot\rangle$ can be extended
canonically to $\Omega_D(E^{(N)})$, and there is then an obvious
notion of ``\ub{unitary connection}'' on $E^{(N)}$: $\nabla$ is
unitary iff
$$
d\langle s,s'\rangle\;=\;\langle\nabla s, s'\rangle\;-\;\langle s,
\nabla s'\rangle\;.
\tag 2.29
$$
This is equivalent to the condition that
$$
\rho_\alpha^\beta\;=\;\bigl( \rho_\beta^\alpha\bigr)^*,
\tag 2.30
$$
where the $\rho_\alpha^\beta$ are the components of $\nabla$ in the
orthonormal basis $\{ s_1,\cdots , s_N\}$ of $E^{(N)}$.

In the examples studied in Sections 3 through 5, we shall consider
unitary connections on trivial vector bundles, in particular on
``\ub{line bundles}'' for which $N=1$. A (unitary) connection $\nabla$
on a line bundle $E^{(1)} \simeq {\Cal A}$ is completely determined
by a (selfadjoint) 1-form $\rho \in \Omega_D^1 ({\Cal A})$.

The data $({\Cal A},\pi,H,D)$ defining a non-commutative space with
differentiable structure is also called a \ub{Fredholm module}.
Following [4], we shall say that the Fredholm module $({\Cal
A},\pi,H,D)$ is $(d,\infty)$-summable if
$$
tr\,\bigl(D^2 + \id\bigr)^{- p/2} \;<\;\infty , \quad \text{for
all}\quad p > d\;.
\tag 2.31
$$
Let $Tr_\omega (\cdot)$ denote the so-called \ub{Dixmier trace} on
$B(H)$ which is a positive, cyclic trace vanishing on trace-class
operators; see [4]. We define a notion of \ub{integration of forms},
$\int\!\!\!\!\!-$, by setting
$$
\int\!\!\!\!\!\!- \alpha\;:=\; Tr_\omega\;\bigl( \alpha\mid
D\mid^{-d}\bigr),
\tag 2.32
$$
for \ $\alpha \in \Omega ({\Cal A}) = \oplus \Omega^n ({\Cal A})$; \
(see (2.2), (2.7)). If $d=\infty$ but \ exp$\;(-\varepsilon D^2)$ is
trace class, for any $\varepsilon > 0 $ (as assumed), we set
$$
\int\!\!\!\!\!\!-
\alpha\;:=\;\displaystyle\mathop{Lim_\omega}_{\varepsilon\downarrow
0}\; \frac{tr \bigl(\alpha \exp (-\varepsilon D^2)\bigr)}{tr
\bigl(\exp (- \varepsilon D^2)\bigr)}\;,
\tag 2.33
$$
(on forms $\alpha$ which are ``
analytic elements'' for the automorphism group determined by the
dynamics \ exp$(it D^2), t\in\Bbb R$; see [12]) and $Lim_\omega$
denotes a limit defined in terms of a kind of ``Cesaro mean''
described in [4]. Then
$$
\int\!\!\!\!\!\!- \alpha\beta\;=\;\int\!\!\!\!\!\!- \beta\alpha,
\tag 2.34
$$
i.e., $\int\!\!\!\!\!-$ is cyclic; it is also a \ub{non-negative}
linear functional on $\Omega ({\Cal A})$. It can thus be used to
define a positive semi-definite inner product on $\Omega({\Cal A})$:
For $\alpha$ and $\beta$ in $\Omega({\Cal A})$, we set
$$
(\alpha,\beta)\;=\;\int\!\!\!\!\!\!- \alpha\beta^*.
\tag 2.35
$$
Then the closure of $\Omega({\Cal A})$ (mod kernel of $(\cdot,\cdot)$)
in the norm determined by $(\cdot,\cdot)$ is a Hilbert space, denoted
by $L^2\bigl(\Omega({\Cal A})\bigr)$. Given an element
$\alpha\in\Omega^n({\Cal A})$, we can now define a canonical
representative, $\alpha^\perp$, in the equivalence class $\alpha$ (mod
$Aux^n) \in \Omega_D^n({\Cal A})$ as the unique (modulo the kernel of
$(\cdot,\cdot)$) operator in $\alpha$ (mod $Aux^n)$ which is
orthogonal to $Aux^n$ in the scalar product $(\cdot,\cdot)$ given by
(2.35); $(Aux^n$ has been defined in eq. (2.3)). Then, for $\alpha$
and $\beta$ in $\Omega_D({\Cal A})$, we set
$$
(\alpha,\beta)\;:=\;(\alpha^\perp,
\beta^\perp)\;\equiv\;\int\!\!\!\!\!\!- \alpha^\perp (\beta^\perp)^*,
\tag 2.36
$$
and this defines a positive semi-definite inner product on
$\Omega_D({\Cal A})$. The closure of $\Omega_D({\Cal A})$ (mod kernel
of $(\cdot,\cdot))$ in the norm determined by $(\cdot,\cdot)$ is the
Hilbert space of ``square-integrable differential forms'', denoted by
$\Lambda_D({\Cal A})$.

In order to define the Chern-Simons forms and Chern-Simons actions in
non-commut\-ative geometry, it is useful to consider a trivial example
of the notions introduced, so far. Let $I$ denote the interval
$[0,1]\subset\Bbb R$. Let ${\Cal A}_1 = C^\infty (I)$ be the algebra
of smooth functions, $f(t)$, on the open
interval (0,1) which, together with all their derivatives in $t$, have
(finite) limits as $t$ tends to 0 or 1. Let $H_1=L^2(I)\otimes\Bbb
C^2$ denote the Hilbert space of square-integrable (with respect to
Lebesgue measure, $dt$, on $I$) two-component spinors, and $D_1 = i
\frac{\partial}{\partial t} \otimes \sigma_1$ the one-dimensional
Dirac operator (with appropriate selfadjoint boundary conditions),
where $\sigma_1,\sigma_2$ and $\sigma_3$ are the usual Pauli matrices.
A representation $\pi_1$ of ${\Cal A}_1$ on $H_1$ is defined by
setting
$$
\pi_1(a)\;=\;a \otimes \id_2,\quad a\in{\Cal A}_1.
\tag 2.37
$$
The geometry of $I$ is then coded into the space $({\Cal A}_1,
\pi_1,H_1,D_1)$. The space of 1-forms is given by
$$
\Omega_{D_1}^1({\Cal A}_1)\;=\;\bigl\{ \omega\otimes\sigma_1 : \omega
= \sum_i a^i \partial_t b^i; a^i, b^i \in {\Cal A}_1\bigr\}.
\tag 2.38
$$
The space, $\Omega_{D_1}^2 ({\Cal A}_1)$, of 2-forms is easily seen to
be \ub{trivial}, and the cohomology groups vanish. The Fredholm module
$({\Cal A}_1,\pi_1, H_1, D_1)$ is $\Bbb Z_2$-graded.
The $\Bbb Z_2$-grading, $\gamma$, is given by
$$
\gamma\;=\;\id \otimes \sigma_3,
\tag 2.39
$$
and $[\gamma, \pi_1 (a)] = 0$, for all $a\in {\Cal A}_1$, while
$$
\bigl\{ \gamma, D_1\bigr\}\;\equiv\;\gamma D_1 + D_1\gamma \;=\;0.
\tag 2.40
$$

Using this trivial example, we may introduce the notion of a \
``\ub{cylinder over a non-} \ub{commutative space}'': Let $({\Cal
A},\pi,H,D)$ be an arbitrary non-commutative space, and let $({\Cal
A}_1,\pi_1,H_1,D_1)$ be as specified in the above example. Then we
define the cylinder over $({\Cal A},\pi,H,D)$ to be given by the
non-commutative space $(\widetilde{\Cal A},\widetilde\pi,\widetilde H,
\widetilde D)$, where
$$
\widetilde H = H\otimes H_1,\quad
\widetilde\pi = \pi \otimes \pi_1,\quad
\widetilde{\Cal A} = {\Cal A}\otimes{\Cal A}_1,
\tag 2.41
$$
and
$$
\widetilde D\;=\;\id \otimes D_1\;+\;D \otimes \gamma,
\tag 2.42
$$
with $\gamma$ as in (2.39). The space $(\widetilde{\Cal
A},\widetilde\pi,\widetilde H, \widetilde D)$ is $\Bbb Z_2$-graded: We
define
$$
\Gamma\;=\;\id \otimes (\id \otimes \sigma_2),
\tag 2.43
$$
$\widetilde D_1 := \id\otimes D_1$, $\widetilde D_2 = D\otimes\gamma$. Then
$$
\{ \Gamma,\widetilde D_1\}\;=\;\{ \Gamma,\widetilde
D_2\}\;=\;\{\Gamma,\widetilde D\}\;=\; \{\widetilde D_1, \widetilde
D_2\}\;=\;0,
\tag 2.44
$$
and
$$
\bigl[ \Gamma, \widetilde\pi (\tilde a)\bigr]\;=\;0,\quad\text{for all}
\quad \tilde a\in\widetilde{\Cal A}.
\tag 2.45
$$
It is easy to show (see [12]) that arbitrary sums of operators of the
form
$$
\tilde a_0 [\widetilde D_{\varepsilon_1}, \tilde a_1] \cdots
[\widetilde D_{\varepsilon_n}, \tilde a_n],\quad \varepsilon_1,
\cdots, \varepsilon_n = 1,2,
\tag 2.46
$$
belong to $\Omega^k(\widetilde{\Cal A})$. Furthermore, if two or more
of the $\varepsilon_i$'s take the value 1 then the operator defined in
(2.46) belongs to $Aux^n$.

We define integration, $\int\!\!\!\!\!\!\sim (\cdot)$, on $(\widetilde{\Cal
A},\widetilde\pi,\widetilde H,\widetilde D)$ by setting, for any
$\alpha\in\Omega (\widetilde{\Cal A})$,
$$
\int\!\!\!\!\!\!\!\sim \alpha\;:=\;\int_0^1 dt \int\!\!\!\!\!\!- Tr_{\Bbb
C^2} \bigl(\alpha (t)\bigr),
\tag 2.47
$$
where $\alpha (t)$ is a 2$\times$2 matrix of elements of $\Omega({\Cal
A})$. The integral $\int\!\!\!\!\!\!\sim (\cdot)$ is positive semi-definite
and cyclic on the algebra $\Omega (\widetilde{\Cal A})$. We are now
prepared to define the Chern-Simons forms and Chern-Simons actions in
non-commutative geometry, (for connections on trivial vector bundles).
Let $({\Cal A},\pi,H,D)$ be a real, compact non-commutative space with
a differentiable structure determined by $D$. Let $E=E^{(N)} \simeq
{\Cal A}^N$ be a trivial vector bundle over ${\Cal A}$, and let
$\nabla$ be a connection on $E$. By (2.19), $\nabla$ is completely
determined by an $N\times N$ matrix $\rho = (\rho_\alpha^\beta)$ of
1-forms. By (2.21), the curvature of $\nabla$ is given by the $N\times
N$ matrix of 2-forms
$$
\theta\;=\;d \rho + \rho^2,
$$
where $d$ is the differential on $\Omega_D ({\Cal A})$ defined in
(2.10). Following Quillen [11], we define the Chern-Simons
(2$n$--1)-form associated with $\nabla$ as follows: Let $\nabla_0$
denote the flat connection on $E$ corresponding to an $N\times N$
matrix $\rho_0$ of 1-forms which, in an appropriate gauge, vanishes.
We set
$$
\rho_t\;=\;t \rho + (1-t)\;\rho_0\;=\; t \rho,
\tag 2.48
$$
for $\rho_0=0$, corresponding to the connection $\nabla_t = t\nabla +
(1-t) \nabla_0$. The curvature of $\nabla_t$ is given by the matrix
$\theta_t$ of 2-forms given by
$$
\theta_t\;=\;d\rho_t + \rho_t^2\;=\;t d\rho + t^2\rho^2.
$$
The Chern-Simons $(2n-1)$-form associated with $\nabla$ is then given
by
$$
\vartheta^{2n-1} (\rho)\;:=\;\frac{1}{(n-1)!} \int_0^1 dt \;\rho\;
\theta_t^{n-1}. \tag 2.49
$$
For $n=2$, we find
$$
\vartheta^3 (\rho)\;=\;\frac 1 2\;\bigl\{ \rho d \rho + \frac 2 3
\;\rho^3 \bigr\},
\tag 2.50
$$
and, for $n=3$,
$$
\vartheta^5 (\rho)\;=\;\frac 1 6 \bigl\{ \rho d\rho d\rho + \frac 3 4
\rho^3 d \rho + \frac 3 4 \rho (d\rho) \rho^2 + \frac 3 5 \rho^5
\bigr\}.
\tag 2.51
$$

In order to understand where these definitions come from and how to
define Chern-Simons actions, we extend $E$ to a trivial vector bundle
over the cylinder $(\widetilde{\Cal A},\widetilde\pi,\widetilde H,
\widetilde D)$ over $({\Cal A}, \pi, H, D)$: We set
$$
\widetilde E\;=\; E \otimes C^\infty (I) \otimes \id_2 \simeq
\widetilde{\Cal A}^N .
\tag 2.52
$$
We also extend the connection $\nabla$ on $E$ to a connection
$\widetilde\nabla$ on $\widetilde E$ by interpolating between $\nabla$
and the flat connection $\nabla_0$: By (2.39), (2.41) and (2.42), a
1-form in $\Omega_{\widetilde D}^1 (\widetilde{\Cal A})$ is given by
$$
\pmatrix
\rho (t) & \quad \phi (t) \\
\phi (t) & - \rho (t)
\endpmatrix ,
$$
where $\rho (t)\in \Omega_D^1({\Cal A})$ for all $t\in I$. Thus
$$
\tilde\rho_\alpha^\beta (t)\;:=\;t\enskip
\pmatrix
\rho_\alpha^\beta & \enskip 0 \\
0 & - \rho_\alpha^\beta
\endpmatrix , \quad
\alpha,\beta = 1, \cdots, N,
\tag 2.53
$$
is an element of $\Omega_{\widetilde D}^1 (\widetilde{\Cal A})$. We
define $\widetilde \nabla$ to be the connection on $\widetilde E$
determined by the matrix $\widetilde \rho (t) = \bigl(
\widetilde\rho_\alpha^\beta (t)\bigr)$ of 1-forms defined in (2.53).
Let $\tilde d$ be the differential on $\Omega_{\widetilde
D}(\widetilde{\Cal A})$ defined as in (2.10), (with ${\Cal A}$
replaced by $\widetilde{\Cal A}$ and $D$ replaced by $\widetilde D$).
By (2.42)
$$
\tilde d \tilde\rho \;=\;
\pmatrix 0 &- i\rho \\
i\rho &\enskip 0
\endpmatrix\;+\;t\;
\pmatrix
d\rho & 0 \\
\enskip 0 & d\rho
\endpmatrix ,
\tag 2.54
$$
with $\rho = (\rho_\alpha^\beta)$. Hence the curvature of
$\widetilde\nabla$ is
given by the matrix of 2-forms $\widetilde\theta$, with
$$
\widetilde\theta (t)\;=\;\theta_t\otimes\id_2 + \rho\otimes\sigma_2,
\tag 2.55
$$
where
$$
\theta_t\;=\;t d \rho + t^2 \rho^2.
$$
Let $\varepsilon$ be an arbitrary bounded operator on $H$ commuting
with $D$ and with all operators in $\pi ({\Cal A})$. Then,
$$
\widetilde\varepsilon\;:=\;\varepsilon\otimes\id_2
$$
commutes with $\widetilde D$ and with $\widetilde\pi (\widetilde{\Cal
A})$ and hence with $\Omega(\widetilde{\Cal A})$. It also commutes
with the $\Bbb Z_2$-grading $\Gamma =\id\otimes (\id \otimes \sigma_2)$, (as
defined in (2.43)). We now define a graded trace $\tau_\epsilon
(\cdot)$ on $\Omega(\widetilde{\Cal A})$ by setting
$$
\tau_\varepsilon (\alpha)\;:=\;\int\!\!\!\!\!\!\!\sim (\tilde\varepsilon
\Gamma \alpha), \quad \alpha \in\Omega (\widetilde{\Cal A}),
\tag 2.56
$$
where $\int\!\!\!\!\!\!\sim (\cdot)$ is given by (2.47). It is easy to show
that
$$
\tau_\varepsilon (\alpha)\;=\;0 \quad \text{if deg } \alpha\quad
\text{is \ub{odd}},
\tag 2.57
$$
and
$$
\tau_\varepsilon\bigl([\alpha,\beta]_*\bigr)\;=\;0,\quad
\text{for all } \alpha,\beta\quad\text{in } \Omega(\widetilde{\Cal
A}), \tag 2.58
$$
where
$$
[\alpha,\beta]_*\;=\;\alpha\cdot\beta - (-1)^{\deg \alpha \deg
\beta}\; \beta\cdot\alpha
$$
is the graded cummutator.

Using the Bianchi identity,
$$
d\theta^n + [\rho, \theta^n] = 0,
$$
which follows from eq.~(2.23) by induction, and the graded cyclicity
of $\tau_\varepsilon$ (see (2.57), (2.58)) one shows that
$$
\tau_\varepsilon \bigl((\widetilde\theta^n)^\perp\bigr)\;=\; n!
\tau_\varepsilon \bigl(\bigl(\tilde d \vartheta^{2n-1}
(\widetilde\rho)\bigr)^\perp\bigr),
\tag 2.59
$$
where $\alpha^\perp$ is the canonical representative in the
equivalence class $\alpha$ (mod $Aux^m) \in \Omega_{\widetilde D}^m
(\widetilde{\Cal A})$ orthogonal to $Aux^m$, for any $m=1,2,\cdots$,
(as explained after eq.~(2.35)). The calculation proving (2.59) is
indicated in [11]; (see also [12] for details). In fact, eq.~(2.59) is
a general identity valid for arbitrary connections on trivial vector
bundles over a non-commutative space and arbitrary graded traces [11].

In the case considered here, the l.h.s. of eq.~(2.59) can be rewritten
in the following interesting way:
$$
\align
\tau_\varepsilon\bigl((\widetilde\theta^n)^\perp\bigr)\;&=\;
\int_0^1 dt \int\!\!\!\!\!\!- Tr_{\Bbb C^2} \bigl(\tilde\varepsilon\; \Gamma
\bigl( \widetilde\theta^n (t)\bigr)^\perp\bigr) \\
&=\;n \int_0^1 dt \int\!\!\!\!\!\!- Tr_{\Bbb C^2}
\bigl((\varepsilon\otimes\id_2)\, \Gamma\, (\rho\otimes\sigma_2)(\theta_t^{n-1}
\otimes \id_2)\bigr).
\tag 2.60
\endalign
$$
This is shown by plugging eq.~(2.55) for $\widetilde\theta (t)$ into
the expression in the middle of (2.60) and noticing that (1) all terms
contributing to $\widetilde\theta^n(t)$ with more than one factor
proportional to $[\widetilde D_1, \tilde a]$, i.e., with more than one
factor of the form $\rho\otimes\sigma_2$, are projected out when
passing from $\widetilde\theta^n(t)$ to $\bigl(\widetilde\theta^n
(t)\bigr)^\perp$, (see the remark following eq.~(2.46)), and \ (2) \
$Tr_{\Bbb C^2} \bigl((\varepsilon\otimes\id_2)\; \Gamma$ \break
$(\theta_t^n\otimes\id_2)\bigr)=0$.

Evaluating the trace, $Tr_{\Bbb C^2}$, on the r.h.s. of (2.60) and
recalling the definition (2.49) of the Chern-Simons form, we finally
conclude that
$$
\align
\tau_\varepsilon\bigl((\widetilde\theta^n)^\perp\bigr)\;
&=\;2n \int_0^1 dt \int\!\!\!\!\!\!-
\bigl(\varepsilon\rho\theta_t^{n-1}\bigr)^\perp \\
&=\;2n! \int\!\!\!\!\!\!- \bigl(\varepsilon\bigl(\vartheta^{2n-1}
(\rho)\bigr)^\perp\bigr).
\tag 2.61
\endalign
$$
\bf\ub{Remark}\rm . The r.h.s. of (2.59) can actually be rewritten as
$$
n! \int\!\!\!\!\!\!- Tr_{\Bbb C^2} \bigl( \id \otimes \sigma_3 \bigl(
\vartheta^{2n-1} \bigl( \widetilde\rho (1)\bigr)\bigr)^\perp\bigr);
$$
see [12].

\ub{Chern-Simons actions}, $I_\varepsilon$, in non-commutative
geometry are defined by setting
$$
I_\varepsilon^{2n-1}(\rho)\;:=\;\kappa\int\!\!\!\!\!\!-
\bigl(\varepsilon\bigl( \vartheta^{2n-1} (\rho)\bigr)^\perp\bigr),
\tag 2.62
$$
where $\kappa$ is a constant. Using (2.50) and (2.51) and using the
properties of $\varepsilon$ and the cyclicity of $\int\!\!\!\!\!-
(\cdot)$, we find
$$
I_\varepsilon^3 (\rho)\;=\;\frac\kappa 2 \int\!\!\!\!\!\!-
\bigl(\varepsilon\bigl( \rho d \rho + \frac 2 3
\rho^3\bigr)^\perp\bigr),
$$
and
$$
I_\varepsilon^5 (\rho)\;=\;\frac\kappa 6 \int\!\!\!\!\!\!-
\bigl(\varepsilon \bigl( \rho d \rho d \rho + \frac 3 2 \rho^3 d \rho
+ \frac 3 5 \rho^5 \bigr)^\perp\bigr).
\tag 2.63
$$
A particularly important special case is obtained by choosing the
operator $\varepsilon$ to belong to $\Omega({\Cal A})$.
Since $\varepsilon$ commutes with $D$ and with $\pi({\Cal A})$, this
implies that $\varepsilon$ belongs to the \ub{centre} of the algebra
$\Omega({\Cal A})$. In the examples discussed in the remainder of this
paper, this property is always assumed.

One point of formula (2.61) and generalizations thereof, discussed in
[12] (and involving ``higher-dimensional cylinders''), is that it
enables us to define \ub{differences} of Chern-Simons actions even
when the underlying vector bundle is \ub{non-trivial}. If $\nabla_0$
denotes a fixed reference connection on a vector bundle $E$ over a
non-commutative space $({\Cal A},\pi,H,D)$ and $\nabla$ is an
arbitrary connection on $E$ we set
$$
\int\!\!\!\!\!\!- \bigl(\varepsilon\bigl(\vartheta^{2n-1}
(\nabla)\bigr)^\perp\bigr)\;:=\;\tau_\varepsilon \bigl(
(\widetilde\theta^n)^\perp\bigr)\;+\;\text{const.,}
\tag 2.64
$$
where $\widetilde\theta$ is the curvature of a connection
$\widetilde\nabla$ on a vector bundle $\widetilde E$ over the cylinder
$(\widetilde{\Cal A},\widetilde\pi,\widetilde H, \widetilde D)$
interpolating between $\nabla$ and $\nabla_0$, and the constant on the
r.h.s. of (2.64) is related to the choice of $\widetilde\nabla$ and of
the Chern-Simons action associated with $\nabla_0$.

Formulas (2.62) and (2.64) are helpful in understanding the
topological nature of Chern-Simons actions.

Next, we propose to discuss various concrete examples and indicate
some applications to theories of gravity.

\vskip 1.5truecm
\bf 3. \underbar{Some ``three-dimensional'' Chern-Simons actions}\rm.
\medskip

We consider a ``Euclidean space-time manifold'' $X$ which is the
Cartesian product of a Riemann surface $M_2$ and a two-point set, i.e.
$X$ consists of two copies of $M_2$. The algebra ${\Cal A}$ used in
the definition of the non-commutative space considered in this section
is given by
$$
{\Cal A}\;=\;C^\infty (M_2) \otimes {\Cal A}_0,
\tag 3.1
$$
where ${\Cal A}_0$ is a finite-dimensional, unital $^*$algebra of
$M\times M$ matrices. The Hilbert space $H$ is chosen to be
$$
H\;=\;H_0\oplus H_0,
\tag 3.2
$$
where
$$
H_0\;=\;\Bbb C^N \otimes L^2 (S) \otimes \Bbb C^M,
\tag 3.3
$$
and $L^2(S)$ is the Hilbert space of square-integrable spinors on
$M_2$ for some choices of a spin structure and of a (Riemannian)
volume form on $M_2$.

The representation $\pi$ of ${\Cal A}$ on $H$ is given by
$$
\pi (a)\;=\;
\pmatrix
\id_N \otimes a & \quad 0 \\
\quad 0 & \id_N\otimes a
\endpmatrix ,
\tag 3.4
$$
for $a \in {\Cal A}$.

We shall work locally over some coordinate chart of $M_2$, but we do
not describe how to glue together different charts (this is standard),
and we shall write ``$M_2$'' even when we mean a coordinate chart of
$M_2$. Let $g=(g_{\mu\nu})$ be some fixed, Riemannian reference metric
on $M_2$, and $(e_\mu^a)$ a section of orthonormal 2-frames, $\mu,a =
1,2$. Let $\gamma^1, \gamma^2$ denote the two-dimensional Dirac
matrices satisfying
$$
\bigl\{ \gamma^a, \gamma^b\bigr\}\;\equiv\;
\gamma^a\gamma^b + \gamma^b\gamma^a\;=\;2\delta^{ab},
\tag 3.5
$$
and
$$
\gamma^5\;=\;\gamma^1 \gamma^2.
\tag 3.6
$$
The matrices $\id_N\otimes\gamma^a$ will henceforth also be denoted by
$\gamma^a$. Let $\partial\!\!\!/$
denote the covariant Dirac operator on $\Bbb C^N \otimes L^2(S)$
corresponding to the Levi-Civita spin connection determined by
$(e_\mu^a)$ and acting trivially on $\Bbb C^N$. Let $K$ denote an
operator of the form
$$
K\;=\;k\otimes\id\otimes\id,
\tag 3.7
$$
where $k$ is some real, symmetric $N\times N$ matrix. The vector space
$\Bbb C^N$ and the matrix $k$ do not play any interesting role in the
present section but are introduced for later convenience. Let $\phi_0$
be a hermitian $M\times M$ matrix $(\neq \id_M)$. The operator $D$ on
$H$ required in the definition of a non-commutative space is chosen as
$$
D\;=\; \pmatrix
\partial\!\!\!/ \otimes \id_M\quad & i\gamma^5 K\otimes \phi_0 \\
- i\gamma^5 K\otimes\phi_0 & \partial\!\!\!/ \otimes \id_M \endpmatrix ,
\tag 3.8
$$
Then (\ub{locally} on $M_2$) the space of 1-forms, $\Omega_D^1 ({\Cal
A})$, (the ``cotangent bundle'') is a free, hermitian ${\Cal
A}$-bimodule of dimension 3, with an orthonormal basis given by
$$
\varepsilon^a\;=\;\pmatrix
\gamma^a \otimes \id_M \quad & \quad 0 \quad\\
\quad 0\quad  & \gamma^a \otimes \id_M \endpmatrix ,
\quad a\;=\;1,2,
\tag 3.9
$$
and
$$
\varepsilon^3\;=\; \pmatrix
\quad 0 \qquad & - \gamma^5 \otimes \id_M \\
\gamma^5 \otimes \id_M & \quad 0 \endpmatrix
\phantom{\quad a\;=\;1,1,\quad }
\tag 3.10
$$
and the hermitian structure is given by the normalized trace, $tr$, on
$\Bbb M_N(\Bbb C)\otimes$ Cliff. Then, for $a,b=1,2,3,$
$$
\langle \varepsilon^a,\varepsilon^b\rangle\;=\;tr\,\bigl(\varepsilon^a
(\varepsilon^b)^*\bigr)\;=\;\delta^{ab}\id_M.
$$
We define a central element $\varepsilon\in\Omega^3({\Cal A})$ by
setting
$$
\varepsilon\;=\;\varepsilon^1\varepsilon^2\varepsilon^3\;=\;\pmatrix
\quad 0\quad & \id\\
- \id & 0\endpmatrix .
\tag 3.11
$$
It is trivial to verify that $\varepsilon $ commutes with the operator
$D$ and with $\pi({\Cal A})$, and, since $\varepsilon^1,\varepsilon^2$
and $\varepsilon^3$ belong to $\Omega^1 ({\Cal A})$, $\varepsilon$
belongs to $\Omega^3({\Cal A})$.

A 1-form $\rho$ has the form
$$
\rho\;=\;\sum_j \pi (a^j) \bigl[ D, \pi (b^j)\bigr], \quad a^j,b^j\in
{\Cal A},
\tag 3.12
$$
and, without loss of generality, we may impose the constraint
$$
\sum_j a^j b^j\;=\;\id .
\tag 3.13
$$
Then
$$
\rho\;=\;\pmatrix
\quad A \qquad & i \gamma^5 K\phi \\
- i \gamma^5 K\phi & \quad A \quad \endpmatrix ,
\tag 3.14
$$
where \ $A = \sum_j a^j (\partial\!\!\!/ b^j)$, and
$\phi+\phi_0=\id_N\otimes \bigl( \sum_j a^j \phi_0 b^j\bigr)$. \
The 1-form $\rho$ given in (3.14) determines a connection, $\nabla$,
on the ``line bundle'' $E \simeq {\Cal A}$. The three-dimensional
Chern-Simons action of $\nabla$ is then given by
$$
\align
I_\varepsilon^3 (\rho)\; &=\;\frac \kappa 2 \;\int\!\!\!\!\!\!- \;
\bigl(\varepsilon (\rho\, d\rho + \frac 2 3\;\rho^3)^\perp\bigr)  \\
&=\;\frac \kappa 2\; Tr_\omega\;\bigl(\varepsilon (\rho\, d\rho\;+\;
\frac 2 3 \;\rho^3)^\perp D^{-2}\bigr),
\tag 3.15
\endalign
$$
as follows from eqs.~(2.63) and (2.32); \ (the Fredholm module \ $({\Cal
A},\pi,H,D)$ \ is \ $(2,\infty)$-summable!).

In order to proceed in our calculation, we must determine the spaces
of ``auxiliary fields'' $Aux^n$, for $n=1,2,3$. Clearly $Aux^1=0$. To
identify $Aux^2$, we consider a 1-form
$$
\rho\;=\;\sum_j\pi (a^j) \bigl[ D,\pi (b^j)\bigr]\;=\;
\pmatrix
\quad A \quad & i\gamma^5 K\phi \\
- i\gamma^5 K\phi & \quad A \quad \endpmatrix .
$$
Then
$$
\align
d\rho\;&=\;\sum_j\bigl[ D,\pi (a^j)\bigr]\bigl[D,\pi\bigr)b^j)\bigr]
\\
&=\;\pmatrix
\frac  1 2 \;\gamma^{\mu\nu}\partial_\mu A_\nu + X, \phantom{Zeichnung}&
-i\gamma^5\gamma^\mu K (\partial_\mu\phi + A_\mu \phi_0 - \phi_0
A_\mu) \\
i\gamma^5\gamma^\mu K (\partial_\mu\phi + A_\mu \phi_0-\phi_0 A_\mu), &
\frac 1 2 \; \gamma^{\mu\nu} \partial_\mu A_\nu + X
\phantom{Zeichnung}\endpmatrix ,
\endalign
$$
where $X=\id_N\otimes\bigl(\sum_j a^j \partial^\mu \partial_\mu
b^j\bigr)+\partial^\mu A_\mu$ is an arbitrary element of $\id_N\otimes
{\Cal A}$, and $\gamma^{\mu\nu} := [\gamma^\mu, \gamma^\nu]$. Hence
$$
Aux^2\;\simeq\;\pi ({\Cal A}).
\tag 3.16
$$
Next, let $\eta\in\Omega^2 ({\Cal A})$. Then one finds that
$$
d \eta\bigm|_{\eta=0} \;=\;\varepsilon\enskip
\pmatrix
\gamma^\mu X_\mu  & i\gamma^5 KX \\
- i\gamma^5 K X & \gamma^\mu X_\mu \endpmatrix ,
\tag 3.17
$$
where $X_\mu$ and $X$ are arbitrary elements of $\id_N \otimes {\Cal
A}$. Thus, in a 3-form $\vartheta$, terms proportional to $\gamma^\mu
\otimes \id_M$ in the off-diagonal elements and terms proportional to
$\gamma^5 K\otimes \id_M$ in the diagonal elements must be discarded
when evaluating $\vartheta^\perp$.

Next, we propose to check under what conditions the Chern-Simons
action $I_\varepsilon^3 (\rho)$ is gauge-invariant. In eq.~(2.25) we
have seen that, under a gauge transformation $M\in\pi({\Cal A})$,
$\rho$ transforms according to
$$
\rho\mapsto \tilde\rho\;=\; M\rho M^{-1} - (dM) M^{-1}\;=\;
g^{-1} \rho g + g^{-1} dg,
\tag 3.18
$$
with $g=M^{-1}$. From this equation and the cyclicity of
``integration'', $\int\!\!\!\!\!- (\cdot)$, we deduce that
$$
\align
I_\varepsilon^3 (\tilde \rho) \;&=\; I_\varepsilon^3 (\rho)\;+\\
&\frac \kappa 2\;\int\!\!\!\!\!\!- \bigl(\varepsilon \bigl\{ dg^{-1}
\rho dg + g^{-1} d\rho dg - \frac 1 3 \; (g^{-1} dg)^3\bigr\}^\perp
\bigr) .
\tag 3.19
\endalign
$$
The second term on the r.h.s. of (3.19) is equal to
$$
\align
\int_{M_2} d\,vol\,tr\;\bigl(\varepsilon\bigl\{[D,g^{-1}]
&\rho [D,g] + g^{-1}  d\rho [D,g] \\
&- \frac 1 3\; \bigl(g^{-1} [D,g]\bigr)^3\bigr\}^\perp\bigr).
\tag 3.20
\endalign
$$
Here, and in the following, $tr(\cdot)$ denotes a \ub{normalized
trace}, $\bigl( tr (\id) =1\bigr)$. A straightforward calculation
shows that
$$
[D,g]\;=\;\pmatrix
\partial\!\!\!/\; g& i\gamma^5 K (\phi_0 g - g\phi_0) \\
- i \gamma^5 K (\phi_0 g - g\phi_0) & \partial\!\!\!/\; g \endpmatrix ,
\tag 3.21
$$
and expression (3.20) is found to be given by
$$
\align
- i \,tr\,K \int_{M_2} \partial_\mu\,
&tr\;\bigl[ g^{-1} \phi \partial_\nu g + A_\nu \bigl( \phi_0 - g
\phi_0 g^{-1}\bigr) \\
&-\;\bigl( g \partial_\nu g^{-1} \phi_0 + g^{-1} \partial_\nu g
\phi_0\bigr)\bigr] \; dx^\mu \wedge dx^\nu,
\tag 3.22
\endalign
$$
which vanishes if $\partial M_2 = \phi$ (i.e., $M_2$ has no boundary).
\bigskip
\bf\ub{Remark}. \rm \ Had we considered a more general setting with \
${\Cal A} = C^\infty (M_2) \otimes {\Cal A}_1 \oplus C^\infty (M_2)
\otimes {\Cal A}_2$, where ${\Cal A}_1$ and ${\Cal A}_2$ are two
independent matrix algebras, and $\pi(a)=\id_N\otimes a$, for $a\in
{\Cal A}$, then, with $\varepsilon$ chosen as above,
$I_\varepsilon^3(\rho)$ would fail to be gauge-invariant.

Thus the condition for $I_\varepsilon^3(\rho)$ to be gauge-invariant
is that $\partial M_2=\phi$ and that the non-commutative space is
invariant under permuting the two copies of $M_2$ (of the space
$X_c$), i.e., the elements of $\pi ({\Cal A})$ commute with the
operator $\varepsilon$ defined in (3.11).

Under this condition one finds, after a certain amount of algebra,
that
$$
I_\varepsilon^3(\rho)\;=\;i\;\kappa \int_{M_2} tr\;(\Phi F),
\tag 3.23
$$
where
$$
\Phi\;=\;K (\phi+\phi_0),
$$
and
$$
F\;=\;\bigl(\partial_\mu A_\nu - \partial_\nu A_\mu + [A_\mu,
A_\nu]\bigr)\;
dx^\mu\wedge dx^\nu .
\tag 3.24
$$
We note that $I_\varepsilon^3(\rho)$ is obviously gauge-invariant and
topological, i.e., metric-independent. Since the ``Dirac operator''
$D$ depends on the reference metric $g$ on $M_2$, the
metric-independence of $I_\varepsilon^3$ is not, a priori, obvious
from its definition (3.15).

Let us consider the special case where
$$
{\Cal A}_o\;=\;\Bbb M_3 (\Bbb R),
\tag 3.25
$$
the algebra of \ub{real} 3$\times$3 matrices. Then
$$
I_\varepsilon^3 (\rho)\;=\;i\,\kappa \int_{M_2} \biggl( \sum_{A=1}^3
\Phi^A F_{\mu\nu}^A\biggr)\; dx^\mu \wedge dx^\nu
\tag 3.26
$$
where $F_{\mu\nu}^A=\partial_\mu A_\nu^A-\partial_\nu A_\mu^A+
\varepsilon^{ABC} A_{[\mu}\, ^B \; A_{\nu]}\,^C$,
with $A=a,3,a=1.2.$ \ Setting
$$
A_\mu^a\;=\;e_\mu^a,\quad A_\mu^3\;=\;\frac 1
2\;\omega_\mu^{ab}\;\varepsilon_{ab} \;\equiv\;\omega_\mu
\tag 3.27
$$
one observes that the action $I_\varepsilon^3$ is the one of
two-dimensional topological gravity introduced in [13]. Varying
$I_\varepsilon^3$ w.r. to $\Phi^A$ one obtains the zero-torsion and
constant-curvature conditions:
$$
\align
&\varepsilon^{\mu\nu} F_{\mu\nu}^a\;\equiv\;
 \varepsilon^{\mu\nu} T_{\mu\nu}^a\;=\;\varepsilon^{\mu\nu}
\bigl( \partial_\mu e_\nu^a+\,\frac 1 2 \sum_b \omega_\mu
\varepsilon^{ab} e_\nu^b\bigr)\;=\;0 \\
&\varepsilon^{\mu\nu} F_{\mu\nu}^3\;=\;\frac 1 2\;
 \varepsilon^{\mu\nu} R_{\mu\nu}^{ab}\, \varepsilon_{ab}\;=\;\frac 1 2
\;\varepsilon^{\mu\nu}\,
\bigl( \partial_\mu \omega_\nu + 2 \;\varepsilon_{ab}\; e_\mu^a\;
e_\nu^b\bigr)\;=\;0.
\tag 3.28
\endalign
$$
Variation of $I_\varepsilon^3$ with respect to $A_\mu^A$ implies that
$\Phi^A$ is covariantly constant, i.e.,
$$
D_\mu \Phi^A\;=\;\partial_\mu\Phi^A + \varepsilon^{ABC} A_\mu^B
\Phi^C\;=\;0.
\tag 3.29
$$
The space of solutions of (3.28) and (3.29) is characterized in [13].

These results suggest that the study of Chern-Simons actions in
non-commutative geometry is worthwhile.
\bigskip

\vskip 1.5truecm
\bf 4. \ub{Some ``five-dimensional'' Chern-Simons actions} \rm .
\medskip
In this section, we consider non-commutative space $({\Cal
A},\pi,H,D)$ with ``cotangent bundles'' $\Omega_D^1({\Cal A})$ that
are free, hermitian ${\Cal A}$-bimodules of dimension 5, and we
evaluate the ``five-dimensional'' Chern-Simons action,
$I_\varepsilon^5(\rho)$, defined in eq.~(2.63), for connections on the
``line bundle'' $E=E^{(1)} \simeq {\Cal A}$. We shall consider
algebras ${\Cal A}$ generated by matrix-valued functions on Riemann
surfaces or on four-dimensional spin manifolds. We start with the
analysis of the latter example.
\medskip

\bf (I)\rm \ We choose $X=M_4\times \{-1,1\}$, where $M_4$ is a
four-dimensional, smooth Riemannian spin manifold. The non-commutative
space $({\Cal A},\pi,H,D)$ is chosen as in Sect.~3, except that $M_2$
is replaced by $M_4$, the 2$\times$2 Dirac matrices
$\gamma^1,\gamma^2$ are replaced by the 4$\times$4 Dirac matrices
$\gamma^1,\gamma^2,\gamma^3,\gamma^4$, and
$\gamma^5=\gamma^1 \gamma^2 \gamma^3\gamma^4$. The definition of the
``Dirac operator'' $D$ is analogous to that in eq.~(3.8).

An orthonormal basis for $\Omega_D^1({\Cal A})$ is then given
(locally on $M_4$) by
$$
\varepsilon^a\;=\;\pmatrix
\gamma^a\otimes \id_M & \quad 0\quad \\
\quad 0\quad & \gamma^a\otimes\id_M \endpmatrix ,
 a=1,\cdots,4, \quad \varepsilon^5\;=\;
\pmatrix
\quad 0\quad & \gamma^5\otimes\id_M \\
- \gamma^5\otimes\id_M & \quad 0\quad \endpmatrix,
\tag 4.1
$$
and
$$
\varepsilon\;=\;\varepsilon^1\;\varepsilon^2\;\varepsilon^3\;
\varepsilon^4\; \varepsilon^5\;=\;\pmatrix
\enskip 0\quad & \id \\
-\id & 0 \endpmatrix ,
\tag 4.2
$$
similarly as in (3.11).

Again, we must determine the spaces $Aux^n$, $n=1,2,3,4,5$, of
``auxiliary fields''. The most important one is $Aux^5$. To determine
it, let us consider a vanishing element $\eta$ of $\Omega^4({\Cal
A})$ and compute $d\eta$, as given by eq.~(2.10). After a certain
amount of labouring one finds that
$$
d\eta\bigm|_{\eta=0}\;=\;\varepsilon \pmatrix
\gamma^{\mu\nu\rho} X_{\mu\nu\rho} + \gamma^\mu (K^2 X_\mu+Y_\mu),
& i \gamma^5(\gamma^{\mu\nu} K X_{\mu\nu} + K^3 X+KY) \\
- i \gamma^5 (\gamma^{\mu\nu} K X_{\mu\nu} + K^3 X+KY),
& \gamma^{\mu\nu\rho} X_{\mu\nu\rho} + \gamma^\mu (K^2 X_\mu+Y_\mu)
\endpmatrix ,
\tag 4.3
$$
where \ $X_{\mu\nu\rho}, X_{\mu\nu}, X_\mu, X, Y_\mu$ and $Y$ are
arbitrary
elements of $\id_N\otimes{\Cal A}$, and $\gamma^{\mu\nu\rho} =
\displaystyle\mathop{\Sigma}_{a,b,c}$ sig \  $\mu\nu\rho \choose a b c $ \
$\gamma^a\gamma^b\gamma^c$.
By (4.3), the passage from an element $\vartheta\in\Omega^5({\Cal A})$
to $\vartheta^\perp$ amounts to discarding all terms proportional to
$\gamma^{\mu\nu\rho} \otimes\id_M$, $K^2\gamma^\mu\otimes\id_M$ and
$\gamma^\mu\otimes\id_M$ from off-diagonal elements of $\vartheta$
and all terms proportional to $K\gamma^5\gamma^{\mu\nu}\otimes\id_M$,
$K^3\gamma^5\otimes \id_M$ and $K\gamma^5\otimes\id_M$ from the
diagonal elements of $\vartheta$. Now we start understanding the
useful role played by the matrix $K$.

It is then easy  to
evaluate $I_\varepsilon^5(\rho)$, with $\rho$ given by
$$
\rho\;=\;\pmatrix \quad A\quad & i\gamma^5 K\phi \\
- i \gamma^5 K \phi &\quad A\quad \endpmatrix ,
\quad A\;=\;\gamma^\mu A_\mu .
\tag 4.4
$$
Using eq.~(2.63), the result is
$$
I_\varepsilon^5 (\rho)\;=\;i\;\frac{3\kappa}{4}\;\int_{M_4} Tr\;
(\Phi\;F\wedge F) ,
\tag 4.5
$$
where $\Phi = K(\phi+\phi_0)$, and $F=F_{\mu\nu} dx^\mu \wedge
dx^\nu$, with $F_{\mu\nu}$ the curvature, or field strength, of
$A_\mu$. Provided that $\partial M_4=\emptyset$, $I_\varepsilon^5$ is
gauge-invariant and topological (metric-independent), as expected. The
field equation obtained by varying $I_\varepsilon^5$ w.r. to $\Phi$ is
$$
\varepsilon^{\mu\nu\rho\sigma}\;F_{\mu\nu}\;F_{\rho\sigma}\;=\;0 .
\tag 4.6
$$
Setting $\Phi$ to a constant, $I_\varepsilon^5$ turns out to be the
action of four-dimensional, topological Yang-Mills theory [16] before
gauge-fixing.
\medskip
\bf (II) \rm \ We choose $X=M_2\times \{-1,1\}$ and ${\Cal A},\pi$ and
$H$ as above, but the operator $D$ is given by
$$
D\;=\;\pmatrix
\quad\partial\!\!\!/\quad & K \gamma^\alpha \phi_{0\alpha} \\
- K \gamma^\alpha \phi_{0\alpha} &\quad\partial\!\!\!/\quad
\endpmatrix ,
\tag 4.7
$$
where $\partial\!\!\!/ = \gamma^1\partial_1+\gamma^2\partial_2$, and
$\alpha = 3,4,5$. The matrices $\gamma^1,\cdots,\gamma^4$ are
antihermitian 4$\times$4 Dirac matrices, and, \ub{only in this
paragraph}, $\gamma^5 = i\,\gamma^1\gamma^2\gamma^3\gamma^4$, so that
$\gamma^5$ is now antihermitian, too, rather than hermitian (as in the
rest of this paper). Locally on $M_2$, the cotangent bundle
$\Omega_D^1 ({\Cal A})$ is a free, hermitian ${\Cal A}$-bimodule of
dimension 5, with an orthonormal basis given by
$$
\varepsilon^a\;=\;\pmatrix
\gamma^a\otimes\id_M &\quad 0\quad\\
\quad 0\quad & \gamma^a\otimes \id_M \endpmatrix,
 a= 1,2,\quad\varepsilon^\alpha\;=\;
\pmatrix
\quad 0\quad & i\gamma^\alpha \otimes \id_M \\
- i\gamma^\alpha \otimes \id_M &\quad 0\quad \endpmatrix ,
\alpha = 3,4,5,
\tag 4.8
$$
and $\varepsilon$ is taken to be
$$
\varepsilon\;=\;\varepsilon^1 \varepsilon^2 \varepsilon^3
\varepsilon^4 \varepsilon^5\;=\; \pmatrix
\enskip 0\quad &\id \\
- \id & 0 \endpmatrix .
\tag 4.9
$$
A 1-form $\rho = \sum_j \pi (a^j) \bigl[ D,\pi (b^j)\bigr]$ has the
form
$$
\rho\;=\;\pmatrix
\quad A\quad & K \gamma^\alpha \phi_\alpha \\
- K \gamma^\alpha \phi_\alpha & \quad A\quad \endpmatrix ,
\tag 4.10
$$
with $A=\sum_j a^j \partial\!\!\!/ b^j$ and $\phi_\alpha +
\phi_{0\alpha} = \sum_j a^j \phi_{0\alpha} b^j$. Evaluating $d\rho$ as
in eq.~(2.10), one finds that
$$
d\rho\;=\;\pmatrix
\gamma^{\mu\nu} \partial_\mu A_\nu - K^2 \gamma^{\alpha\beta}
L_{\alpha\beta} + X - K^2 L_\alpha^\alpha, &
K \gamma^\mu \gamma^\alpha D_\mu^0 \phi_\alpha \\
- K \gamma^\mu \gamma^\alpha D_\mu^0 \phi_\alpha, &
\gamma^{\mu\nu} \partial_\mu A_\nu - K^2\gamma^{\alpha\beta}
L_{\alpha\beta} + X - K^2 L_\alpha^\alpha \endpmatrix ,
\tag 4.11
$$
where
$$
L_{\alpha\beta}\;=\; \phi_{0\alpha} \phi_\beta + \phi_\alpha
\phi_{0\beta} + \sum_j a^j \bigl[ b^j, \phi_{0\alpha}
\phi_{0\beta}\bigr],
$$
$$
X\;=\; -\;\sum_j a^j \;\partial\!\!\!/\,^2\;b^j\;+\;\partial^\mu A_\mu
,
$$
and
$$
D_\mu^0 \phi_\alpha\;=\; \partial_\mu \phi_\alpha\;+\;A_\mu
\phi_{0\alpha}\;-\;\phi_{0\alpha} A_\mu .
\tag 4.12
$$
For simplicity we assume that
$$
\bigl[ \phi_{0\alpha}, \phi_{0\beta}\bigr]\;=\; 0, \quad \text{and }
\phi_{0\alpha} \phi_0^\alpha\;=\; 1 .
\tag 4.13
$$
Since we may assume that $\sum a^j b^j = 1$, we then have that
$L_{\alpha\beta} = \phi_{0\alpha} \phi_\beta + \phi_\alpha
\phi_{0\beta}$, for $L_{[\alpha\beta]}$ and $L_\alpha^{\enskip\alpha}$
appearing in (4.11), which is not an auxiliary field. A tedious calculation
then yields the formula
$$
\align
I_\varepsilon^5 (\rho)\;&=\; 2\kappa \int_{M_3} \varepsilon^{\mu\nu}
\varepsilon^{\alpha\beta\gamma} tr K^3 \bigl[\bigl( L_{\alpha\beta}
\phi_\gamma + \phi_\alpha L_{\beta\gamma}\bigr)\;\partial_\mu A_\nu
\\
&-\;\phi_{0\alpha} D_\mu^0 \phi_\beta D_\nu^0 \phi_\gamma + A_\mu
L_{\alpha\beta} D_\nu^0 \phi_\gamma + A_\mu D_\nu^0 \phi_\alpha
L_{\beta\gamma} \\
&+\; \frac 3 2 \; \phi_\alpha A_\mu A_\nu L_{\beta\gamma} +\frac 3 2\;
\phi_\alpha \phi_\beta\phi_\gamma\partial_\mu A_\nu \\
&-\;\frac 3 2 \; A_\mu \phi_\alpha A_\nu L_{\beta\gamma} + \frac 3 2 \;
A_\mu A_\nu \phi_\alpha L_{\beta\gamma} \\
&-\;\frac 3 2\; \phi_\alpha A_\mu \phi_\beta D_\nu^0 \phi_\gamma
+ \frac 3 2\; \phi_\alpha \phi_\beta A_\mu D_\nu^0 \phi_\gamma \\
&+\;\frac 3 2\; A_\mu \phi_\alpha\phi_\beta D_\nu^0 + 3\; A_\mu A_\nu
\phi_\alpha\phi_\beta\phi_\gamma \\
&-\;3\;\phi_\alpha A_\mu \phi_\beta A_\nu \phi_\gamma \bigr]\; d^2 x .
\tag 4.14
\endalign
$$
If $\partial M_2=\emptyset$, and after further algebraic manipulations,
the action (4.14) can be shown to have the manifestly gauge-invariant
form
$$
\align
I_\varepsilon^5 (\rho)\;=\; 2\kappa &\int_{M_2}
\varepsilon^{\alpha\beta\gamma} tr\;\bigl[ - \Phi_\alpha \bigl( D_\mu
\Phi_\beta\bigr) \bigl(D_\nu \Phi_\gamma\bigr) \\
& +\;2\;\Phi_\alpha \Phi_\beta\Phi_\gamma\;\bigl(\partial_\mu A_\nu +
A_\mu A_\nu\bigr)\bigr]\; dx^\mu \wedge dx^\nu ,
\tag 4.15
\endalign
$$
where $\Phi_\alpha = K (\phi_\alpha + \phi_{0\alpha})$.

If the constraints (4.13) are not imposed then one must explicitly
determine $Aux^5$, in order to derive an explicit expression for
$I_\varepsilon^5$. The result is that (4.15) still holds.

It is remarkable that all the Chern-Simons actions derived in
eqs.~(3.23), (4.5) and (4.15) can be obtained from Chern-Simons
actions for connections on vector bundles over classical, commutative
manifolds by \ub{dimensional reduction}. For example, setting $M_3 =
M_2\times S^1$ and $\phi := A_3$, and assuming that $A_1, A_2$ and
$A_3$ are indpendent of the coordinate (angle) parametrizing $S^1$, we
find that
$$
\align
I^3 (A)\;&=\;i \kappa' \int_{M_3} tr\;\bigl( A \wedge dA + \frac 2 3
\; A \wedge A \wedge A \bigr) \\
&=\;i \kappa' \int_{M_2} tr\; (\phi \;F) ,
\tag 4.16
\endalign
$$
where \ $F = \bigl(\partial_1 A_2 - \partial_2 A_1 + [ A_1,
A_2]\bigr)\;dx^1\wedge dx^2.$ \ Setting $\kappa' =\kappa\; tr K$, (4.16)
reduces to (3.23). Similarly, reducing a classical, five-dimensional
Chern-Simons action to four dimensions, with $M_5 = M_4 \times S^1$,
results in
$$
\align
I^5(A)\;&=\;i \kappa' \int_{M_5} tr\;\bigl( A\wedge dA \wedge dA +\;
\frac 3 2 \; A\wedge A\wedge A\wedge dA \\
&\phantom{Zeichnung} +\; \frac 3 5\; A \wedge A\wedge A\wedge A\wedge A\bigr)
\\
&=\;i\;\frac{3\kappa'}{4}\; \int_{M_4} tr\;(\phi \;F\wedge F),
\endalign
$$
with $\phi := A_5$, and $A_1,\cdots,A_5$ independent of the angle
parametrizing $S^1$. Thus we recover (4.5). Finally, dimensionally
reducing $I^5(A)$ to a two-dimensional surface (setting $M_5 = M_2
\times S^1 \times S^1 \times S^1)$ reproduces the action (4.15).

The advantage of the non-commutative formulation is that it
automatically eliminates all excited modes corresponding to a
non-trivial dependence of the gauge potential $A$ on angular
variables.

\vfill\eject
\bf 5. \ub{Relation to four-dimensional gravity and supergravity} \rm.
\medskip
Chern-Simons actions are topological actions. In order to obtain
dynamical actions from Chern-Simons actions, one would have to
impose  constraints on the field configuration space.
In this section, we explore this possibility. As a result, we are able
to derive some action functionals of four-dimensional gravity and
supergravity theory.

We propose to impose a constraint on the scalar multiplet $\Phi$
appearing in the Chern-Simons action (4.5). The non-commutative space
$({\Cal A}, \pi, H, D)$ is chosen as in example (I) of Sect.~4; (see
also Sect.~3). Let us compute the curvature 2-form,
$$
\theta\;=\; ( d\rho + \rho^2)^\perp ,
$$
of a connection $\nabla$ on the line bundle $E\simeq A$ given by a
1-form $\rho$ as displayed in eq.~(4.4). Then
$$
\theta\;=\;\pmatrix
\frac 1 2\;\gamma^{\mu\nu} F_{\mu\nu}+(K^2)^\perp
\bigl((\phi+\phi_0)^2-\phi_0^2\bigr),
& - K i \gamma^5\gamma^\mu D_\mu (\phi +\phi_0) \\
K i \gamma^5 \gamma^\mu D_\mu (\phi +\phi_0),
&\frac 1 2\;\gamma^{\mu\nu} F_{\mu\nu} + (K^2)^\perp
\bigl((\phi+\phi_0)^2 - \phi_0^2\bigr) \endpmatrix ,
\tag 5.1
$$
where $(K^2)^\perp = K^2 - (tr\;K^2) \id$; (recall that $tr (\cdot)$
is normalized: $tr (\id) = 1)$. The appearance of $(K^2)^\perp$ is due
to the circumstance that when passing from $d\rho+\rho^2$ to $(d\rho
+\rho^2)^\perp$ terms proportional to $\id_N$ must be removed. Let
$p\,tr (\cdot)$ denote the partial trace over the Dirac-Clifford
algebra. Then
$$
p\,tr\;(\theta)\;=\;(K^2)^\perp \bigl((\phi+\phi_0)^2-\phi_0^2\bigr),
\tag 5.2
$$
and we shall impose the constraint
$$
p\,tr\;(\theta)\;=\;0 .
\tag 5.3
$$
Choosing $\phi_0$ to satisfy $\phi_0^2=\id$, and renaming
$\phi+\phi_0$ to read $\phi$, the constraint (5.3) becomes
$$
\phi^2\;=\;\id,
\tag 5.4
$$
provided $(K^2)^\perp \;\neq\; 0. $

As our matrix algebra ${\Cal A}_0$ (see eq.~(3.1)) we choose
$$
{\Cal A}_0\;=\enskip\text{real part of Cliff } \bigl(SO(4)\bigr) .
\tag 5.5
$$
We propose to show that, for this choice of ${\Cal A}_0$ and assuming
that the constraint (5.4) is satisfied, the Chern-Simons action (4.5)
is the action of the metric-independent (first-order) formulation of
four-dimensional gravity theory.

Let $\Gamma_1,\cdots,\Gamma_4$ denote the usual generators of ${\Cal
A}_0$, (i.e., 4$\times$4 Dirac matrices in a real representation), and
$\Gamma_5 = \Gamma_1\Gamma_2\Gamma_3\Gamma_4$. Then
$$
\bigl\{ \Gamma_a, \Gamma_b\bigr\}\;=\;- 2 \delta_{ab},\;
\Gamma_a^*\;=\;- \Gamma_a,\;a,b\;=\;1,\cdots,4,
$$
and $\Gamma_5^* = \Gamma_5$. A basis for ${\Cal A}_0$ is then given by
$\id_4, \Gamma_a, a=1,\cdots,4, \Gamma_5, \Gamma_{ab}, a,b
=1,\cdots,4$, and $\Gamma_a\Gamma_5$. For a 1-form $\rho$ as in
eq.~(4.4), we may expand the gauge potential $A$ and the scalar field
$\phi$ in the basis of ${\Cal A}_0$ just described:
$$
A\;=\;\gamma^\mu \;\bigl( A_\mu^0 \id + A_\mu^a \Gamma_a + A_\mu^{ab}
\Gamma_{ab} + A_\mu^5 \Gamma_5 + A_\mu^{a5} \Gamma_a \Gamma_5 \bigr) ,
\tag 5.6
$$
and
$$
\phi\;=\;\bigl( \phi^0 \id + \phi^a \Gamma_a + \phi^{ab} \Gamma_{ab} +
\phi^5 \Gamma_5 + \phi^{a5} \Gamma_a \Gamma_5\bigr) .
\tag 5.7
$$
In this section, we only consider \ub{unitary} connections on $E\equiv
E^{(1)} \simeq {\Cal A}$; see eq.~(2.29). By (2.30), this is
equivalent to \ub{hermiticity} of $\rho$. This implies that
$$
A_\mu^0=- \overline{A_\mu^0}, A_\mu^a = \overline{A_\mu^a},
A_\mu^{ab} = \overline{A_\mu^{ab}}, A_\mu^5= - \overline{A_\mu^5},
A_\mu^{a5} = - \overline{A_\mu^{a5}},
\tag 5.8
$$
and
$$
\phi^0 = \overline{\phi^0}, \phi^a = - \overline{\phi^a},
\phi^{ab} = - \overline{\phi^{ab}}, \phi^5 = \overline{\phi^5},
\phi^{a5} = \overline{\phi^{a5}},
\tag 5.9
$$
where $\bar z$ denotes the complex conjugate of $z$. Since ${\Cal
A}_0$ is chosen to be real, the coefficients of $A$ and $\phi$ should
be chosen to be real. It then follows from (5.8) and (5.9) that
$$
A\;=\;\gamma^\mu \biggl(\frac{1}{2\kappa}\;e_\mu^a \Gamma_a + \frac 1
4\; \omega_\mu^{ab} \Gamma_{ab}\biggr)
\tag 5.10
$$
and
$$
\phi\;=\;\phi^0+\phi^5\Gamma_5 + \phi^{a5} \Gamma_a \Gamma_5,
\tag 5.11
$$
where we have set $A_\mu^a =: \frac{1}{2\kappa}\;e_\mu^a$, and
$A_\mu^{ab} =: \frac 1 4 \; \omega_\mu^{ab}$, and $\kappa^{-1}$ is the
Planck scale.

Imposing the constraint that \ $tr_{{\Cal A}_0} (\varepsilon \rho) =
0$ implies that
$$
\phi^0\;=\; 0 .
\tag 5.12
$$
Constraints (5.12) and (5.4) then yield the condition
$$
(\phi^5)^2\;+\;(\phi^{a5})^2\;=\;1 .
\tag 5.13
$$
Under a gauge transformation $M\equiv g^1$, $\rho$ transforms according to
$$
\rho\;\mapsto\; M\rho M^{-1} - (dM) M^{-1},\quad M\in\pi ({\Cal A}),
$$
see (2.25), which implies the transformation law
$$
\phi \;\mapsto\; g^{-1}  \phi g, \quad g\;=\;\exp \;\frac 1 2\; \bigl(
\Lambda^a \Gamma_a + \Lambda^{ab} \Gamma_{ab}\bigr),
\tag 5.14
$$
where $\Lambda^a$ and $\Lambda^{ab}$ are smooth functions on $M_4$.
The infinitesimal form of (5.14) reads
$$
\align
\delta \phi^5\;&=\;-\;\sum_a \Lambda^a \phi^{a5} , \\
\delta \phi^{a5}\;&=\;-\;\Lambda^a \phi^5 \;-\sum_b \Lambda^{ab}
\phi^{b5} .
\tag 5.15
\endalign
$$
 From this it follows that, locally, we can choose a gauge such that
$$
\phi^{a5}\;=\;0 .
\tag 5.16
$$
In this gauge, the constraint (5.13) has the solutions
$$
\phi^5\;=\; \pm \;1 .
\tag 5.17
$$
The action (4.5) then becomes
$$
I_\varepsilon^5 (\rho)\;=\; \pm\; k \int_{M_4} tr\; (\Gamma_5 F\wedge
F) ,
\tag 5.18
$$
(with $k=i \frac{3\kappa}{4}$, in the notation of Sect.~4). Next, we
expand the field strength $F_{\mu\nu}$ in our Clifford algebra basis
which yields
$$
F_{\mu\nu}\;=\;\frac{1}{2\kappa}\; F_{\mu\nu}^a\; \Gamma_a \;+\;\frac 1
4\; F_{\mu\nu}^{ab} \;\Gamma_{ab} ,
\tag 5.19
$$
where
$$
F_{\mu\nu}^a\;=\;\partial_\mu\; e_\nu^a + \omega_{\mu\;b}^a\;
e_\nu^b\;-\;(\mu\leftrightarrow \nu) ,
\tag 5.20
$$
$$
F_{\mu\nu}^{ab}\;=\;\partial_\mu \;\omega_\nu^{ab} +
\omega_{\mu\;c}^a\;\omega_\nu^{c\;b} + \;\frac{1}{\kappa^2}\;
e_\mu^a\; e_\nu^b \;-\;(\mu\leftrightarrow \nu) ,
\tag 5.21
$$
and the indices $a,b,\cdots$ are raised and lowered with the flat
metric $\eta_{ab} = - \delta_{ab}$.

The only non-vanishing contribution to (5.18) comes from the trace \
$tr (\Gamma_5\Gamma_{ab}\Gamma_{cd}) = \varepsilon_{abcd}$, and
$I_\varepsilon^5$ is found to be given by
$$
\align
I_\varepsilon^5\;=\;\pm\;k \int_{M_4} \varepsilon_{abcd}\;\bigl(
R_{\mu\nu}^{ab} +\;\frac{2}{\kappa^2}\; e_\mu^a\;&e_\nu^b\bigr)
\bigl( R_{\rho\sigma}^{cd} + \;\frac{2}{\kappa^2}\;e_\rho^c\;
e_\sigma^d \bigr) \\
&\times\; dx^\mu \wedge dx^\nu \wedge dx^\rho \wedge dx^\sigma ,
\tag 5.22
\endalign
$$
where
$$
R_{\mu\nu}^{ab}\;=\;\partial_\mu\;\omega_\nu^{ab} +
\omega_{\mu\;c}^a\;\omega_\nu^{c\;b}\;-\; (\mu \leftrightarrow \nu) .
\tag 5.23
$$
Interpreting $\omega_\mu^{ab}$ as the components of a connection on
the spinor bundle over $M_4$, $R_{\mu\nu}^{ab}$ are the components of
its curvature, and $F_{\mu\nu}^a$ are the components of its torsion,
as is well known from the Cartan structure equations.

Setting the variation of $I_\varepsilon^5$ with respect to
$\omega_\mu^{ab}$ to zero, we find that the torsion of $\omega$
vanishes:
$$
F_{\mu\nu}^a\;=\;0,\quad\text{for all }\mu,\nu\text{ and } a.
\tag 5.24
$$
If the frame $\bigl(e_\mu^a\bigr)$ is invertible, (5.24) can be solved
for $\omega_\mu^{ab}$:
$$
\omega_\mu^{ab}\;=\;\frac 1 2 \; \bigl( \Omega_{\mu a b} -
\Omega_{ab\mu} + \Omega_{b\mu a}\bigr),
\tag 5.25
$$
where
$$
\Omega_{ab}\;^c\;=\; e_a^\mu\;e_b^\nu\;\bigl(
\partial_\mu\;e_\nu\,^c\;-\;\partial_\nu\; e_\mu\,^c \bigr).
$$
Substituting (5.25) back into (5.22) yields a functional that depends
only on the metric
$$
g_{\mu\nu}\;=\;e_\mu^a\; e_{\nu a},
\tag 5.26
$$
and is given by
$$
\align
I_\varepsilon^5\;=\;\pm\;k \int_{M_4} d^4 x\, \sqrt{g}\; \bigl[\bigl(\;
&4\,R_{\mu\nu\rho\sigma} R^{\mu\nu\rho\sigma} \;-\;4\, R_{\mu\nu} R^{\mu\nu}
+ R^2 \bigr) \\
&+\;\frac{16}{\kappa^2}\;R\;+\;\frac{96}{\kappa^4}\;\bigr]
 \tag 5.27
\endalign
$$
where $R_{\mu\nu\rho\sigma}$ is the Riemann curvature tensor,
$R_{\mu\nu}$ is the Ricci tensor, and $R$ is the scalar curvature
determined by the metric $g_{\mu\nu}$ given in (5.26). The term in
round brackets on the r.h.s. of (5.27) yields the topological
Gauss-Bonnet term for $M_4$, the second term yields the
Einstein-Hilbert action, and the last term is a cosmological constant.

Next, we show how to derive a metric-independent formulation of \
\ub{four-dimensional} \ub{supergravity} from the action $I_\varepsilon^5$
given in eq.~(4.5). For this purpose we choose the algebra ${\Cal
A}_0$ in (3.1) to be a graded algebra [18]:
$$
{\Cal A}_0\;=\;\text{real part of SU}(4\mid 1) .
\tag 5.28
$$
This algebra is generated by graded 5$\times$5 matrices preserving the
quadratic form
$$
(\vartheta^\alpha)^*\; C_{\alpha\beta} \;\vartheta^\beta\;-\;z^* \;z ,
\tag 5.29
$$
where $C_{\alpha\beta}$ is an antisymmetric matrix and
$\vartheta^\alpha$ is a Dirac spinor. At this point, one must note
that we are leaving the conventional framework of non-commutative
geometry, since, for ${\Cal A}_0$ as in (5.28), the algebra ${\Cal A}$
is not a $^*$algebra of operators. But let us try to proceed and find
out what the result is.

Let $\rho$ be a 1-form as in eq.~(4.4). Then the matrix elements
$A_\mu$ and $\phi$ of $\rho$ have the graded matrix representation
$$
\phi\;=\;\pmatrix
\Pi_\alpha^\beta & \lambda_\alpha \\
\bar\lambda^\alpha & \Pi_1 \endpmatrix ,
\tag 5.30
$$
and
$$
A_\mu\;=\; \pmatrix
M_{\mu \alpha}^{\enskip\beta} & \sqrt{\kappa}\;\psi_{\mu\alpha} \\
- \sqrt{\kappa}\;\bar\psi_\mu^\alpha &\quad B_\mu \endpmatrix .
\tag 5.31
$$
The reality conditions for $\phi$ and $A_\mu$ imply that
$\lambda_\alpha$ and $\psi_{\mu\alpha}$ are Majorana spinors:
$$
\lambda_\alpha\;=\;C_{\alpha\beta}\;\bar\lambda^\beta,
\;\psi_{\mu\alpha}\;=\;C_{\alpha\beta}\;\bar\psi_\mu^\beta .
$$
Furthermore, one finds that
$$
\align
\Pi_\alpha^\beta\;&=\;\biggl( \frac 1 4\; \Pi^0 \id \;+\; \Pi^5 \Gamma_5 \;+\;
\Pi^{a5} \Gamma_a\Gamma_5\biggr)_\alpha^\beta , \\
M_{\mu\alpha}^\beta\;&=\;\biggl(
\frac{1}{2\kappa}\;e_\mu^e\;\Gamma_a \;+\;\frac 1 4
\;\omega_\mu^{ab}\;\Gamma_{ab}\biggr)_\alpha^\beta ,
\tag 5.32
\endalign
$$
and
$$
B_\mu\;=\;0 .
$$

We shall now impose the constraints
$$
Str\;(\varepsilon\;\rho)\;=\;0 ,
\tag 5.33
$$
$$
Str\;(\theta)\;=\;0 ,
\tag 5.34
$$
and
$$
Str \;\biggl(\varepsilon\;\bigl( \rho d\rho\;+\;\frac 2
3\;\rho^3\bigr)^\perp \biggr)\;=\;0 ,
\tag 5.35
$$
along with $Str \;\phi_0^2 = 1$, and $Str\;\phi_0 = 0$. Here \ $Str
(\cdot)$ denotes the graded trace on ${\Cal A}_0$. Renaming $\phi +
\phi_0$ to read $\phi$, these constraints
imply that
$$
\Pi^0\;=\;\Pi_1 ,
\tag 5.36
$$
$$
-\;\frac 3 4 \; \Pi_1^2 + 4\bigl((\Pi^5)^2 - \sum_a (\Pi^{a5})^2\bigr)
+ \bar\lambda \lambda \;=\;1 ,
\tag 5.37
$$
and
$$
(K^3)^\perp\;S tr (\phi^3)\;=\; 0,
\tag 5.38
$$
where $(K^3)^\perp$ is defined so as to satisfy \ $tr
\bigl(K(K^3)^\perp\bigr) = 0$.

In order to determine the dynamical contents of a theory with an
action $I_\varepsilon^5$ given by (4.5), ${\Cal A}_0$ as in (5.28) and
constraints (5.36) through (5.38), it is convenient to work in a
special gauge, the unitary gauge. Consider a gauge transformation
$$
g\;=\;\exp\; \pmatrix
\bigl(\Lambda_\alpha^\beta\bigr)\quad &\sqrt{\kappa}\;\varepsilon_\alpha \\
- \sqrt{\kappa}\;\bar\varepsilon^\alpha & 0 \endpmatrix ,
\tag 5.39
$$
where $\Lambda_\alpha^\beta = \frac 1 2$ $\bigl( \Lambda^a \Gamma_a +
\Lambda^{ab} \Gamma_{ab}\bigr)_\alpha^\beta $. The transformation law of
$\phi$ is then given by $\phi \mapsto g^{-1} \phi g$. From this we
find the infinitesimal gauge transformations of the fields $\Pi$ and
$\lambda$:
$$
\align
&\delta\; \Pi_1\;=\;2\;\sqrt{\kappa}\; \bar\varepsilon \lambda, \\
&\delta\; \Pi^5\;=\;+\; \Lambda^a \Pi^{a5} + \frac{\sqrt{\kappa}}{2}\;
\bar\varepsilon \;\Gamma^5 \lambda , \\
&\delta\;\Pi^{a5}\;=\; -\; \Lambda^{ab} \Pi^{b5} - \Lambda^a \Pi^5
- \frac 1 2\;\bar\varepsilon \;\Gamma^a\Gamma^5\lambda, \\
&\delta\;\lambda_\alpha\;=\;\sqrt{\kappa}\;\bigl(- \frac 3 4\; \Pi_1 +
\Pi^5 \Gamma_5 + \Pi^{a5}
\Gamma_a\Gamma_5\bigr) \varepsilon_\alpha - \frac 1 2\;
\bigl(\Lambda^a\Gamma_a+\Lambda^{ab}\Gamma_{ab}\bigr)_\alpha^\beta\;
\lambda_\beta .
\tag 5.40
\endalign
$$
Thus, locally, we can choose the gauge
$$
\Pi^{a5}\;=\;0, \text{ and } \lambda_a\;=\; 0 .
\tag 5.41
$$
The constraints (5.37) and (5.38) then reduce to
$$
\align
-\;\frac 3 4\;\Pi_1^2 + 4\; (\Pi^5)^2\;&=\;1 , \\
\Pi_1\bigl(-\frac{5}{16}\;\Pi_1^2 + (\Pi^5)^2\bigr)\;&=\; 0 .
\tag 5.42
\endalign
$$
These equations have the solutions
$$
\Pi_1\;=\;0,\quad \Pi^5\;=\;\pm\;\frac 1 2 ,
\tag 5.43
$$
and
$$
\Pi_1\;=\;\pm\;\sqrt{2}\;, \;\Pi^5\;=\;\pm\;\sqrt{\frac{5\,}{8\,}} .
\tag 5.44
$$
We further study the first solution. Inserting it into the action
(4.5), we arrive at the expression
$$
I_\varepsilon^5\;=\;\pm\;\frac k 2 \int_{M_4} Str\;\biggl( {\Gamma_5
\quad 0 \choose \enskip 0\quad 0}\; F_{\mu\nu} F_{\rho\sigma}\biggr)\; dx^\mu
\wedge dx^\nu\wedge dx^\rho \wedge dx^\sigma ,
\tag 5.45
$$
where
$$
F_{\mu\nu}\;=\;\pmatrix
\frac 1 4 \; F_{\mu\nu}^{ab}\;\Gamma_{ab} +
\frac{1}{2\kappa}\;F_{\mu\nu}^a\;\Gamma_a, &
\sqrt{\kappa}\;\psi_{\mu\nu} \\
-\;\sqrt{\kappa}\;\bar\psi_{\mu\nu}\phantom{ZeichnungZ}, &\quad 0
\endpmatrix ,
\tag 5.46
$$
with
$$
\align
F_{\mu\nu}^a\;&=\;\partial_\mu\;e_\nu^a + \omega_{\mu\; b}^a\;e_\nu^b
-\;\frac{\kappa^2}{2}\;\bar\psi_\mu \Gamma^a\psi_\nu -
(\mu\leftrightarrow \nu) , \\
F_{\mu\nu}^{ab}\;&=\;R_{\mu\nu}^{ab} + \frac{1}{\kappa^2}\; \bigl(
e_\mu^a\;e_\nu^b - e_\nu^a \; e_\mu^b\bigr) + \kappa \bar\psi_\mu
\Gamma^{ab} \psi_\nu , \\
\psi_{\mu\nu\alpha}\;&=\;\partial_\mu\;\psi_{\nu\alpha} + \frac 1 4\;
\omega_\mu^{ab} \;\bigl( \Gamma_{ab}\psi_\nu\bigr)_\alpha +
\frac{1}{2\kappa}\; e_\mu^a \bigl( \Gamma_a\psi_\nu\bigr)_\alpha \\
&\phantom{ZeichnungZeichnung} -\;(\mu\leftrightarrow \nu) .
\tag 5.47
\endalign
$$
After some further manipulations and evaluating all the traces, one
obtains the elegant result that the action reduces to that proposed in
[17], namely
$$
\align
I_\varepsilon^5\;=\;\pm\;k\int_{M_4} \biggl[ \frac 1 4\;
\varepsilon_{abcd} F_{\mu\nu}^{ab} &F_{\rho\sigma}^{cd} + \alpha \kappa
\; \bar\psi_{\mu\nu} \Gamma_5 \psi_{\rho\sigma}\biggr] \\
& \times\; dx^\mu \wedge dx^\nu \wedge dx^\rho \wedge dx^\sigma ,
\tag 5.48 \
\endalign
$$
where $\alpha$ is some constant introduced for later convenience, but
here $\alpha=1$. Substituting eqs.~(5.47) into (5.48), one obtains
that
$$
\align
I_\varepsilon^5\;=\;&\pm\;k \biggl\{ \int_{M_4} \varepsilon_{abcd}\;
\frac 1 4 \;\bigl[ R_{\mu\nu}^{ab} R_{\rho\sigma}^{cd} + 2\kappa\;
R_{\mu\nu}^{ab} \bigl( \bar\psi_\rho \Gamma^{cd} \psi_\sigma\bigr)\\
&\phantom{Zeich} + \kappa^2 \bigl(\bar\psi_\mu \Gamma^{ab}
\psi_\nu\bigr) \bigl(\bar\psi_\rho \Gamma^{cd}
\psi_\sigma\bigr)\bigr]\; dx^\mu\wedge dx^\nu\wedge dx^\rho\wedge
dx^\sigma \\
&+\;\frac{4}{\kappa^2}\;\int_{M_4} d^4x\; \sqrt{g} \;e_a^\mu e_b^\nu
\bigl(R_{\mu\nu}^{ab} + \kappa \bigl(\bar\psi_\mu \Gamma^{ab} \psi_\nu
\bigr)\bigr) \\
&+\;4 \alpha \kappa \int_{M_4} \bigl( D_\mu \bar\psi_\nu\bigr)
\Gamma^5\; \bigl(D_\rho \psi_\sigma\bigr) \; dx^\mu \wedge \cdots
\wedge dx^\sigma \\
&+\; 4\alpha \int_{M_4} \bigl( \bar\psi_\mu \Gamma_\nu \Gamma^5 D_\rho
\psi_\sigma\bigr) \; dx^\mu \wedge \cdots \wedge dx^\sigma \\
&+\;\frac{2\alpha}{\kappa}\;\int_{M_4} d^4x\; \sqrt{g}\; \bar\psi_\mu
\Gamma^{\mu\nu} \psi_\nu + \frac{24}{\kappa^4}\;\int_{M_4} d^4 x\,
\sqrt{g} \biggr\} ,
\tag 5.49
\endalign
$$
where
$$
D_\mu \;\psi_\nu\;=\;\partial_\mu\; \psi_\nu \;+\;\frac 1 4 \;
\omega_\mu^{ab}\; \Gamma_{ab}\; \psi_\nu .
$$
After Fierz reshuffling, the term quartic in the gravitino field
$\psi_\mu$ disappears. The remaining terms describe massive
supergravity with a Gauss-Bonnet term. It is an interesting fact that
the action (5.48), with $\alpha = 2$ (!), is invariant under
the same supersymmetry transformation obtained form the variation of
$\Pi(\rho)$, except for $\delta \omega_\mu^{ab}$ which is chosen to
preserve the constraint [18]:
$$
F_{\mu\nu}^{\enskip a}\;=\; 0 .
\tag 5.50
$$
The supersymmetry transformations can be read by substituting (5.10)
into eq.~(3.18):
$$
\align
&\delta\;e_\mu^a\;=\;\kappa\;\bar\varepsilon\;\Gamma^a \psi_\mu,\\
&\delta\;\psi_\mu\;=\;\bigl( \partial_\mu + \frac 1 4\; \omega_\mu^{ab}
\Gamma_{ab} + \frac{1}{2\kappa}\;e_\mu^a \Gamma_a\bigr) \varepsilon,
\tag 5.51
\endalign
$$
and, for $F_{\mu\nu}^{ab}$ and $\psi_{\mu\nu}$, they are
$$
\align
&\delta\;F_{\mu\nu}^{ab}\;=\;\kappa\;\bar\varepsilon \,\Gamma^{ab}
\psi_{\mu\nu}, \\
&\delta\;\psi_{\mu\nu}\;=\;-\;\frac 1 4\;
F_{\mu\nu}^{ab}\;(\Gamma_{ab}\; \varepsilon ).
\tag 5.52
\endalign
$$
When $\alpha=2$ the action (5.48) becomes invariant under the
transformations (5.51) with the constraint (5.50), and the action
corresponds to de Sitter supergravity where the cosmological constant
and the gravitino mass-like term are fixed with respect to each other.
In this case the action (5.48) simplifies to
$$
\align
I_{sg}\;=\;&-\;\biggl[\int_{M_4} d^4x\; \varepsilon^{\mu\nu\rho\sigma}
\,\bigl(\frac 1 4\;\varepsilon_{abcd} R_{\mu\nu}^{\enskip ab}
R_{\rho\sigma}^{\enskip cd} + 8\, \bar\psi_\mu \Gamma_\nu \Gamma_5
D_\rho \psi_\sigma\bigr) \\
&+\;4 \int d^4x\,e\;\bigl(e_a^\mu\, e_b^\nu\,R_{\mu\nu}^{\enskip ab} +
\frac 2 \kappa\;\bar\psi_\mu\; \Gamma^{\mu\nu} \psi_\nu +
\frac{6}{\kappa^4}\bigr)\biggr] .
\tag 5.52
\endalign
$$
The first term in (5.52) is a topological invariant and can be removed
from the action without affecting its invariance. After rescaling
$$
\align
e_\mu^a\;&\to\;r\;e_\mu^a \\
\psi_{\mu\alpha}\;&\to\;\sqrt{r} \; \psi_{\mu\alpha} \\
I_{sg}\;&\to\;8\,r^2\;I_{sg}
\tag 5.53
\endalign
$$
and taking the limit $r\to0$ the action (5.52) reduces to that of
$N$=1 supergravity [19]:
$$
I_{sg}\;=\;-\;\frac{1}{2\kappa^2} \int_{M_4} d^4x\,
e\;e_a^\mu\,e_b^\nu\,R_{\mu\nu}^{\;ab} - \int_{M_4}
d^4x\,\varepsilon^{\mu\nu\rho\sigma} \,\bar\psi_\mu \Gamma_5
\Gamma_\nu D_\rho \psi_\sigma .
\tag 5.54
$$
The significance of the constraint (5.50) and the choice $\alpha=2$ in
the non-commutative construction is not clear to us. It would be
helpful to better understand  this point.

If we had worked instead with the solution (5.44), then additional
terms which are dynamically trivial will be present. We shall not
present the details for this case.

\vskip 1.5truecm

\bf 6. \ub{Conclusions and outlook} \rm .
\medskip
In this paper, we have shown how to construct Chern-Simons forms and
Chern-Simons actions in real, non-commutative geometry; (more detailed
results will appear in [12]). We have illustrated the general,
mathematical results of Sect.~2 by discussing a number of examples.
These examples involve non-commutative spaces described by
$^*$algebras of matrix-valued functions over even-dimensional spin
manifolds. As expected, the Chern-Simons actions associated with these
spaces are manifestly topological (metric-independent). By imposing
constraints on the field configurations on which these action
functionals depend (and choosing convenient gauges) we have been able
to derive the metric-independent, first-order formulation of
four-dimensional gravity theory from a Chern-Simons action over a
``five-dimensional'' non-commutative space. By extending the
mathematical framework, formally, to allow for graded algebras, we
have also recovered an action functional for supergravity.

It would appear to be of interest to study Chern-Simons actions for
more general non-commutative spaces, e.g. those considered in [8], and
to derive from them theories of interest to physics. In this regard,
one should recall that a rather profound theory has the form of a
Chern-Simons theory: Witten's open string field theory [15]. We are
presently attempting to formulate that theory within Connes'
mathematical framework of non-commutative geometry, using a variant of
the formalism developed in Sect. 2.

On the mathematical side, it appears to be of interest to better
understand the topological nature of Chern-Simons actions over general
non-commutative spaces, to understand the connection between the
material presented in Sect.~2 and the theory of characteristic classes
in non-commutative geometry and cyclic cohomology, see [3,10], and, most
importantly, to learn how to quantize Chern-Simons theories in
non-commutative geometry, in order to construct new topological field
theories.

\vskip 1.5truecm
\bf \ub{References}\rm\;:
\medskip

\item{[1]} E. Witten, ``Quantum field theory and the Jones
polynomial'', Comm. Math. Phys. \ub{121}, (1989) 351.

\item{[2]} J. Fr\"ohlich ``Statistics of fields, the Yang-Baxter
equation, and the theory of knots and links'', 1987 Carg\`ese
lectures, in: ``Non-Perturbative Quantum Field Theory'', eds.: G. 't~Hooft
et al., (Plenum Press, New York, 1988).

\item{[3]} M. Dubois-Violette, C.R. Acad. Sc. Paris, \ub{307}, I, (1988)
403.
\item{} J. Fr\"ohlich and C. King, Comm. Math. Phys. \ub{126} (1989)
187.
\item{} E. Guadagnini, M. Martellini and M. Mintchev, Nucl. Phys.
B\ub{330} (1990) 575.

\item{[4]} A. Connes, Publ. Math. I.H.E.S. \ub{62} (1985) 41-144,
``Non-Commutative Geometry'', Academic Press, to appear (1994).

\item{[5]} A. Connes and J. Lott, Nucl. Phys. B (Proc.~Supp.) \ub{18}B
(1990) 29; in Proc. 1991 Summer Carg\`ese Conference, eds.: J.
Fr\"ohlich et al., (Plenum Press, New York 1992).

\item{[6]} R. Coquereaux, G. Esposito-Far\`ese, G. Vaillant, Nucl.
Phys. B\ub{353} (1991) 689;
\item{} M. Dubois-Violette, R. Kerner, J. Madore, J. Math. Phys.
\ub{31} (1990) 316;
\item{} B. Balakrishna, F. G\"ursey and K.C. Wali, Phys. Lett.
\ub{254}B (1991) 430; Phys. Rev. D\ub{46} (1991) 6498.
\item{} R. Coquereaux, G. Esposito-Far\`ese and F. Scheck,
Int.~J.~Mod.~Phys.~A\ub{7} (1992) 6555.

\item{[7]} D. Kastler, ``A detailed account of Alain Connes' version
of the standard model in non-commutative geometry'' I, II and III, to
appear in Rev. Math. Phys.;
\item{} D. Kastler and M. Mebkhout, ``Lectures on non-commutative
differential geometry and the standard model'', World Scientific, to
be published;
\item{} D. Kastler and T. Sch\"ucker, Theor. Math. Phys. \ub{92}
(1992) 522.

\item{[8]} A.H. Chamseddine, G. Felder and J. Fr\"ohlich, Phys. Lett.
B\ub{296} (1992) 109; Nucl. Phys. B\ub{395} (1993) 672.

\item{[9]} A.H. Chamseddine, G. Felder and J. Fr\"ohlich, Commun. Math.
Phys. \ub{155} (1993) 205.

\item{[10]} A.H. Chamseddine and J. Fr\"ohlich ``Some elements of
Connes' non-commutative geometry and space-time geometry'', to appear
in Yang-Festschrift.

\item{[11]} D. Quillen, ``Chern-Simons forms and cyclic cohomology'',
in: ``The Interface of Mathematics and Particle Physics'', D. Quillen,
G. Segal and S. Tsou (eds.), Oxford University Press, Oxford 1990.

\item{[12]} A.H. Chamseddine, J. Fr\"ohlich and O. Grandjean, in
preparation.

\item{[13]} A.H. Chamseddine and D. Wyler, Phys. Lett. B\ub{228} (1989)
75; Nucl. Phys. B\ub{340} (1990) 595;
\item{} E. Witten, ``Surprises with topological field theories'' in
Proc. ``Strings 90'', eds R. Arnowitt et al.

\item{[14]} E. Witten, Nucl. Phys. B\ub{311} (1988) 96; B\ub{323}
(1989) 113;
\item{} A.H. Chamseddine, Nucl. Phys. B\ub{346} (1990) 213.

\item{[15]} E. Witten, Nucl. Phys. B\ub{268} (1986) 253; Nucl. Phys.
B\ub{276} (1986) 291.

\item{[16]} E. Witten, Commun. Math. Phys. \ub{117} (1988) 353.

\item{[17]} A.H. Chamseddine, Ann. Phys. \ub{113} (1978) 219; Nucl.
Phys. B\ub{131} (1977) 494;
\item{} K. Stelle and P. West, J. Phys. A\ub{12} (1979) 1205.

\item{[18]} P. van Nieuwenhuizen, Phys. Rep. \ub{68} (1981) 189.

\item{[19]} D. Freedman, P. van Nieuwenhuizen and S. Ferrara, Phys.
Rev. D\ub{13} (1976) 3214;
\item{} S. Deser and B. Zumino, Phys. Lett. B\ub{62} (1976) 335.

\bye